\documentclass[12pt]{article}
\usepackage{amsmath}
\usepackage{graphicx}
\usepackage{enumerate}
\usepackage{natbib}
\usepackage{url}
\usepackage{amsfonts}
\usepackage{graphicx}
\usepackage{mathrsfs}
\usepackage{amsmath}
\usepackage{amssymb,verbatim}
\usepackage{float}
\usepackage{rotating,color}
\usepackage{graphicx}
\usepackage{amsthm}
\usepackage{har2nat,bm,amsmath,amssymb,amsfonts,amsthm,mathabx,mathrsfs,amsmath}
\newcommand{\blind}{1}
\allowdisplaybreaks[4]

\newcommand{\cp}{\stackrel{\mathcal{P}}{\rightarrow}}

\newcommand{\be}{\begin{equation}}
\newcommand{\ee}{\end{equation}}
\newcommand{\beaa}{\begin{eqnarray*}}
\newcommand{\eeaa}{\end{eqnarray*}}
\newcommand{\bea}{\begin{eqnarray}}
\newcommand{\eea}{\end{eqnarray}}
\newcommand{\lbl}{\label}

\newcommand{\bd}{\bold}
\newtheorem{Rem}{\bf Remark}

\newtheorem{theorem}{Theorem}
\newtheorem{prop}{Proposition}
\newtheorem{coro}{Corollary}
\newtheorem{lemma}{Lemma}

\def\proof {{\noindent\bf Proof.}\quad}
\def\var{\mathrm {var}}

\def\cov{\mathrm {cov}}
\def\cor{\mathrm {cor}}
\def\diag{\mathrm {diag}}
\def\bxi{\bm \xi}

\def\tr{\mathrm {tr}}
\def\U{{\bf U}}
\def\A{{\bf A}}
\def\V{{\bm V}}
\def\M{{\bf M}}

\def\z{{\bm z}}

\def\I{{\bf I}}

\def\bmu{{\bm \mu}}

\def\X{{\bm X}}
\def\x{{\bm x}}

\def\W{{\bm W}}

\def\O{{\bf \Omega}}

\def\tr{\mathrm {tr}}

\def\bms{{\bm\Sigma}}

\def\cp{\mathop{\rightarrow}\limits^{p}}
\def\cd{\mathop{\rightarrow}\limits^{d}}

\def\bmv{\bm \varepsilon}

\def\squarebox#1{\hbox to #1{\hfill\vbox to #1{\vfill}}}
\def\boxit#1{\vbox{\hrule\hbox{\vrule\kern6pt\vbox{\kern6pt#1\kern6pt}\kern6pt\vrule}\hrule}}

\usepackage{setspace}
\begin{document}

\def\spacingset#1{\renewcommand{\baselinestretch}%
{#1}\small\normalsize} \spacingset{1}


\if1\blind
{
  \title{\bf Testing for high-dimensional white noise}
  \author{Long Feng\\
  School of Statistics and Data Science, Nankai University\\
  and\\
    Binghui Liu\\
    School of Mathematics and Statistics \& KLAS, Northeast Normal University\\
    and \\
    Yanyuan Ma \\
    Department of Statistics, Pennsylvania State University}
  \maketitle
} \fi

\if0\blind
{
  \bigskip
  \bigskip
  \bigskip
  \begin{center}
    {\LARGE\bf Testing for high-dimensional white noise}
\end{center}
  \medskip
} \fi

\bigskip
\begin{abstract}
Testing for multi-dimensional white noise is an important subject in statistical inference.
Such test  in the high-dimensional case becomes an open problem waiting to be solved,
especially when the dimension of a time series is comparable to or even
greater than the sample size.
To detect an arbitrary form of departure from high-dimensional
white noise, a few tests have been developed. Some of these tests are
based on max-type statistics, while others are based on sum-type ones. Despite the
progress,  an urgent issue awaits to be resolved: none of these tests is robust
to the sparsity of the serial correlation structure.
 Motivated by this, we propose a
Fisher's combination test by combining the max-type and the sum-type statistics, based on
the established asymptotically independence between them. This combination test
can achieve robustness to the sparsity of the serial correlation structure, and combine the
advantages of the two types of tests. We demonstrate the advantages of
the proposed test over some existing tests through extensive numerical results and
an empirical analysis.
\end{abstract}

\noindent%
\vfill

\newpage
\spacingset{1.5} 
\section{Introduction}

Testing for white noise or serial correlation is an important problem
in statistical modeling and
inference, especially in diagnostic checking for linear regression and
linear time series modeling.
In recent years, researchers are increasingly interested in modeling
high-dimensional time series data,
which are becoming one of the most common data types, and frequently
appear in many applications,
including meteorology, genomics, chemometrics, biological and
environmental research, finance and econometrics,
etc. This brings further challenge to diagnostic checking, as we need
to perform test
for high-dimensional white noise, where the dimension of time series is comparable to or
even greater than the sample size,
i.e. the observed length of the time series.

For univariate time series, many widely used white noise tests have
been proposed in the literature
\citep{Li2004}. Some of these tests have been extended for testing multivariate time series
\citep{Hosking1980,1981Distribution}, which are, however, only suitable for the case that
the dimension of the time series is small compared to the sample size.
Specifically, for univariate time series, the celebrated Box-Pierce portmanteau test and its
variations are considered to be among the most popular omnibus tests
for detecting non-specific forms of deviation from  white noise. These
tests are particularly convenient in practical applications,
due to the fact that they are asymptotically distribution-free and
$\chi^2$-distributed under the
null hypothesis \citep{Li2004,Lutkepohl2005}. However, it is widely
known that when extended to the multivariate cases, these tests
suffer the issue of slow convergence
to their asymptotic null distributions \citep{2019On}.

Recently, multivariate white noise tests have undergone rapid
development. Some new omnibus tests,
such as the tests proposed by \citet{asw066}, \citet{2019On} and
\citet{2020Testing} respectively,
can even deal with high-dimensional time series, where the
dimension of the time series is
comparable to or even greater than the sample size.
Specifically, \citet{asw066} proposed a max-type test for high-dimensional white noise,
using the maximum absolute auto-correlations and cross-correlations of
the component series.
Based on an approximation by the $L_\infty$-norm of a normal random
vector, the critical value
of the max-type test can be evaluated by bootstrapping from a
multivariate normal distribution.
Subsequently, \citet{2020Testing} proposed a rank-based max-type test
for high-dimensional white noise
by using Spearman's rank correlation, and established the limiting
null distribution based on the theory of extreme values.
On the other hand, \citet{2019On} proposed a sum-type test for
high-dimensional white noise, using
the sum of the squared singular values of several lagged sample autocovariance matrices.
Using the random matrix theory, the asymptotic normality
for the test statistic under the null is established under the
Mar\v{c}enko-Pastur asymptotic regime.

In general, the max-type test performs well in the case of sparse
correlations, i.e. there is a small amount of large absolute auto-
or cross-correlations at any nonzero lag. In contrast, the sum-type test
performs well in the case of non-sparse correlations, which
encapsulates the serial correlations within and across all component
series. These two types of tests have their own applicability,
but neither of them can perform well in both cases. In other words, neither of
these two types of tests is applicable in the case of sparse serial
correlations.
 This motivates us to establish a new test, which can
combine the advantages of both types
and is therefore applicable to sparse and non-sparse serial correlations. To this
end, we reconsider the max-type test and
the sum-type test, establish their asymptotic independence, then
combine them to construct a combination test.
{\color{black}It should be noted that the general idea of constructing
  combination tests by
establishing the asymptotic independence of max-type and sum-type statistics
has appeared in independence tests and covariance matrix tests for high-dimensional random
vectors, for example, in \citet{LiXue2015} and \citet{YuLiXue2020}.}

To combine the asymptotically independent tests, we employ the
{\color{black}framework of combining the p-values of independent tests \citep{Ramon1973Asymptotic}.}
In a great deal of early literatures, the problem of combining
independent tests of hypotheses has been widely considered, such as in
\citet{1938The}, \citet{Fisher95}, \citet{1968The} and
\citet{Naik1969The}, to name but a few. Among these methods, the well
known Fisher's combination test proposed in \citet{Fisher95} is usually
regarded as one of the best choices, whose advantages were discussed
in \citet{Ramon1973Asymptotic}.
{\color{black}It should be noted that in addition to the way of combining p-values of independent tests,
there are other ways for combining independent tests. For example, if all test statistics asymptotically
follow Gaussian distributions, a linear combination of the statistics can be used
to construct a combined test statistic. However, in situation where the test statistics have different
types of asymptotic distributions, such as normal distribution and Gumble distribution, it is usually
difficult to directly combine these statistics, hence the way of combining p-values becomes more practical.}

In this paper, for testing high-dimensional white noise, we propose a Fisher's combination
test by combining the p-values of the max-type and sum-type tests,
which is suitable to detect sparse and non-sparse serial correlations.
Using the extreme value theory and the martingale's central limit
theorem, we establish the limiting null distributions of the max-type
and sum-type statistics, respectively. Then, we establish the
asymptotically independence between the two statistics under the null
hypothesis, which enables us to use Fisher's framework of combining independent tests.
Furthermore, we demonstrate the advantages of the proposed Fisher's
combination test over its competitors through extensive numerical
results. In the empirical application, we demonstrate the robust
performance of the proposed Fisher's combination test.


The main contributions of this paper are threefold as follows.
\begin{itemize}
\item First, we established the limiting null distribution of the max-type statistic for testing high-dimensional white noise,
proved that this max-type test is rate-optimal and investigated its local power function in some special cases.
\item Then, we proposed a new sum-type test for testing high-dimensional white noise, where the relationship between the sample size and the dimension is not constrained,
 while the existing sum-type test needs to impose the Mar$\breve{\textrm{c}}$enko-Pastur regime, i.e. the ratio of the sample size to the dimension goes to a constant.
\item Finally, we proved the asymptotic independence between the above max-type and sum-type test statistics
{\color{black}under both Gaussian and non-Gaussian distributions}, which is {\color{black} the most important contribution} of this paper.
Based on this asymptotic independence, we constructed the Fisher's combination test that is suitable to detect either sparse or non-sparse serial correlations.
In particular, we can eliminate the Gaussianity requirement of the error distribution in
  establishing the asymptotic independence. In contrast, all the existing literatures on
establishing such asymptotic independence are limited by the Gaussianity requirement.
Indeed, the proof of such asymptotic independence is more challenging than that of the asymptotic independence between the max-type and sum-type test statistics proposed for
many other important high-dimensional testing problems, such as the high-dimensional cross-sectional independence testing problem \citep{FJLX2020}, the
high-dimensional location testing problem \citep{Xu2016} and the
high-dimensional covariance matrix testing problem \citep{LiXue2015,YuLiXue2020},
which are limited by the Gaussianity requirement.
\end{itemize}

The rest of this paper is organized as follows. In Section \ref{Meth},
we describe the problem of testing for high-dimensional white noise,
reconsider the max-type and sum-type tests, establish their
asymptotic independence and then construct the Fisher's combination
test. In Section \ref{Nume}, we present extensive numerical results of the
proposed test in comparison with some of its competitors,
followed by an empirical application in Section \ref{Appl}. Then, we
conclude the paper with some discussions in Section \ref{Conc},
and relegate the technical proofs to the supplementary.

\section{Methodology}\label{Meth}

\subsection{Testing problem}

Let $\{\bmv_t\}$ be a $p$-dimensional weakly stationary time series with mean zero,
where $\bmv_t=(\varepsilon_{t1},\cdots,\varepsilon_{tp})^\top\in \mathbb{R}^p$.
Let $\bm\Sigma(k)=\left\{{\sigma}_{i j}(k)\right\}_{1 \leqslant i, j
  \leqslant p}\doteq \cov(\bmv_{t+k},\bmv_t)$ denote the
autocovariance
of $\bmv_t$ at lag $k$, and let $\bm\Gamma(k)=\left\{{\rho}_{i
    j}(k)\right\}_{1 \leqslant i, j \leqslant
  p}\doteq\diag\{\bm\Sigma(0)\}^{-1/2}\bm\Sigma(k)\diag\{\bm\Sigma(0)\}^{-1/2}$
denote the autocorrelation of $\bmv_t$ at lag $k$, where for any
matrix $\M$, $\diag(\M)$ denotes the diagonal matrix consisting of
the diagonal elements of $\M$ only. Let $\bms=\{\sigma_{ij}\}_{1
  \leqslant i, j \leqslant p}=\bm\Sigma(0)$ and
$\bm\Gamma=\{\rho_{ij}\}_{1 \leqslant i, j \leqslant
  p}=\bm\Gamma(0)$.
Let $\sigma_i^2=\sigma_{ii}$, for each $i\in \{1,\cdots,p\}$.
$\{\bmv_t\}$ is white noise, if $\bm\Sigma(k)\equiv0$ for all $k\neq 0$.

With the observations $\{\bmv_1, \ldots, \bmv_n\}$, let
$$
\hat{\bm\Gamma}(k) =\left\{\hat{\rho}_{i j}(k)\right\}_{1 \leqslant i,
  j \leqslant p}\doteq\operatorname{diag}\{\hat{\bm\Sigma}(0)\}^{-1 /
  2} \hat{\bm\Sigma}(k) \operatorname{diag}\{\hat{\bm\Sigma}(0)\}^{-1
  / 2}
$$
denote the sample autocorrelation matrix at lag $k$, where
$$
\hat{\bm\Sigma}(k)=\left\{\hat{\sigma}_{i j}(k)\right\}_{1 \leqslant
  i, j \leqslant p}\doteq\frac{1}{n} \sum_{t=1}^{n-k} \bmv_{t+k}
\bmv_{t}^{\mathrm{T}}
$$
denotes the sample autocovariance matrix at lag $k$.

We consider the following testing problem:
\begin{align}\label{problem}
H_0: \{\bmv_t\} \mbox{~is white noise} ~~v.s.~~ H_1: \{\bmv_t\} \mbox{~is not white noise,}
\end{align}
where the dimension of time series $p$ is comparable to or even
greater than the sample size $n$.

\subsection{The max-type test}

Before proposing the Fisher's combination test for testing high-dimensional
white noise, we need to re-examine the max-type and sum-type tests,
which will be combined to construct the combination test.

Since $\bm\Gamma(k)= 0$ for any $k \geqslant 1$ under $H_{0}$, the
max-type test statistic $T_{\textrm{MAX}}$ is defined as
\begin{align}\label{max}
T_{\textrm{MAX}}\doteq\max _{1 \leqslant k \leqslant K} T_{n, k},
\end{align}
which was first proposed by \citet{asw066}, where $T_{n, k}\doteq\max
_{1 \leqslant i, j \leqslant p} n^{1 / 2}\left|\hat{\rho}_{i
    j}(k)\right|$
and $K \geqslant 1$ is an integer. For this max-type test statistic,
\citet{asw066} evaluated the critical value by bootstrapping from a
multivariate normal distribution, which is a
widely recognized practice in the case of sparse correlations.

To establish the Fisher's combination test in this paper, we first derive
the limiting null distribution of the max-type test statistic,
which will be presented in the following Theorem \ref{maxnull}.
Specifically, Theorem \ref{maxnull} states that
$T_{\textrm{MAX}}^2-2\log(Kp^2)+\log \log (Kp^2)$ has an asymptotic
extreme-value distribution when both $n$ and $p$ go to infinity. Hence,
a level-$\alpha$ test with $\alpha\in(0,1)$ will be performed by
rejecting $H_{0}$ when $T_{\textrm{MAX}}^2-2\log(Kp^2)+\log \log
(Kp^2)$ is larger than $q_{\alpha}$, i.e. the $1-\alpha$ quantile of
$G(y)\doteq\exp \left\{-\pi^{-1 / 2} \exp (-y / 2)\right\}$.

In deriving Theorem \ref{maxnull}, we impose the following three conditions.
\begin{itemize}
\item[(C1)] $\varepsilon_{ti}$'s have one of the following two types
  of tails: (i) sub-gaussian-type tails, i.e. there exist some
  constant $\eta>0$ and $M>0$, such that
  $\mathbb{E}e^{\eta\varepsilon_{ti}^2/\sigma_i^2}\le M$ for all $i\in
  \{1,\cdots,p\}$ and $t\in \{1,\cdots,n\}$, where $p$ satisfies $\log
  p=o(n^{1/5})$; (ii) polynomial-type tails, i.e. for some $\gamma_0$
  and $c_1>0$, $p\le c_1n^{\gamma_0}$ and for some $\epsilon>0$ and
  $M>0$,
  $\mathbb{E}|\varepsilon_{ti}/\sigma_i|^{4\gamma_0+4+\epsilon}\le M$
  for all $i\in \{1,\cdots,p\}$ and $t\in \{1,\cdots,n\}$.
\item[(C2)] There exists a positive constant $C$ such that $C^{-1}\le
  \min_{1\le i\le p}\sigma_i^2\le \max_{1\le i\le p}\sigma_i^2\le C$.
\item[(C3)] There exists $\varrho\in (0,1)$ s.t. $|\rho_{ij}|\leq \varrho$ for all $1\leq i<j \leq p$
 with $p\geq 2$.  $|C_p|/p^2\to 0$ as $p \to\infty$ if (C1)-(i)
 holds; and $|C_p|/n^{\epsilon/8}\to 0$ if (C1)-(ii) holds.
Here $C_p\doteq\big\{(i,j): |B_{p,(i,j)}|\geq p^{\kappa_p}\big\}$ and
$B_{p,(i,j)}\doteq\big\{(s,l):|\rho_{ij}\rho_{sl}|\geq \delta_p\big\}$
for $1\leq i,j \leq p$ with $\delta_p,\kappa_p>0$, $\delta_p=o(1/\log
p)$ and $\kappa_p=o(1)$ as $p\to\infty$.
\end{itemize}

\begin{Rem}
Condition (C1) requires that the tail of the distributions of
$\varepsilon_{ti}$'s is sub-gaussian-type or polynomial-type,
which is the same as Condition (C2) or (C2$^*$) used in
\citet{clx2013}. It is a more general moment condition
than the normal distribution assumption. Condition (C2) requires that
all the variances of $\varepsilon_{ti}$'s are bounded.
Condition (C3) requires that the number of variable pairs with strong
correlation cannot be too large.
\end{Rem}

\begin{theorem}\label{maxnull}
Suppose Conditions (C1)-(C3) hold. Then, under $H_0$, for any $y \in \mathbb{R}$, we have
\begin{align*}
\operatorname{P}\left\{T_{\textrm{MAX}}^2-2\log(Kp^2)+\log \log (Kp^2) \leqslant y\right\}\to G(y)
\end{align*}
as $n,p \to\infty$, where $G(y)=\exp \left\{-\pi^{-1 / 2} \exp (-y / 2)\right\}$.
\end{theorem}

We recall that all technical proofs of the theorems and the proposition are relegated to the supplementary.

Let $\mathcal{U}(c)$ be a set of matrices indexed by a constant $c$, which is given by
$$
\left[\big\{\bm\Gamma(1),\cdots,\bm\Gamma(K)\big\} \in \mathbb{R}^{p
    \times Kp}: \max _{1\leq k\leq K,1 \leqslant i<j \leqslant
    p}\left|\rho_{ij}(k)\right| \geqslant c(\log p / n)^{1 /
    2}\right].
$$ Consider the following sparse alternative
\begin{align}\label{HR}
H_{\mathrm{a}}^{R}(c)\doteq\left[F(\bmv_{1},\cdots,\bmv_{n}):
  \big\{\cor_F(\bmv_{t+1},\bmv_t),\cdots,\cor_F(\bmv_{t+K},\bmv_t)\big\}\in
  \mathcal{U}(c)\right],
\end{align}
where $F(\bmv_{1},\cdots,\bmv_{n})$ denotes the joint distribution of
$\{\bmv_{1},\cdots,\bmv_{n}\}$ and $\cor_F$ denotes the
autocorrelation matrix under the joint distribution $F$.
Let $\mathcal{T}_{\alpha}$ denote the set of all measurable size-$\alpha$ tests.

Further, the following theorem characterizes the conditions under
which the power of the proposed max-type test
$\mathbb{I}\{T_{\textrm{MAX}}-2\log(Kp^2)+\log \log (Kp^2)\geq
q_{\alpha}\}$ tends to 1 as $n\to\infty$, under the alternative
$H_{\mathrm{a}}^{R}\left(b_{0}\right)$ for some constant $b_{0}$.

\begin{theorem}\label{opr0} Assume that $\{\bm \varepsilon_t\}$ is a
  strictly stationary time series and the long-run variance
  $\gamma_i^L\doteq
  \lim_{n\to\infty}\var\left(n^{-1/2}\sum_{t=1}^n\varepsilon_{it}^2\right)$
  is bounded for all $1\le i\le p$.
Then, we have
$$\inf_{F(\bmv_{1},\cdots,\bmv_{n}) \in
  H_{\mathrm{a}}^{R}\left(b_{0}\right)}
\operatorname{P}\left[\mathbb{I}\{T_{\operatorname{MAX}}-2\log(Kp^2)+\log
  \log (Kp^2)\geq q_{\alpha}\}=1\right]=1-o(1),$$
for all $b_{0}>3$, where the infimum is taken over the
joint distribution family $H_{\mathrm{a}}^{R}\left(b_{0}\right)$ of
$\{\bmv_{1},\cdots,\bmv_{n}\}$ defined in \eqref{HR}.
\end{theorem}

Theorem \ref{opr0} indicates that the above max-type test can detect
alternatives of order $(\log p/n)^{1/2}$. In Theorem \ref{opr}, we
will show that this test is
rate-optimal, i.e. the rate of the signal gap, $(\log p/n)^{1/2}$,
cannot be further relaxed.

\begin{theorem}\label{opr}
Suppose $c_{0}<1$ is a positive constant, and let $\beta$ be a
positive constant satisfying $\alpha+\beta<1$. If $\log p / n=o(1)$,
we have
$$
\inf_{T_{\alpha} \in \mathcal{T}_{\alpha}} \sup
_{F(\bmv_{1},\cdots,\bmv_{n}) \in
  H_{\mathrm{a}}^{R}\left(c_{0}\right)}
\operatorname{P}\left(T_{\alpha}=0\right) \geqslant 1-\alpha-\beta
$$
as $n,p\to \infty$, where the supremum is taken over the joint
distribution family $H_{\mathrm{a}}^{R}\left(c_{0}\right)$ of
$\{\bmv_{1},\cdots,\bmv_{n}\}$ defined in \eqref{HR}.
\end{theorem}

Theorem \ref{opr} indicates that any measurable size-$\alpha$
test cannot differentiate between the null hypothesis $H_0$ and the
sparse alternative, when
$${\color{black}\max _{1\leq k\leq K,1 \leqslant i<j \leqslant
  p}\left|\rho(k)_{ij}\right| < c_0(\log p / n)^{1 / 2}}$$ for
some constant $c_0<1$.

We now consider a special case to investigate the
  power function of the max-type test when $c_0\in[1,3]$.
Let $\varepsilon_{t1}=z_{t1}+\rho z_{t-1,1}$, where $z_{t1}\sim
\mathcal{N}(0,1)$ and $\rho=O(\sqrt{\log p/n})$. For $i\in
\{2,\cdots,p\}$, $\varepsilon_{ti}$'s are all i.i.d. from
$\mathcal{N}(0,1)$. $\{\varepsilon_{t1}\}_{t=1,\cdots, n}$ are
independent of
$\{\varepsilon_{ti}\}_{t=1,\cdots, n},~i=2,\cdots,p$. By the
Central Limit Theorem, we have that $\sqrt{n-k}\hat{\rho}_{ij}(k)\cd
\mathcal{N}(0,1)$ for $k>1$; $\sqrt{n-1}\hat{\rho}_{ij}(1)\cd
\mathcal{N}(0,1)$ for $i\not=j$ and
${\color{black}\frac{\sqrt{n-1}}{\sqrt{1-\frac{3\rho^2+2\rho^4+3\rho^6}{(1+\rho^2)^4}}}\{\hat{\rho}_{11}(1)-\frac{\rho}{1+\rho^2}\}\cd
\mathcal{N}(0,1)}$ for $k=1$.
Define
$x_{\alpha}=2\log(Kp^2)-\log \log (Kp^2)+q_{\alpha}$. Define
$\mathcal{A}=\{(i,j,k)|1\le i,j\le p, 1\le k\le K\}$. Thus,
\begin{align*}
&\textrm{P}\left\{T_{\operatorname{MAX}}^2-2\log(Kp^2)+\log \log (Kp^2)\geq q_{\alpha}\right\}\\
=&\textrm{P}\left\{\max_{1\le i,j\le p,1\le k\le K}(n-k)\hat{\rho}^2_{ij}(k)\geq x_{\alpha}\right\}\\
\ge&\textrm{P}\left\{(n-1)\hat{\rho}^2_{11}(1)\ge x_{\alpha}\right\}\\
=&\textrm{P}\left\{\left|\mathcal{N}\left(\frac{\sqrt{n}\rho}{{1+\rho^2}},1-{\color{black}\frac{3\rho^2+2\rho^4+3\rho^6}{(1+\rho^2)^4}}\right)\right|\ge \sqrt{x_{\alpha}}\right\}+o(1)\\
=&\textrm{P}\left\{\left|\mathcal{N}\left(\sqrt{n}\rho,1\right)\right|\ge \sqrt{x_{\alpha}}\right\}+o(1)\\
=&
\Phi(\sqrt{n}\rho-\sqrt{x_{\alpha}})+\Phi(-\sqrt{n}\rho-\sqrt{x_{\alpha}})+o(1)
\end{align*}
{\color{black}for sufficiently small $\rho$ and diverging $p$,}
where
$\mathcal{N}(\mu,\sigma^2)$
denotes a random variable
that follows the normal distribution with mean
$\mu$ and variance
$\sigma^2$.

Additionally,
\begin{align*}
&\textrm{P}\left\{T_{\operatorname{MAX}}^2-2\log(Kp^2)+\log \log (Kp^2)\geq q_{\alpha}\right\}\\
=&\textrm{P}\left\{\max_{1\le i,j\le p,1\le k\le K}(n-k)\hat{\rho}^2_{ij}(k)\geq x_{\alpha}\right\}\\
\le&\textrm{P}\left\{ (n-1)\hat{\rho}^2_{11}(1)\ge x_{\alpha}\right\}+\textrm{P}\left\{\max_{(i,j,k)\in \mathcal{A}/{(1,1,1)}}(n-k)\hat{\rho}^2_{ij}(k)\geq x_{\alpha}\right\}\\
=&\Phi(\sqrt{n}\rho-\sqrt{x_{\alpha}})+\Phi(-\sqrt{n}\rho-\sqrt{x_{\alpha}})+\alpha+o(1).
\end{align*}
In the last equality, we used the following derivation.
By the proof of Theorem 1 in \citet{c2018}, we have
\begin{align*}
\textrm{P}\left\{\max_{(i,j,k)\in \mathcal{A}/{(1,1,1)}}(n-k)\hat{\rho}^2_{ij}(k)\geq x_{\alpha}\right\}-\textrm{P}\left(\max_{1\le l\le Kp^2-1}\xi_l^2\geq x_{\alpha}\right)\to 0,
\end{align*}
where $(\xi_1,\cdots,\xi_{Kp^2-1})$ follows the multivariate normal distribution with zero mean and  the same correlation matrix of $\{(n-k)\hat{\rho}^2_{ij}(k)\}_{(i,j,k)\in \mathcal{A}/{(1,1,1)}}$. After some calculations, we have $\cor\left\{(n-k)\hat{\rho}_{11}(k),(n-k-1)\hat{\rho}_{11}(k+1)\right\}\to 2\rho$, $\cor\left\{(n-k)\hat{\rho}_{11}(k),(n-k-2)\hat{\rho}_{11}(k+2)\right\}\to \frac{\rho^2}{1+\rho^2}$ and the other correlations between $(n-k)\hat{\rho}^2_{ij}(k)$ are all zeros. By Theorem 1 in the supplementary and condition $\rho=O(\sqrt{\log p/n})$, we have $\textrm{P}\left(\max_{1\le l\le Kp^2-1}\xi_l^2\geq x_{\alpha}\right)\to \alpha$.


Hence, the power function of the max-type test
is
\begin{align*}
\lim_{n,p\to
  \infty}\Phi(\sqrt{n}\rho-\sqrt{x_{\alpha}})&+\lim_{n,p\to
  \infty}\Phi(-\sqrt{n}\rho-\sqrt{x_{\alpha}})\le \beta_{\textrm{MAX}}(\rho)\\
  \le & \alpha+\lim_{n,p\to
  \infty}\Phi(\sqrt{n}\rho-\sqrt{x_{\alpha}})+\lim_{n,p\to
  \infty}\Phi(-\sqrt{n}\rho-\sqrt{x_{\alpha}}).
  \end{align*}
{\color{black}Note that $x_\alpha\sim 2\sqrt{\log p}$.}
If $\rho=c_0\sqrt{\log p/n}$, we have: (1)
if $0<c_0<2$, $c_0\sqrt{\log p}-\sqrt{x_{\alpha}}\to -\infty$,
$-c_0\sqrt{\log p}-\sqrt{x_{\alpha}}\to -\infty$ and
$\beta_{\textrm{MAX}}(\rho)\in (0,\alpha)$; (2)
if $c_0>2$, $c_0\sqrt{\log p}-\sqrt{x_{\alpha}}\to \infty$,
$-c_0\sqrt{\log p}-\sqrt{x_{\alpha}}\to -\infty$ and
$\beta_{\textrm{MAX}}(\rho)=1$.
On the other hand, if $\sqrt{n}\rho=\sqrt{4\log p+c_1\sqrt{\log p}}$,
we have $\sqrt{n}\rho-\sqrt{x_{\alpha}}\to c_1/4$ and
$-\sqrt{n}\rho-\sqrt{x_{\alpha}}\to -\infty$, then
$\beta_{\textrm{MAX}}(\rho)\in \left(\Phi(c_1/4),\alpha+\Phi(c_1/4)\right)$.

{\color{black}
\begin{Rem}
As mentioned in  \citet{Chang2017Comparing}, the max-type tests based on asymptotic Gumble distributions
usually have conservative size performance. The proposed MAX test also has such limitation.
To solve this problem, some resampling methods can be employed, such as the bootstrap procedure used in the max-type test proposed by \citet{asw066}.
The resampling methods can also relax the conditions imposed on $\bmv_t$'s \citep{asw066}. Unfortunately, such methods generally need to pay a much heavier computational cost,
especially in high-dimensional situations. Hence, whether to use a test based on asymptotic distribution or based on resampling strongly depends on the requirement of computational efficiency.
\end{Rem}
}

\subsection{The sum-type test}

In this subsection, we reconsider the sum-type test, with the test statistic defined as
\begin{align}\label{sum}
T_{\operatorname{SUM}}\doteq\frac{1}{n(n-1)}\sum_{l=1}^K\underset{t\not=s}{\sum\sum}\bmv_t^\top
  \bmv_s\bmv_{t+l}^\top \bmv_{s+l}.
\end{align}
It can be seen from the following Theorem \ref{thsum} and Proposition
\ref{prop1} that under $H_{0}$,
$T_{\operatorname{SUM}}/\hat{\sigma}_S$  has an asymptotically
standard normal distribution when both $n$ and $p$ go to infinity,
where $\hat{\sigma}_S^2\doteq\frac{2K}{n(n-1)}\widehat{\tr(\bms^2)}^2$
and
$\widehat{\tr(\bms^2)}\doteq\frac{1}{n(n-1)}\underset{t\not=s}{\sum\sum}(\bmv_t^\top
\bmv_s)^2$. Hence, a level-$\alpha$ test will be performed by
rejecting $H_{0}$ when $T_{\operatorname{SUM}}/\hat{\sigma}_S$ is
larger than $z_{\alpha}$, i.e. the $1-\alpha$ quantile of the
standard normal distribution.

Note that the test statistic in \eqref{sum} is similar to the
sum-type test statistic proposed by \citet{2019On}.
The differences between them are twofold: first, the test statistic in
\eqref{sum} removes the diagonal elements
$\bmv_t^\top \bmv_t\bmv_{t+l}^\top \bmv_{t+l}$ from the summation to
{\color{black}reduce the requirement of $\bms$}; second, we
use the martingale's central limit theorem to establish the limiting
null distribution of the test statistic,
while \citet{2019On} used the random matrix theory.

In deriving the asymptotic properties of $T_{\operatorname{SUM}}$, we
impose the following two conditions.

\begin{itemize}
\item[(C4)] Let $\bmv_t=\bms^{1/2}\z_t$ under $H_0$, where $\left\{{\z}_{t}\right\}$ with
${\z}_{t}=(z_{t1},\cdots,z_{tp})^\top$ is a sequence of
$p$-dimensional independent random vectors with
independent components $z_{ti}$'s, satisfying $\mathbb{E} z_{ti}=0$,
$\mathbb{E} z_{ti}^{2}=1$ and $\mathbb{E} z_{ti}^{4}<\infty$.
\item[(C5)] $\tr(\bms^4)=o\{\tr^2(\bms^2)\}$.
\end{itemize}

\begin{Rem}
Condition (C5) is mild and holds automatically
if all the eigenvalues of $\bms$ are bounded, {\color{black}i.e. Condition (C5)
is weaker than the condition of bounded eigenvalues of $\bms$ imposed in \citet{2019On},
which indeed reduces the requirement of $\bms$. Note that Condition (C5) is also commonly adopted in the literature of testing
high-dimensional covariance matrices, such as in \citet{czz2010}}.
\end{Rem}

\begin{theorem}\label{thsum}
Suppose Conditions (C4)-(C5) hold. Then, under $H_0$, we have
$T_{\operatorname{SUM}}/\sigma_S\cd \mathcal{N}(0,1)$, where
$\sigma_S^2\doteq\frac{2K}{n(n-1)}\tr^2(\bms^2)$.
\end{theorem}

Following the result in Proposition \ref{prop1}, we use the above
$\widehat{\tr(\bms^2)}$ to estimate $\tr(\bms^2)$.

\begin{prop}\label{prop1}
If $\bmv_t=\bms^{1/2}\z_t$ and $\tr(\bms^4)=o\{\tr^2(\bms^2)\}$,
 then under $H_0$, $\widehat{\tr(\bms^2)}/\tr(\bms^2)\cp 1$.
\end{prop}

Further, we present the asymptotic power function of the sum-type test
$\mathbb{I}(T_{\operatorname{SUM}}/\hat{\sigma}_S\geq z_{\alpha})$,
when an alternative hypothesis $H_{1}$ is specified. Here, we assume
that under $H_{1}$, the observations $\{\bmv_{1}, \cdots, \bmv_{n}\}$
follow a $p$-dimensional first-order vector moving average process,
abbreviated as VMA(1), of the form
\begin{align}\label{h1}
  H_{1}: \bmv_{t}=\bm A_{0} \z_{t}+\bm A_{1} \z_{t-1},
\end{align}
where $\bm A_{0}, \bm A_{1}\in \mathbb{R}^{p \times p}$ are the
coefficient matrices. We consider the asymptotic distribution of
$T_{\operatorname{SUM}}$ in the case of $K=1$ in the following Theorem
\ref{sh1}.

\begin{theorem}\label{sh1}
Under $H_1$ in \eqref{h1} with $K=1$, we have
$(T_{\operatorname{SUM}}-\mu_S)/\sigma_{S1}\cd \mathcal{N}(0,1)$,
where
\begin{align*}
\mu_S\doteq&\tr(\tilde{\bms}_0\tilde{\bms}_1)+\frac{2}{T}\tr^2(\tilde{\bms}_{01}),
             \quad \tilde\bms_{0}\doteq\A_{0}^{\top} \A_{0}, \quad
             \tilde\bms_{1}\doteq\A_{1}^{\top} \A_{1}, \quad
             \tilde{\bms}_{01}\doteq\A_{0}^{\top} \A_{1},
\end{align*}
\begin{align*}
&\sigma_{S1}^2\\
\doteq&\frac{2}{T^2}\tr^2(\tilde\bms_0^2+\tilde\bms_1^2)+\frac{6}{T^2}\tr^2(\tilde\bms_0\tilde\bms_1)\\
&+\frac{4}{T}\left[2
                     \operatorname{tr}(\widetilde{\bms}_{0}
                     \tilde{\bms}_{1})^{2}+(\nu_{4}-3)
                     \operatorname{tr}\left\{D^{2}(\tilde{\bms}_{0}
                     \tilde{\bms}_{1})\right\}\right]\\
&+\frac{8}{T^2}\tr(\tilde\bms_{01}\tilde{\bms}_{01}^\top)\tr(\tilde\bms_0^2+\tilde\bms_1^2)+\frac{16}{T^2}\tr(\tilde\bms_{01}\tilde\bms_1)\tr(\tilde\bms_{01}\tilde\bms_0)\\
&+\frac{16}{T^2}\tr(\tilde\bms_0+\tilde{\bms}_1)\left\{\tr(\tilde\bms_{01}^\top\tilde\bms_{01}\tilde\bms_0)+\tr(\tilde\bms_{01}\tilde\bms_{01}^\top\tilde\bms_1)\right\}\\
&+\frac{16}{T^2}\tr(\tilde\bms_{01}
  )\left\{\operatorname{tr}(\tilde{\bms}_{0}^{2}
  \tilde{\bms}_{01}^{\top})+\operatorname{tr}(\tilde{\bms}_{1}^{2}
  \tilde{\bms}_{01})+2 \operatorname{tr}(\tilde{\bms}_{1}
  \tilde{\bms}_{01} \tilde{\bms}_{0})\right\}\\
&+\frac{4}{T} \operatorname{tr}(\tilde{\bms}_{01}^{\top}
  \tilde{\bms}_{01} \tilde{\bms}_{0}^{2}+\tilde{\bms}_{01}
  \tilde{\bms}_{01}^{\top} \tilde{\bms}_{1}^{2}+2
  \tilde{\bms}_{01}^{\top} \tilde{\bms}_{1} \tilde{\bms}_{01}
  \tilde{\bms}_{0})\\
&+\frac{4}{T} \operatorname{tr}(\widetilde{\bms}_{01}
  \tilde{\bms}_{01}^{\top} \tilde{\bms}_{01}^{\top}
  \tilde{\bms}_{01})+\frac{12}{T^{2}}
  \operatorname{tr}^{2}(\widetilde{\bms}_{01}
  \tilde{\bms}_{01}^{\top})+\frac{16}{T^{2}}
  \operatorname{tr}(\tilde{\bms}_{01})
  \operatorname{tr}(\tilde{\bms}_{01} \tilde{\bms}_{01}^{\top}
  \tilde{\bms}_{01}^{\top})\\
&+\frac{4}{T^{2}} \operatorname{tr}^{2}(\tilde{\bms}_{0}
  \tilde{\bms}_{01})+\frac{4}{T^{2}}
  \operatorname{tr}^{2}(\tilde{\bms}_{1} \tilde{\bms}_{01})+r_n,
\end{align*}
and the remainder $r_n=o(\sigma_{S1}^2)$. Here, for each square matrix $A$, $D(A)$ denotes the diagonal matrix consisting of the main diagonal elements of $A$.
\end{theorem}

Similar to Proposition \ref{prop1}, we have $\widehat{\tr(\bms^2)}/\xi_0\cp 1$ under $H_1$ in \eqref{h1}, where
$\xi_0\doteq\tr(\tilde\bms_0^2+\tilde\bms_1^2)+2\tr(\tilde\bms_{01}^\top \tilde\bms_{01})$.
As a result, the asymptotic power function of the proposed sum-type test $\mathbb{I}(T_{\operatorname{SUM}}/\hat{\sigma}_S\geq z_{\alpha})$ under $H_1$ in \eqref{h1} is approximately equal to
\begin{align}\label{pw}
\beta_{\operatorname{SUM}}\doteq\Phi\left(\frac{\mu_S}{\sigma_{S1}}-z_{\alpha}\frac{\sqrt{2}n^{-1}\xi_0}{\sigma_{S1}}\right).
\end{align}

\subsection{Fisher's combination test}

To combine the proposed max-type and sum-type tests, we propose the
Fisher's combination test,
based on the asymptotic independence between $T_{\operatorname{MAX}}$
and $T_{\operatorname{SUM}}$, to be provided in the following Theorem
\ref{ms}.
Specifically, let $$p_{\operatorname{MAX}}\doteq
1-G\left\{T_{\operatorname{MAX}}^2-2\log(Kp^2)+\log \log
  (Kp^2)\right\}$$ and
$$p_{\operatorname{SUM}}\doteq
1-\Phi\left(T_{\operatorname{SUM}}/\widehat{\sigma}_{S}\right)$$
denote the $p$-values with respect to the test statistics
$T_{\operatorname{MAX}}$ and $T_{\operatorname{SUM}}$
respectively. Based on $p_{\operatorname{MAX}}$ and
$p_{\operatorname{SUM}}$, the proposed Fisher's combination test
rejects $H_0$ at the significance level $\alpha$, if
$$T_{\textrm{FC}}\doteq -2 \log p_{\operatorname{MAX}}-2\log p_{\operatorname{SUM}}$$
is larger than $c_{\alpha}$, i.e. the $1-\alpha$ quantile of the
chi-squared distribution with 4 degrees of freedom
\citep{Fisher95,Ramon1973Asymptotic}.

In deriving the asymptotic independence between
$T_{\operatorname{MAX}}$ and $T_{\operatorname{SUM}}$, we need to
impose an additional condition as follows.
Let $\lambda_{\min}(\bms)$ and $\lambda_{\max}(\bms)$ denote the
minimum and maximum eigenvalues of $\bms$, respectively.


{\color{black}
\begin{itemize}
\item[(C6)] ${{\tr^{-1}(\bms^2)}(\log p)^{\gamma}\max\{\lambda_{\max}(\bms)M_p,M_p^2,M_p^{3/2}\lambda_{\max}^{1/2}(\bms)\}}\to 0$ for
  some positive constant $\gamma>1$, where
    $M_p\doteq\max_{1\leq i \leq p}\sum_{j\not=i}^p\sigma_{ij}^2$.
\end{itemize}

\begin{Rem}
Condition (C6) requires that the covariance between each pair of variables can not be very large, which
holds in many common cases. For example, it automatically holds when all the variables are independent, i.e. $M_p=0$;
if all the eigenvalues of $\bms$ are bounded, $M_p$ is also bounded, hence Condition (C6) holds. In addition, if $\bms$ is a bandable covariance matrices, i.e. $\sigma_{ij}=0$ if $|i-j|>k$ for some fixed integer $k$, and all nonzero $\sigma_{ij}$'s are bounded by $c$, then Condition (C6) holds when $\lambda_{\max}(\bms)(\log p)^{\gamma}{\tr^{-1}(\bms^2)} \to 0$.

\end{Rem}

Note that Condition (C6) is significantly different from Assumption 1 (ii) of \citet{YuLiXue2020}
for testing high-dimensional covariance matrix. For example, when $M_p=0$, we do not need to impose conditions on $\lambda_i(\bms)$'s.
In fact, these two type of conditions cannot contain each other.}

\begin{theorem}\lbl{ms} Suppose $\bmv_t\sim \mathcal{N}(\bm 0,\bms)$
  for $t=1,\cdots,n$ and  Conditions (C2), (C3), (C5) and (C6)
  hold. Then, we have
\begin{align}\label{ai}
\operatorname{P}\left\{T_{\operatorname{MAX}}^2-2\log(Kp^2)+\log \log
  (Kp^2) \leq x,T_{\operatorname{SUM}} / \widehat{\sigma}_{S} \leq
  y\right\} \rightarrow G(x)\cdot \Phi(y),
\end{align}
as $n, p\to \infty$, i.e.  $T_{\operatorname{MAX}}$ and
$T_{\operatorname{SUM}}$ are asymptotically independent.
\end{theorem}

{\color{black}Further, we relax the assumption of Gaussian distribution of $\bmv_t$ in
Theorem \ref{ms} to non-Gaussian distributions, with sub-gaussian-type or polynomial-type
tails. To establish the theoretical result under
non-Gaussian distributions, Condition (C1) is modified as follows.

\begin{itemize}
\item[(C1$'$)] $\varepsilon_{ti}$'s have one of the following two types
  of tails: (i) sub-gaussian-type tails, i.e. there exist some
  constant $\eta>0$ and $M>0$, such that
  $\mathbb{E}e^{\eta\varepsilon_{ti}^2/\sigma_i^2}\le M$ for all $i\in
  \{1,\cdots,p\}$ and $t\in \{1,\cdots,n\}$, where $p$ satisfies $\log
  p=o(n^{1/6})$; (ii) polynomial-type tails, i.e. for some $\gamma_0$
  and $c_1>0$, $p\le c_1n^{\gamma_0}$ and for some $\epsilon>0$ and
  $M>0$,
  $\mathbb{E}|\varepsilon_{ti}/\sigma_i|^{6\gamma_0+6+\epsilon}\le M$
  for all $i\in \{1,\cdots,p\}$ and $t\in \{1,\cdots,n\}$.
\end{itemize}

\begin{theorem}\lbl{ms2} Assume Conditions (C1$'$) and (C2)-(C6)
  hold.  Then, we have
\begin{align}\label{ai2}
\operatorname{P}\left\{T_{\operatorname{MAX}}^2-2\log(Kp^2)+\log \log
  (Kp^2) \leq x,T_{\operatorname{SUM}} / \widehat{\sigma}_{S} \leq
  y\right\} \rightarrow G(x)\cdot \Phi(y),
\end{align}
as $n, p\to \infty$, i.e.  $T_{\operatorname{MAX}}$ and
$T_{\operatorname{SUM}}$ are asymptotically independent.
\end{theorem}
}

\begin{Rem}
{\color{black}Note that relaxing the assumption of Gaussian distribution in establishing
the asymptotically independence between the max-type and the sum-type statistics is
an important contribution of this paper, since all the existing literatures on
establishing such asymptotic independence, including \citet{LiXue2015},
\citet{Xu2016}, \citet{FJLX2020}  and \citet{YuLiXue2020},
are limited by the assumption of Gaussian distribution.
In this paper, we have developed a novel theoretical tool to weaken the Gaussian distribution to
non-Gaussian distributions with sub-gaussian-type or polynomial-type tails.
Its theoretical framework is enlightening, which can be generalized to analogous studies.

}
\end{Rem}

Based on Theorem \ref{ms} {\color{black}or Theorem \ref{ms2}}, we immediately have the following result
for $T_{\textrm{FC}}$.

\begin{coro}
Assume the same conditions as in Theorem \ref{ms} {\color{black}or Theorem \ref{ms2}}, then we have
$T_{\textrm{FC}}\cd \chi_4^2$ as $n,p\to \infty$.
\end{coro}

{\color{black}
Under the alternative hypothesis (\ref{HR}), we have $p_{{\rm MAX}}\to 0$ under the sparse alternatives due to Theorem \ref{opr0}.
On the other hand, under the dense alternative hypothesis (\ref{h1}), we have $p_{{\rm SUM}}\to 0$ if $\mu_S/\sigma_{S1}\to \infty$ due to  Theorem \ref{sh1}.
}

According to the definition of $T_{\textrm{FCP}}$, if $p_{{\rm
    MAX}}\to 0$ or $p_{{\rm SUM}}\to 0$, we have $T_{\textrm{FC}}\to
\infty$, hence we reject the null hypothesis.

\begin{Rem}
{\color{black}Note that Conditions (C2), (C3), (C5) and (C6) are all about $\bms$, which hold automatically if
all the eigenvalues of $\bms$ are bounded. This indicates that these conditions are compatible and the
intersection of these conditions is very conventional, which means that the scope of application of the proposed Fisher's
combination test is relatively broad.}
\end{Rem}

Next, we show that $T_{{\rm SUM}}$ is still asymptotically independent
of $T_{\rm MAX}$ under a specific alternative hypothesis. Based on
this result, we obtain a low bound of the power function of $T_{\rm
  FC}$.

\begin{theorem}\label{th8}
 Assume Conditions (C1$'$) and (C2)-(C5)
  hold. Assume that all eigenvalues of $\bms=\cov(\bmv_t)$ are
  bounded. Let $K=1$. Then, under the alternative hypothesis
  (\ref{h1}) with
$$
\A_0=\left(
\begin{array}{cc}
  {\bf A}_{011} & {\bf 0} \\
  {\bf 0} & {\A_{022}}
\end{array}
\right),~
\A_1=\left(
\begin{array}{cc}
  {\A}_{111} & {\bf 0} \\
  {\bf 0} & {\bf 0}
\end{array}
\right),
$$
$\A_{011}\in\mathbb{R}^{d\times d}$, $\A_{022}\in \mathbb{R}^{(p-d)\times(p-d)}$, ${\bf A}_{111}\in \mathbb{R}^{d\times d}$ and $d=o(p)$, we have
\begin{align}\label{ai3}
&\operatorname{P}\left\{T_{\operatorname{MAX}}^2-2\log(Kp^2)+\log \log
  (Kp^2) \leq x,T_{\operatorname{SUM}} / \widehat{\sigma}_{S} \leq
  y\right\} \nonumber\\
& \rightarrow \operatorname{P}\left\{T_{\operatorname{MAX}}^2-2\log(Kp^2)+\log \log
  (Kp^2) \leq x\right\}\operatorname{P}\left\{T_{\operatorname{SUM}} / \widehat{\sigma}_{S} \leq
  y\right\},
\end{align}
as $n, p\to \infty$, i.e.  $T_{\operatorname{MAX}}$ and
$T_{\operatorname{SUM}}$ are still asymptotically independent.
\end{theorem}

Define a minimal p-values test
$T_{\min}=\min(p_{\operatorname{SUM}},p_{\operatorname{MAX}}
)$. According to Theorem \ref{ms2}, we reject the null hypothesis if
$p_{\operatorname{SUM}} \le 1-\sqrt{1-\alpha}\approx \alpha/2$ or
$p_{\operatorname{MAX}} \le 1-\sqrt{1-\alpha}\approx \alpha/2$.
According to the results in \cite{Ramon1973Asymptotic}, we have that
the power of Fisher's combination test $\beta_{T_{FC} }$ is comparable to the power of the minimal p-values test $\beta_{T_{\min} }$. Thus,
we have
\begin{align*}
\beta_{T_{FC} }\approx \beta_{T_{\min} }\ge\beta_{T_{\operatorname{SUM}} ,
  \alpha/2}+\beta_{T_{\operatorname{MAX}}
  ,\alpha/2}-\beta_{T_{\operatorname{SUM}} ,
  \alpha/2}\beta_{T_{\operatorname{MAX}} ,\alpha/2},
\end{align*}
where the last inequality is based on the inclusion-exclusion
principle and the result of Theorem \ref{th8}, and
$\beta_{T_{\operatorname{SUM}} , \alpha},
\beta_{T_{\operatorname{MAX}} , \alpha}, $ are respectively the power functions of the
sum-type test $T_{\operatorname{SUM}} $ and
max-type test $T_{\operatorname{MAX}} $
 at significant level
$\alpha$.

\section{Numerical results}\label{Nume}

We now present some numerical results to demonstrate the
performance of the proposed max-type test, sum-type test and Fisher's
combined probability test, abbreviated as MAX, SUM and FC
respectively,
as well as their comparison with the sum-type test proposed by
\citet{2019On}, abbreviated as LY.
Note that
we exclude \citet{asw066} and \citet{2020Testing}'s max-type tests from
comparison because they have almost the same performance
as the proposed max-type test in the case of linear correlations.

\subsection{Size performance}

For the cases under the null hypothesis, we let $\bmv_t=\A\z_t$ with
$\z_t=(z_{t1},\cdots,z_{tp})^\top$ and $\A=\{a_{ij}\}_{1\le i,j\le
  p}$. We consider the following two distributions of $\z_t$: (i)
$\z_t\stackrel{i.i.d}{\sim} \mathcal{N}(\bm 0, \I_p)$; (ii)
$z_{ti}\stackrel{i.i.d}{\sim} Ga(4,0.5)-2$,
and the following three settings of $\A$:
\begin{itemize}
\item[(I)] $\A=\bms^{1/2}$, $\bms=\{\sigma_{ij}\}_{1 \leqslant i, j
    \leqslant p}$, $\sigma_{ii}=1$, $i=1,\cdots,p$,
  $\sigma_{ij}=0.5(i-j)^{-2}$ with $i\neq j$;
\item[(II)] $\A=\bms^{1/2}$, $\bms=\{\sigma_{ij}\}_{1 \leqslant i, j
    \leqslant p}$, $\sigma_{ii}=1$, $i=1,\cdots,p$,
  $\sigma_{ij}=0.5\mathbb{I}(|i-j|< 5)$  with $i\neq j$;
\item[(III)] $a_{ij}\stackrel{i.i.d}{\sim} U(-1,1)$.
\end{itemize}

Tables \ref{tab:t1} and \ref{tab:t2} summarize the empirical size
performance of MAX, LY, SUM and FC under settings (I)-(III) with the
distributions (i) and (ii), respectively.
Both of these tables suggest that in terms of size performance, SUM is
the best and MAX is the most conservative, while FC and LY are in between.
In addition, we find that FC has much better size performance than LY in situations
where $p/n$ is relatively large or $K>1$.

\begin{table}[!th]
\begin{center}
\footnotesize
\caption{\label{tab:t1} Size performance in the case of $\bmv_t=\A\z_t$ with distribution (i): $\z_t\stackrel{i.i.d}{\sim} \mathcal{N}(\bm 0, \I_p)$.}
                     \vspace{0.5cm}
                    \renewcommand\tabcolsep{2.0pt}
                     \renewcommand{\arraystretch}{1.2}
                     {
\begin{tabular}{cc|cccc|cccc|cccc}
\hline \hline
&&\multicolumn{4}{c}{$K=1$}&\multicolumn{4}{c}{$K=2$}&\multicolumn{4}{c}{$K=3$}\\ \hline
$n$&$p$&  MAX &LY &SUM &FC & MAX&LY &SUM &FC & MAX&LY &SUM &FC \\\hline
&&\multicolumn{12}{c}{Setting (I)}\\\hline
100&30 &0.016&0.039&0.046&0.043&0.010&0.025&0.043&0.025&0.016&0.026&0.068&0.050\\
100&60 &0.007&0.036&0.050&0.030&0.004&0.012&0.042&0.021&0.006&0.004&0.043&0.024\\
100&90 &0.004&0.022&0.048&0.024&0.010&0.012&0.043&0.025&0.006&0.002&0.047&0.015\\
100&120&0.008&0.016&0.043&0.021&0.006&0.004&0.040&0.016&0.005&0.000&0.048&0.025\\
200&30 &0.012&0.054&0.062&0.052&0.020&0.044&0.052&0.053&0.014&0.040&0.053&0.038\\
200&60 &0.017&0.048&0.056&0.039&0.016&0.035&0.049&0.032&0.020&0.032&0.056&0.047\\
200&90 &0.016&0.032&0.049&0.027&0.011&0.030&0.054&0.037&0.017&0.021&0.056&0.050\\
200&120&0.014&0.031&0.049&0.036&0.017&0.016&0.043&0.036&0.012&0.015&0.065&0.039\\ \hline
&&\multicolumn{12}{c}{Setting (II)}\\\hline
100&30 &0.014&0.038&0.050&0.043&0.009&0.037&0.052&0.038&0.005&0.022&0.051&0.032\\
100&60 &0.003&0.030&0.049&0.021&0.010&0.017&0.050&0.025&0.009&0.011&0.049&0.025\\
100&90 &0.008&0.027&0.045&0.031&0.011&0.011&0.056&0.040&0.006&0.007&0.046&0.021\\
100&120&0.008&0.017&0.042&0.028&0.002&0.008&0.052&0.022&0.004&0.001&0.035&0.022\\
200&30 &0.018&0.045&0.053&0.043&0.014&0.045&0.051&0.043&0.010&0.030&0.048&0.035\\
200&60 &0.016&0.041&0.052&0.033&0.016&0.031&0.056&0.039&0.013&0.017&0.047&0.027\\
200&90 &0.016&0.028&0.040&0.023&0.017&0.023&0.049&0.043&0.017&0.009&0.050&0.031\\
200&120&0.015&0.035&0.050&0.037&0.011&0.013&0.043&0.028&0.011&0.008&0.052&0.030\\ \hline
&&\multicolumn{12}{c}{Setting (III)}\\\hline
100&30 &0.014&0.038&0.050&0.043&0.009&0.037&0.052&0.038&0.005&0.022&0.051&0.032\\
100&60 &0.003&0.030&0.049&0.021&0.010&0.017&0.050&0.025&0.009&0.011&0.049&0.025\\
100&90 &0.008&0.027&0.045&0.031&0.011&0.011&0.056&0.040&0.006&0.007&0.046&0.021\\
100&120&0.008&0.017&0.042&0.028&0.002&0.008&0.052&0.022&0.004&0.001&0.035&0.022\\
200&30 &0.018&0.045&0.053&0.043&0.014&0.045&0.051&0.043&0.010&0.030&0.048&0.035\\
200&60 &0.016&0.041&0.052&0.033&0.016&0.031&0.056&0.039&0.013&0.017&0.047&0.027\\
200&90 &0.016&0.028&0.040&0.023&0.017&0.023&0.049&0.043&0.017&0.009&0.050&0.031\\
200&120&0.015&0.035&0.050&0.037&0.011&0.013&0.043&0.028&0.011&0.008&0.052&0.030\\
\hline
\hline
\end{tabular}}
\end{center}
\end{table}

\begin{table}[!th]
\begin{center}
\footnotesize
\caption{\label{tab:t2} Size performance in the case of $\bmv_t=\A\z_t$ with distribution (ii): $z_{ti}\stackrel{i.i.d}{\sim} Ga(4,0.5)-2$.}
                     \vspace{0.5cm}
                    \renewcommand\tabcolsep{2.0pt}
                     \renewcommand{\arraystretch}{1.2}
                     {
\begin{tabular}{cc|cccc|cccc|cccc}
\hline \hline
&&\multicolumn{4}{c}{$K=1$}&\multicolumn{4}{c}{$K=2$}&\multicolumn{4}{c}{$K=3$}\\ \hline
$n$&$p$&  MAX &LY &SUM &FC & MAX&LY &SUM &FC & MAX&LY &SUM &FC \\\hline
&&\multicolumn{12}{c}{Setting (I)}\\\hline
100&30 &0.011&0.050&0.058&0.049&0.008&0.038&0.060&0.036&0.012&0.018&0.053&0.031\\
100&60 &0.010&0.023&0.043&0.027&0.009&0.014&0.034&0.020&0.009&0.007&0.040&0.020\\
100&90 &0.006&0.026&0.044&0.022&0.006&0.013&0.058&0.031&0.008&0.002&0.058&0.027\\
100&120&0.011&0.017&0.046&0.028&0.006&0.012&0.046&0.019&0.009&0.001&0.047&0.022\\
200&30 &0.018&0.046&0.052&0.054&0.012&0.047&0.064&0.041&0.013&0.027&0.039&0.031\\
200&60 &0.014&0.047&0.060&0.037&0.013&0.038&0.054&0.035&0.012&0.029&0.045&0.041\\
200&90 &0.015&0.034&0.049&0.039&0.015&0.026&0.055&0.035&0.010&0.009&0.042&0.025\\
200&120&0.022&0.040&0.059&0.038&0.014&0.010&0.043&0.022&0.016&0.010&0.057&0.029\\ \hline
&&\multicolumn{12}{c}{Setting (II)}\\\hline
100&30 &0.010&0.049&0.061&0.048&0.003&0.032&0.053&0.035&0.011&0.021&0.044&0.032\\
100&60 &0.012&0.023&0.042&0.029&0.013&0.022&0.053&0.038&0.008&0.016&0.063&0.037\\
100&90 &0.007&0.029&0.053&0.028&0.008&0.008&0.051&0.024&0.003&0.002&0.055&0.031\\
100&120&0.005&0.017&0.052&0.025&0.005&0.006&0.040&0.016&0.009&0.002&0.051&0.023\\
200&30 &0.016&0.045&0.057&0.044&0.019&0.045&0.059&0.045&0.019&0.044&0.060&0.051\\
200&60 &0.013&0.042&0.055&0.038&0.007&0.023&0.044&0.030&0.015&0.019&0.050&0.029\\
200&90 &0.015&0.033&0.046&0.037&0.013&0.027&0.070&0.040&0.016&0.015&0.049&0.031\\
200&120&0.009&0.024&0.048&0.029&0.018&0.027&0.050&0.042&0.008&0.013&0.053&0.028\\ \hline
&&\multicolumn{12}{c}{Setting (III)}\\\hline
100&30 &0.010&0.049&0.061&0.048&0.003&0.032&0.053&0.035&0.011&0.021&0.044&0.032\\
100&60 &0.012&0.023&0.042&0.029&0.013&0.022&0.053&0.038&0.008&0.016&0.063&0.037\\
100&90 &0.007&0.029&0.053&0.028&0.008&0.008&0.051&0.024&0.003&0.002&0.055&0.031\\
100&120&0.005&0.017&0.052&0.025&0.005&0.006&0.040&0.016&0.009&0.002&0.051&0.023\\
200&30 &0.016&0.045&0.057&0.044&0.019&0.045&0.059&0.045&0.019&0.044&0.060&0.051\\
200&60 &0.013&0.042&0.055&0.038&0.007&0.023&0.044&0.030&0.015&0.019&0.050&0.029\\
200&90 &0.015&0.033&0.046&0.037&0.013&0.027&0.070&0.040&0.016&0.015&0.049&0.031\\
200&120&0.009&0.024&0.048&0.029&0.018&0.027&0.050&0.042&0.008&0.013&0.053&0.028\\
\hline
\hline
\end{tabular}}
\end{center}
\end{table}

\subsection{Power comparison}
In this subsection, we compare the empirical power performance of the above four tests.
For the cases under the alternative hypothesis, we only consider the above distribution
(i) to avoid redundancy and consider the following three new settings of $\bmv_t$:
\begin{itemize}
\item[(IV)] VAR(1) model: $\bmv_t=\A\bmv_{t-1}+\z_t$;
\item[(V)] VMA(1) model: $\bmv_t=\z_t+\A\z_{t-1}$;
\item[(VI)] VARMA(1) model: $\bmv_t=0.5\A\bmv_{t-1}+\z_t+0.5\A\z_{t-1}$.
\end{itemize}
Here ``VAR(1)", ``VMA(1)" and ``VARMA(1)" are the abbreviations of
1-order vector autoregressive process, vector moving average process and
vector autoregressive moving average process, respectively.
Let $\A=\{a_{ij}\}_{1\le i,j\le p}$. For the alternative
hypothesis, we let the first $a_{ij}\not=0$ for $1\le i,j\le m$ and
$a_{ij}=0$ otherwise. Note that $m$ controls the signal strength and
sparsity of $\A$. For the VAR(1) model, if $m=1$, $a_{ij}\sim
U(0.4,0.8)$; if $2\le m\le 10$, $a_{ij}\sim U(-1.4/m,1.4/m)$. For the
VMA(1) model, if $m=1$, $a_{ij}\sim U(0.4,0.9)$; if $2\le m\le 10$,
$a_{ij}\sim U(-1.8/m,1.8/m)$. For the VARMA(1) model, if $m=1$,
$a_{ij}\sim U(0.4,0.8)$; if $2\le m\le 10$, $a_{ij}\sim
U(-1.6/m,1.6/m)$.
Specifically, as $m$ decreases, both the signal strength and sparsity
of $\A$ increase. Let $n=200$, $p\in \{60,90\}$ and $K\in \{1,2,3\}$.

\begin{figure}[!th]
\centering
\caption{\label{fig1}{Power curves of the involved tests with
    $m=1,2,\cdots,10$ and $(n,p)=(200,60)$.}}
\includegraphics[width=1\textwidth]{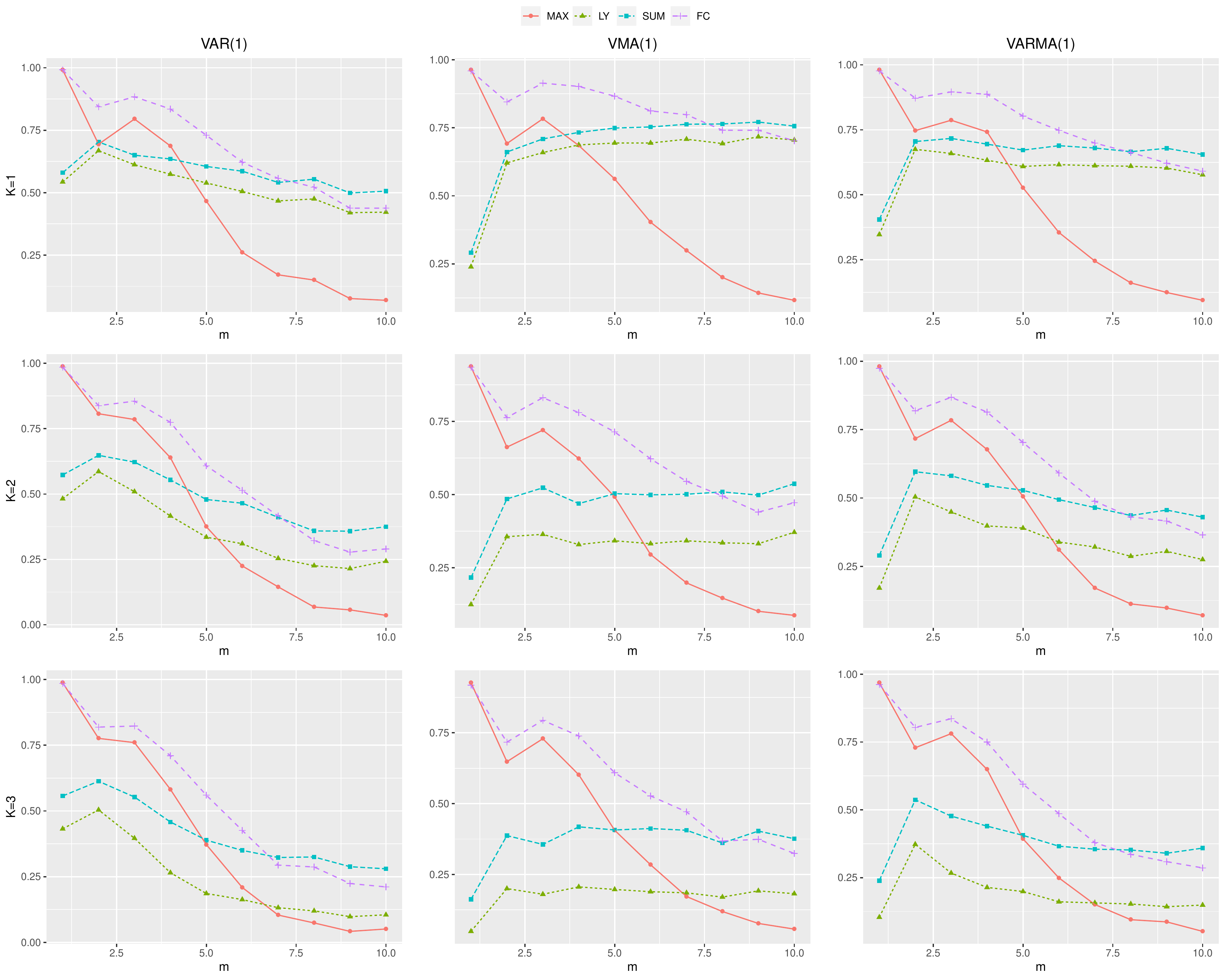}
\end{figure}

\begin{figure}[!th]
\centering
\caption{\label{fig2}{Power curves of the involved tests with
    $m=1,2,\cdots,10$ and $(n,p)=(200,90)$.}}
\includegraphics[width=1\textwidth]{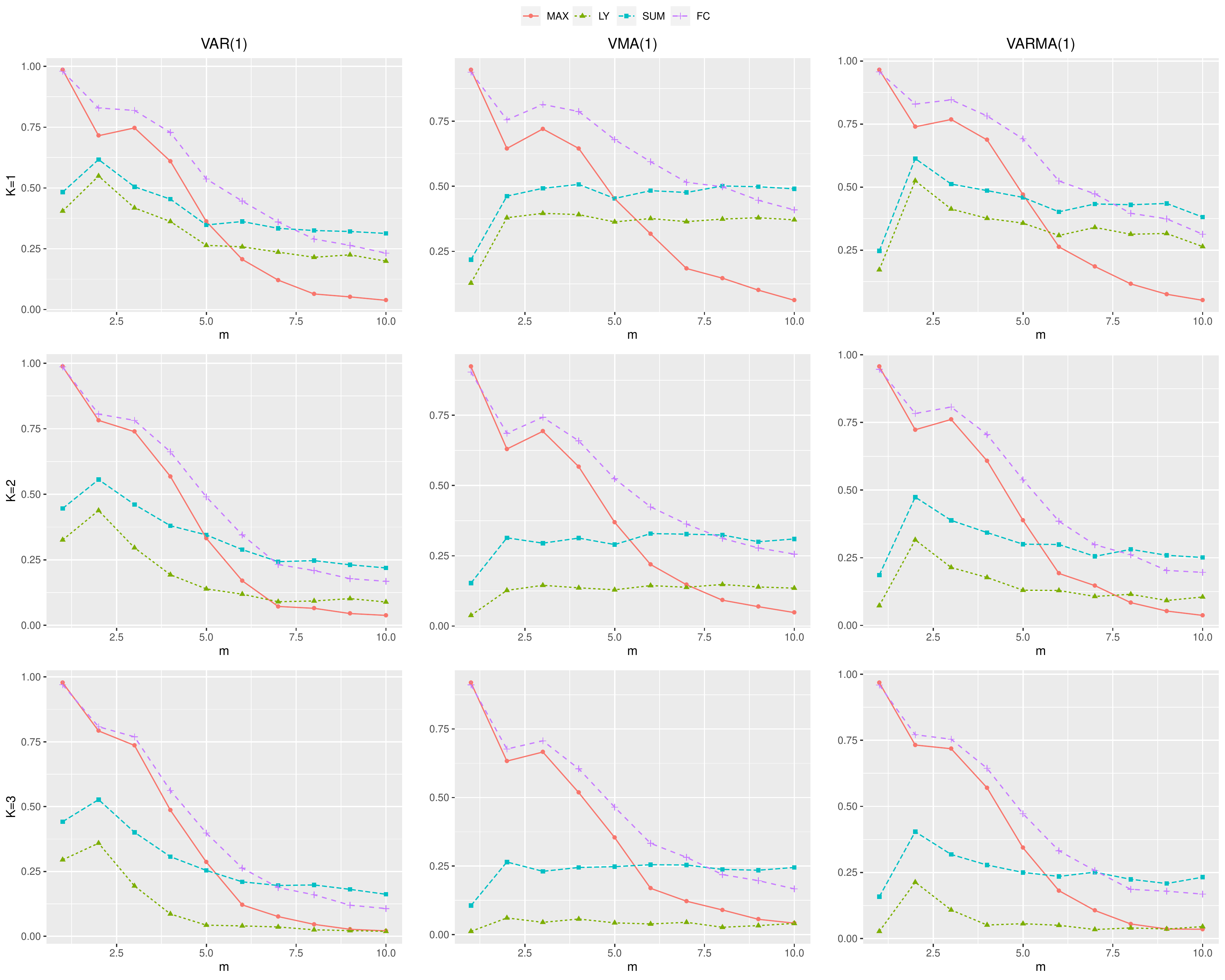}
\end{figure}

Figures \ref{fig1} and \ref{fig2} present the empirical power curves
of MAX, LY, SUM and FC under settings (IV)-(VI) and distribution (i)
for $(n,p)=(200,60)$ and $(200,90)$, respectively. In each panel of
these figures, the abscissa $m$ varies between 1 and 10, corresponding
to the power performance of the involved tests with different
signal strength and sparsity of $\A$. Both figures suggest that in
terms of empirical power performance, FC is better than its
competitors in most cases whether $\A$ is sparse or non-sparse, which has
robust performance due to the combination of the advantages of both
MAX and SUM. {\color{black}Although FC cannot outperform its competitors in all cases,
it is indeed applicable to both sparse and non-sparse cases of $\A$.
 In contrast, MAX generally fails when $\A$ is dense enough,
while SUM and LY generally fail when $\A$ is very sparse.

Note that in all the above simulation studies, the ways of
setting up $\bms$ are very common, all of which satisfy the condition of
bounded eigenvalues of $\bms$. Hence, these simulation settings for $\bms$ satisfy all
the conditions imposed on $\bms$ in the theoretical results established above.}

\section{Application}\label{Appl}

In this section, we are interested in testing whether the error series
$\{\bm \varepsilon_t\}$ under the Fama-French three-factor model \citep{Fama1993Common} is
white noise, i.e.
\begin{align}\label{app}
H_0:\{\bm \varepsilon_t\} \mbox{~is white noise versus}~H_1:
  \{\bm \varepsilon_t\} \mbox{~is not white noise},
\end{align}
where $\bm
\varepsilon_t=(\varepsilon_{t1},\cdots,\varepsilon_{tp})^\top$ and $p$
is the number of securities.
The Fama-French three-factor model is one of the most popular factor
pricing models in finance, which has the explicit form
\[Y_{ti}=r_{ti}-r_{ft}=\alpha_i+\beta_{i1}
  (r_{mt}-r_{ft})+\beta_{i2}SMB_t+\beta_{i3} HML_t+\varepsilon_{ti} \]
for $t\in \{1,\cdots,n\}$ and $i\in \{1,\cdots,p\}$, where $r_{ti}$ is
the return of the $i$-th security at time $t$, $r_{ft}$ is the
risk free rate at time $t$,
$Y_{ti}=r_{ti}-r_{ft}$ is the excess return of the $i$-th security at
time $t$ and $r_{mt}$ is the market return at time $t$.

We collected the return data of the securities in the S\&P 500 index and
considered two forms of data compilation. First, we compiled the monthly
returns on all the securities that constitute the S\&P 500 index
each month over the period from January 2005 to November 2018. Because the
securities that make up the index change over time,
we only consider $p=374$ securities that were included in the S\&P
500 index during the entire period. A total of $T=165$ consecutive
observations were obtained. The time series data on the safe rate of return, and the market
factors are obtained from Ken French's data library web page. The
one-month US treasury bill rate is chosen as the risk-free rate
($r_{ft}$). The value-weighted return on all NYSE, AMEX, and NASDAQ
stocks from CRSP is used as a proxy for the market return ($r_{mt}$).
The average return on the three small portfolios minus the average
return on the three big portfolios ($SMB_t$), and the average return
on two value portfolios minus the average return on two growth
portfolios ($HML_t$) are calculated based on the stocks listed on
the NYSE, AMEX and NASDAQ.

Second, we compiled the weekly returns on all the securities that constitute
the S\&P 500 index over the period from January 2005 to
November 2018. The weekly data were calculated using the security prices on Fridays.
Similar to the monthly data, we only considered a total of $p=381$ stocks that were
included in the S\&P 500 index during the entire period.
We formed a total of $T=716$ weekly return rates for each stock during this
period after excluding the Fridays that happened to be holidays.

Under these two forms of data compilation, we test the hypotheses in \eqref{app}
using the proposed Fisher's combination test as well as its competitors, respectively.
Specifically, we let
\[\hat{\varepsilon}_{ti}\doteq Y_{ti}-\hat{\alpha}_i-\hat{\beta}_{i1}
  (r_{mt}-r_{ft})-\hat{\beta}_{i2}SMB_t-\hat{\beta}_{i3} HML_t, \]
where $\hat{\alpha}_i$, $\hat{\beta}_{i1}$, $\hat{\beta}_{i2}$ and $\hat{\beta}_{i3}$
are the ordinary least squares (OLS) estimators of ${\alpha}_i$,
${\beta}_{i1}$, ${\beta}_{i2}$ and ${\beta}_{i3}$,
respectively. To demonstrate the usefulness of the proposed test, we
treat
$\hat{\bm\varepsilon}_{t}=(\hat{\varepsilon}_{t1},\cdots,\hat{\varepsilon}_{tp})^\top$
as the observation of $\bm\varepsilon_{t}$, instead of considering the testing problem
within the Fama-French three-factor model.

We use the sliding window method for the subsequent application. Given a fixed length $n$,
for each $\tau\in \{1,\cdots, T-n\}$, we implement each of the
involved tests on the data compiled
from the period from $\tau$ to $\tau+n-1$, where $\{\tau,\cdots, \tau+
n-1\}$ is the sliding window of length $n$.
Then, we record the rate of rejecting the null hypothesis in these
$T-n$ testing results corresponding to
the $T-n$ sliding windows.

Due to the great complexity and diversity of the financial market, the Fama-French
three-factor model is only an approximation and the three included
factors may often fail to accurately describe
the generating mechanism of the excess returns of a large number of
securities. Nevertheless, it has
played an important role in pricing analysis of securities. This
certainly motivates the investigation on whether
a certain factor pricing model is sufficient and whether more advanced
factor pricing models with more explanatory
factors are needed. It is not irrational to suspect that the
Fama-French three-factor model is not sufficient
hence the null hypothesis may not be true, especially in the
high-dimensional situations. To this end, of course a testing
method with more tendency of rejection may be considered to perform
better, as long as the test can control the effect size.

Table \ref{tab:t3} summarizes the rejecting rates for each $n\in \{40,50,60,70\}$
and each form of data compilation, where the prescribed integer $K\in
\{2,3\}$ is used to establish the proposed test statistics.
It suggests that for the weekly data, FC, MAX and SUM are more
inclined to reject the null hypothesis than LY,
where FC is the most powerful and MAX is the second.
This may be due to the stronger dependence of securities on time series of weekly data,
compared with the monthly data, which may lead to some correlation
matrices with larger signal strength.
In such a circumstance, both the max-type and sum-type of tests can
perform well, and the Fisher's combined probability test FC
outperform them as a combination of them.

Compared with weekly time series, the time dependence of monthly time
series is much weaker,
which leads to some correlation matrices with much weaker signal
strength. This may be the reason why MAX fails to deal with
the monthly data, while SUM, FC and LY have good performance. In
particular, SUM outperform all the remaining methods in such
circumstance.

Overall, SUM, LY and MAX can only have good performance in their
respective suitable situations, while FC can have robust performance
in both situations.

\begin{table}[!h]
\begin{center}
\footnotesize
\caption{\label{tab:t3} Rejecting rates for the weekly and monthly data respectively.}
                     \vspace{0.5cm}
                    \renewcommand\tabcolsep{5.0pt}
                     \renewcommand{\arraystretch}{1.2}
                     {
\begin{tabular}{cc|cccc|cccc}
\hline \hline
&&\multicolumn{4}{c}{$K=2$}&\multicolumn{4}{c}{$K=3$}\\ \hline
$n$&$p$& MAX&LY &SUM &FC & MAX&LY &SUM &FC \\\hline
&&\multicolumn{8}{c}{Weekly data}\\\hline
40&381&0.74&0.19&0.83&0.93&0.72&0.24&0.63&0.86\\
50&381&0.86&0.26&0.79&0.98&0.83&0.32&0.63&0.94\\
60&381&0.88&0.34&0.73&0.98&0.87&0.39&0.57&0.97\\
70&381&0.89&0.46&0.58&0.98&0.90&0.41&0.64&0.99\\\hline
&&\multicolumn{8}{c}{Monthly data}\\\hline
40&374&0.00&0.45&0.64&0.36&0.00&0.45&0.86&0.67\\
50&374&0.00&0.55&0.68&0.55&0.01&0.62&0.90&0.75\\
60&374&0.00&0.50&0.70&0.49&0.01&0.59&0.95&0.84 \\
70&374&0.21&0.56&0.69&0.53&0.18&0.69&0.99&0.93\\
\hline \hline
\end{tabular}}
\end{center}
\end{table}

\section{Conclusion}\label{Conc}

Driven by the task of testing for high-dimensional white noise, we adopt the strategy of
combining independent tests of hypotheses. In particular, we employ the well known Fisher's
combination test to combine the max-type and sum-type statistics,
which is guaranteed to be valid by
the asymptotic independence between the two types of statistics. Through extensive numerical
results, we demonstrate that the proposed test has clear advantages in power comparison,
due to its robustness to sparsity of the serial correlation structure.
Furthermore, via an empirical application, we demonstrate
the robust performance of the proposed test in testing white noise of the return data of the
S$\&$P 500 securities under the Fama-French three-factor model.


\newpage


\begin{center}
    {\LARGE\bf Supplementary Material of ``Testing for high-dimensional white noise"}
\end{center}
\doublespacing
\section{Technical proofs}

\subsection{Proof of Theorem \ref{maxnull}}

First, we present some technical results for the proof of Theorem \ref{maxnull}.

We restate Theorem 2 in \citet{FJLLX2021} as the following Proposition \ref{theorem_2}, in which the following condition is imposed.

\begin{itemize}
\item[(CA1)]
Let $\bms=\{\sigma_{ij}\}_{1\le i,j\le p}$. For some $\varrho\in (0,1)$, assume $|\sigma_{ij}|\leq \varrho$ for all $1\leq i<j \leq p$
 and $p\geq 2$.  Suppose $\{\delta_p: p\geq 1\}$ and $\{\kappa_p: p\geq 1\}$ are  positive constants with
 $\delta_p=o(1/\log p)$  and $\kappa=\kappa_p\to 0$   as $p\to\infty$. For $1\leq i \leq p$, define
 $B_{p,i}=\big\{1\leq j \leq p: |\sigma_{ij}|\geq \delta_p\big\}$  and $C_p=\big\{1\leq i \leq p: |B_{p,i}|\geq p^{\kappa}\big\}$.
 Assume that $|C_p|/p\to 0$ as $p \to\infty$.
\end{itemize}

\begin{prop}\lbl{theorem_2} Suppose $(Z_1,\cdots,Z_p)^\top \sim \mathcal{N}(\bm 0,\bms)$ and Condition (CA1) holds. Then, we have $\max_{1\leq i \leq p}Z_i^2-2\log p +\log\log p$
converges to a Gumbel distribution with cdf $G(x)=\exp(-\frac{1}{\sqrt{\pi}}e^{-x/2})$ as $p\to \infty$.
\end{prop}

The following lemma is from the proof of Theorem 2 in \citet{FJLLX2021}.

\begin{lemma}\label{atd}
Suppose $(Z_1,\cdots,Z_p)^\top \sim \mathcal{N}(\bm 0,\bms)$ and Condition (CA1) holds. For any $x\in \mathbb{R}$ and any $1\leq t \leq p$, let
\begin{align*}\lbl{Chicago_Seattle}
\alpha_t= \sum \operatorname{P}(|Z_{i_1}|>z, \cdots, |Z_{i_t}|>z),~z=\big(2\log p - \log \log p+x\big)^{1/2},
\end{align*}
where the sum runs over all $i_1<\cdots < i_t$ with $i_1\cdots, i_t\in D_p=\{1,\cdots,p\}\setminus C_p$. Then,
\begin{align}
\lim_{p\to\infty}\alpha_t=\frac{1}{t!}\pi^{-t/2}e^{-tx/2}.
\end{align}
\end{lemma}

\begin{lemma}\label{bern} (Bernstein's inequality) Let $X_{1}, \ldots, X_{n}$ be independent centered random variables a.s. bounded by $A<\infty$ in absolute value. Let $\sigma^{2}=n^{-1} \sum_{i=1}^{n} \mathbb{E}\left(X_{i}^{2}\right).$ Then for all $x>0$,
$$
\operatorname{P}\left(\sum_{i=1}^{n} X_{i} \geq x\right) \leq \exp \left(-\frac{x^{2}}{2 n \sigma^{2}+2 A x / 3}\right).
$$
\end{lemma}

Define $\hat{\sigma}_{i}^2=\frac{1}{n}\sum_{t=1}^n\varepsilon_{ti}^2$ and $\sigma_{i}^2=\var(\varepsilon_{ti})$.
\begin{lemma}\label{vari}
Suppose Condition (C2) holds. Then, under $H_0$,
\begin{itemize}
\item[(1)] if (C1)-(i) holds, we have
\begin{align}
\operatorname{P}\left(\max_{1\le i\le p}\left|\hat{\sigma}_i^2-\sigma_i^2\right|\ge C\frac{\epsilon_n}{\log p}\right)=O(p^{-1}),
\end{align}
\item[(2)] if (C1)-(ii) holds,
\begin{align}
\operatorname{P}\left(\max_{1\le i\le p}\left|\hat{\sigma}_i^2-\sigma_i^2\right|\ge C\frac{\epsilon_n}{\log p}\right)=O(n^{-\epsilon/8}),
\end{align}
\end{itemize}
as $\epsilon_{n}\doteq\max \left\{(\log p)^{1 / 6} / n^{1 / 2},(\log p)^{-1}\right\} \rightarrow 0$.
\end{lemma}
\proof We first assume that (C1)-(i) holds. It suffices to show that, for any $\delta>0$,
\begin{align}\label{le1e}
\mathrm{P}\left\{\max _{1\le i\le p}\left|\frac{1}{n} \sum_{k=1}^{n}\left(\varepsilon_{ki}^{2}-\mathbb{E}\varepsilon_{ki}^{2}\right)\right| \geq C \sqrt{\frac{\log p}{n}}\right\}=O\left(p^{-{\delta}}\right).
\end{align}
Define
$$
\tilde{\varepsilon}_{ki}\doteq\varepsilon_{ki} \mathbb{I}\left\{\left|\varepsilon_{ki}\right| \leq \tau \sqrt{\log (p+n)}\right\},
$$
where $\tau$ is sufficiently large. We have
\begin{align}\label{s6}
\left|\mathbb{E}\varepsilon_{ki}^{2}-\mathbb{E}\tilde{\varepsilon}_{ki}^{2}\right| & \leq C\left(\mathbb{E}\varepsilon_{ki}^{4} \mathbb{E}\left[I\left\{\left|\varepsilon_{ki}\right| \geq \tau \sqrt{\log (p+n)}\right\}\right]\right)^{1 / 2} \nonumber\\
& \leq C(n+p)^{-\tau^{2} \eta / 2}\left\{\mathbb{E} \varepsilon_{ki}^{4} \exp \left(2^{-1} \eta \varepsilon_{ki}^{2}\right)\right\}^{1 / 2} \nonumber\\
& \leq C(n+p)^{-\tau^{2} \eta / 2},
\end{align}
where $C$ does not depend on $n$ and $p$. Thus, it follows that
\begin{align*}
 &\mathrm{P}\left\{\max _{1\le i\le p}\left|\frac{1}{n} \sum_{k=1}^{n}\left(\varepsilon_{ki}^{2}-\mathbb{E}\varepsilon_{ki}^{2}\right)\right| \geq C \sqrt{\frac{\log p}{n}}\right\} \\
 \leq & \mathrm{P}\left\{\max _{1\le i\le p}\left|\frac{1}{n} \sum_{k=1}^{n}\left(\tilde{\varepsilon}_{ki}^{2}-\mathbb{E}\tilde{\varepsilon}_{ki}^{2}\right)\right| \geq 2^{-1} C \sqrt{\left.\frac{\log p}{n}\right)}\right\}\\
 & +n p \mathrm{P}\left\{\left|\varepsilon_{ki}\right| \geq \tau \sqrt{\log (p+n)}\right\},
\end{align*}
where
$$
n p \mathrm{P}\left(\left|\varepsilon_{ki}\right| \geq \tau \sqrt{\log (p+n)}\right) \leq n p(n+p)^{-\tau^{2} \eta} \mathbb{E} \exp \left(\eta \varepsilon_{ki}^{2}\right)=O\left(p^{-{\delta}}\right).
$$
Let $t=\eta\left(8 \tau^{2}\right)^{-1} \sqrt{\log p / n}$ and $\tilde{Z}_{ki}=  \tilde{\varepsilon}_{ki}^{2}-\mathbb{E}  \tilde{\varepsilon}_{ki}^{2}$. Then, we have
\begin{align*}
& \mathrm{P} \left\{\frac{1}{n} \sum_{k=1}^{n}\left(  \tilde{\varepsilon}_{ki}^{2}-\mathbb{E}  \tilde{\varepsilon}_{ki}^{2}\right) \geq C \sqrt{\frac{\log p}{n}}\right\}\\
 \leq &\exp \left(-C t \sqrt{n \log p}\right) \prod_{k=1}^{n} \mathbb{E} \exp \left(t \tilde{Z}_{ki}\right) \\
 \leq &\exp \left(-C t \sqrt{n \log p}\right) \prod_{k=1}^{n}\left\{1+\mathbb{E} t^{2} \tilde{Z}_{ki}^{2} \exp \left(t\left|\tilde{Z}_{ki}\right|\right)\right\} \\
 \leq &\exp \left\{-C t \sqrt{n \log p}+\sum_{k=1}^{n} \mathbb{E} t^{2} \tilde{Z}_{ki}^{2} \exp \left(t\left|\tilde{Z}_{ki}\right|\right)\right\} \\
 \leq & \exp \left\{-C \eta\left(8 \tau^{2}\right)^{-1} \log p+c_{\tau, \eta} \log p\right\}  \leq  C p^{-{\delta}},
\end{align*}
where $c_{\tau, \eta}$ is a positive constant depending only on $\tau$ and $\eta$. Similarly, we can show that
$$
\mathrm{P}\left\{\frac{1}{n} \sum_{k=1}^{n}\left(  \tilde{\varepsilon}_{ki}^{2}-\mathbb{E} \tilde{\varepsilon}_{ki}^{2}\right) \leq-C \sqrt{\frac{\log p}{n}}\right\} \leq C p^{-{\delta}},
$$
which leads to (\ref{le1e}).

It remains to prove this lemma under (C2)-(ii). Define
$$\hat{\varepsilon}_{ki}^2=\varepsilon_{ki}^2 \mathbb{I}\left\{\left|\varepsilon_{ki}^2\right| \leq n /(\log p)^{8}\right\}.$$
Then, as in $(\ref{s6})$, we can show that $\left|\mathbb{E} \varepsilon_{ki}^2-\mathbb{E} \hat{\varepsilon}_{ki}^2\right| \leq C n^{-\gamma_{0} / 4}.$
It follows that
\begin{align*}
&\mathrm{P} \left(\max _{1\le i\le p}\left|\sum_{k=1}^{n}\left(\varepsilon_{ki}^2-\mathbb{E} \varepsilon_{ki}^2\right)\right| \geq \frac{n \epsilon_{n}}{\log p}\right) \\
 \leq &\mathrm{P}\left(\max _{1\le i\le p}\left|\sum_{k=1}^{n}\left(\hat{\varepsilon}_{ki}^2-\mathbb{E} \hat{\varepsilon}_{ki}^2\right)\right| \geq 2^{-1}\frac{n \epsilon_{n}}{\log p}\right)+\mathrm{P}\left(\max _{i, k}\left|\varepsilon_{ki}^2\right| \geq \frac{n}{(\log p)^{8}}\right) \\
 \leq &C p^{2} \exp \left\{-C(\log p)^{4}\right\}+C n^{-\epsilon / 8},
\end{align*}
where the last inequality follows from Lemma \ref{bern} and (C2)-(ii). Then, the proof of this lemma is completed. \hfill $\square$

Define $\tilde{\Gamma}(k) =\left\{\tilde{\rho}_{i j}(k)\right\}_{1 \leqslant i, j \leqslant p}\doteq\operatorname{diag}\{{\bm\Sigma}(0)\}^{-1 / 2} \hat{\bm\Sigma}(k) \operatorname{diag}\{{\bm\Sigma}(0)\}^{-1 / 2}.$ We have the following results for $\tilde{\rho}_{ij}(k)$.
\begin{lemma}\label{st}
Under $H_0$, we have
\begin{itemize}
\item[(1)] if (C1)-(i) holds, we have
\begin{align*}
\operatorname{P}\left\{\max _{(i,j,k)\in \Lambda}n\tilde{\rho}_{ij}^2(k)\ge x^2\right\}\le C|\Lambda|\{1-\Phi(x)\}+O(p^{-2}).
\end{align*}
\item[(2)] if (C1)-(ii) holds, we have
\begin{align*}
\operatorname{P}\left\{\max _{(i,j,k)\in \Lambda}n\tilde{\rho}_{ij}^2(k)\ge x^2\right\}\le C|\Lambda|\{1-\Phi(x)\}+{O(n^{-\epsilon/8}})
\end{align*}
\end{itemize}

uniformly for $0\le x\le \sqrt{8\log p}$ and $\Lambda\subset\{(i,j,k): 1\le i,j\le p,1\le k\le K\}$.
\end{lemma}
\proof Rewrite
\begin{align*}
n\tilde{\rho}_{ij}^2(k)=&\frac{1}{n}\left(\sum_{t=1}^{n-k}\sigma_i^{-1}\sigma_{j}^{-1}\varepsilon_{ti}\varepsilon_{t+k,j}\right)^2\\
=&\frac{\left(\sum_{t=1}^{n-k}\sigma_i^{-1}\sigma_{j}^{-1}\varepsilon_{ti}\varepsilon_{t+k,j}\right)^2}
{\sum_{t=1}^{n-k}\sigma_i^{-2}\sigma_{j}^{-2}\varepsilon_{ti}^2\varepsilon_{t+k,j}^2}
\times\frac{1}{n}\sum_{t=1}^{n-k}\sigma_i^{-2}\sigma_{j}^{-2}\varepsilon_{ti}^2\varepsilon_{t+k,j}^2.
\end{align*}
By the self-normalized large deviation theorem for independent random variables
(Theorem 1 in \citet{jsq2003}), we can get
\begin{align}\label{le22}
\max _{1 \leq i \leq j \leq p, 1\le k\le K} \mathrm{P}\left\{\frac{\left(\sum_{t=1}^{n-k}\sigma_i^{-1}\sigma_{j}^{-1}\varepsilon_{ti}\varepsilon_{t+k,j}\right)^2}
{\sum_{t=1}^{n-k}\sigma_i^{-2}\sigma_{j}^{-2}\varepsilon_{ti}^2\varepsilon_{t+k,j}^2} \geq x^{2}\right\} \leq C\{1-\Phi(x)\}
\end{align}
uniformly for $0 \leq x \leq(8 \log p)^{1 / 2}$. By Lemma 3 in \citet{clx2013}, we have
\begin{itemize}
\item[(1)] if (C1)-(i) holds, we have
\begin{align}\label{le211}
\operatorname{P}\left(\left|\frac{1}{n}\sum_{t=1}^{n-k}\sigma_i^{-2}\sigma_{j}^{-2}\varepsilon_{ti}^2\varepsilon_{t+k,j}^2-1\right|\ge C\frac{{\color{black}\epsilon_n}}{\log p}\right)=O(p^{-\delta}),
\end{align}
\item[(2)] if (C1)-(ii) holds,
\begin{align}\label{le212}
\operatorname{P}\left(\left|\frac{1}{n}\sum_{t=1}^{n-k}\sigma_i^{-2}\sigma_{j}^{-2}\varepsilon_{ti}^2\varepsilon_{t+k,j}^2-1\right|\ge C\frac{{\color{black}\epsilon_n}}{\log p}\right)=O(n^{-{\color{black}\epsilon/8}}),
\end{align}
\end{itemize}
for any $\delta>0$. Thus, together (\ref{le22}), (\ref{le211}) with (\ref{le212}), we can complete the proof of this lemma. \hfill $\square$

Now, we are ready to present the proof of Theorem \ref{maxnull}.

\noindent{\bf Proof of Theorem \ref{maxnull}}
Define $
\tilde{T}_{n}\doteq\max _{1 \leqslant k \leqslant K} \tilde{T}_{n, k},
$
where $\tilde T_{n, k}\doteq\max _{1 \leqslant i, j \leqslant p} n^{1 / 2}\left|\tilde{\rho}_{i j}(k)\right|$. Conditional on the event $\{\max_{1\le i\le p}\left|\hat{\sigma}_i^2-\sigma_i^2\right|\ge C\frac{\epsilon_n}{\log p}\}$, we have
\begin{align*}
|T_n^2-\tilde{T}_n^2|\le C\tilde{T}_n\frac{\epsilon_n}{\log p}.
\end{align*}
Thus, by Lemma \ref{vari}, we only need to show that
$$
\operatorname{P}\left\{\tilde T_n-2\log(Kp^2)+\log \log (Kp^2) \leqslant y\right\}\to\exp \left\{-\pi^{-1 / 2} \exp (-y / 2)\right\}.
$$
Restate that $\tilde{\rho}_{ij}(k)=\frac{1}{n}\sum_{t=1}^{n-k}\sigma_i^{-1}\sigma_{j}^{-1}\varepsilon_{ti}\varepsilon_{t+k,j}$. Without loss of generality, we assume that $\sigma_i=1$ for all $i$. After some simply calculation, we have $\cov\left\{n^{1/2}\tilde{\rho}_{ij}(k),n^{1/2}\tilde{\rho}_{sw}(l)\right\}=\rho_{is}\rho_{jw}\mathbb{I}(k=l)$. Thus, we rearrange $\{n^{1/2}\tilde{\rho}_{ij}(k)\}_{1\le i,j\le p, 1\le k\le K}$ as $\{\nu_1,\cdots,\nu_{N}\}$ with $N=Kp^2$. Let $\bm \nu=(\nu_1,\cdots,\nu_{N})^\top$, then
\begin{align}
\cov(\bm \nu)=\{a_{ij}\}_{1\le i,j\le N}\doteq\diag\{{\color{black}\bm\Gamma(0)\otimes\bm\Gamma(0),\cdots,\bm\Gamma(0)\otimes\bm\Gamma(0)}\},
\end{align}
where $\otimes$ denotes the Kronecker product and $\diag\{\bm\Gamma(0)\otimes\bm\Gamma(0),\cdots,\bm\Gamma(0)\otimes\bm\Gamma(0)\}$ denotes the block diagonal matrix
composed by $\bm\Gamma(0)\otimes\bm\Gamma(0),\cdots,\bm\Gamma(0)\otimes\bm\Gamma(0)$.
For $1\leq i \leq N$, define  $B_{N,i}=\big\{1\le j\le N: |a_{ij}|\geq \delta_p\big\}$
and $C_N=\big\{i: |B_{N,i}|\geq p^{\kappa}\big\}$. By the definition of $a_{ij}$, we have $|C_{N}|=|C_p|$.

Define $z=\{2\log(N)-2\log\log(N)+y\}^{1/2}$.
By Lemma \ref{st}, if (C1)-(i) holds, we have
\begin{align*}
\textrm{P}(|\nu_i| \geq z)\le C\{1-\Phi(z)\}+O(p^{-2})\le C \frac{1}{\sqrt{\pi}}\frac{e^{-y/2}}{N}+O(p^{-2});
\end{align*}
 if (C1)-(ii) holds, we have
\begin{align*}
\textrm{P}(|\nu_i| \geq z)\le C\{1-\Phi(z)\}+O({\color{black}n^{-\epsilon/8}})\le C \frac{1}{\sqrt{\pi}}\frac{e^{-y/2}}{N}+O(n^{-\epsilon/8}).
\end{align*}
Thus, by Condition (C3),
$$
\textrm{P}\left(\max _{i \in C_{N}}\left|\nu_{i}\right|>z\right) \leq\left|C_{N}\right| \cdot \textrm{P}(|\nu_i| \geq z) \rightarrow 0
$$
{\color{black}as $n$, $p \rightarrow \infty.$} Set $D_{N}=\left\{1 \leq i \leq N: \left|B_{N, i}\right|<p^{\kappa}\right\}.$ By  Condition (C3), $\left|D_{N}\right| / N \rightarrow 1$ as $N \rightarrow \infty$.
Obviously,
\begin{align*}
\textrm{P}\left(\max _{i \in D_{N}}\left|\nu_{i}\right|>z\right) & \leq \textrm{P}\left(\max _{1 \leq i \leq N}\left|\nu_{i}\right|>z\right) \leq \textrm{P}\left(\max _{i \in D_{N}}\left|\nu_{i}\right|>z\right)+\textrm{P}\left(\max _{i \in C_{N}}\left|\nu_{i}\right|>z\right).
\end{align*}
Therefore, to prove this theorem, it is enough to show
$$
\lim _{N \rightarrow \infty} \textrm{P}\left(\max _{i \in D_{N}}\left|\nu_{i}\right|>z\right)=1-\exp \left(-\frac{1}{\sqrt{\pi}} e^{-x / 2}\right)
$$
as $N \rightarrow \infty$.

We redefine $\nu_{s}=n^{1/2}\tilde{\rho}_{ij}(k)=n^{-1/2}\sum_{t=1}^{n-k}\varepsilon_{ti}\varepsilon_{t+k,j}= n^{-1/2}\sum_{t=1}^{n-k} Z_{ts}$. Let
$\breve{Z}_{ts}={Z}_{ts}\mathbb{I}({Z}_{ts}\le \tau_n)-\mathbb{E}\{{Z}_{ts}I({Z}_{ts}\le \tau_n)\}$. Here, $\tau_n=\eta^{-1}8M\log(p+n)$, if (C1)-(i) holds, and $\tau_n=\sqrt{n}/(\log p)^8$, if (C1)-(ii) holds. Define $\breve{\nu}_i= n^{-1/2}\sum_{t=1}^{n-k} \breve Z_{ts}$ and $q=|D_N|$.
If (C1)-(i) holds, then
\begin{align*}
&\max _{1 \leq k \leq q} \frac{1}{\sqrt{n}} \sum_{l=1}^{n} \mathbb{E}\left|Z_{lk}\right| \mathbb{I}\left\{\left|Z_{lk}\right| \geq \eta^{-1} 8 M \log (p+n)\right\} \\
\leq & C \sqrt{n} \max _{1 \leq l \leq n} \max _{1 \leq k \leq q} \mathbb{E}\left|Z_{lk}\right| \mathbb{I}\left\{\left|Z_{lk}\right| \geq \eta^{-1} 8 M \log (p+n)\right\} \\
\leq & C \sqrt{n}(p+n)^{-4} \max _{1 \leq l \leq n} \max _{1 \leq k \leq q} \mathbb{E}\left|Z_{lk}\right| \exp \left\{\eta\left|Z_{lk}\right| /\left(2 M\right)\right\} \\
\leq & C \sqrt{n}(p+n)^{-4}.
\end{align*}
If (C1)-(ii) holds, then
\begin{align*}
&\max _{1 \leq k \leq q} \frac{1}{\sqrt{n}} \sum_{l=1}^{n} \mathbb{E}\left|Z_{lk}\right| \mathbb{I}\left\{\left|Z_{lk}\right| \geq \sqrt{n} /(\log p)^{8}\right\} \\
\leq & C \sqrt{n} \max _{1 \leq l \leq n} \max _{1 \leq k \leq q} \mathbb{E}\left|Z_{lk}\right| \mathbb{I}\left\{\left|Z_{lk}\right| \geq \sqrt{n} /(\log p)^{8}\right\}
\leq C n^{-\gamma_{0}-\epsilon / 8}.
\end{align*}
Thus, we have
\begin{align*}
&\operatorname{P}\left\{\max_{1\le k\le q}|\nu_k-\breve{\nu}_k|\ge (\log p)^{-1}\right\}\\
\le& \operatorname{P}\left(\max_{1\le k\le q}\max_{1 \le l\le n}|Z_{lk}|\ge \tau_n\right)\\
\le &n \operatorname{P}(\max_{1\le i,j\le p}\max_{1\le s\le K}|\varepsilon_{ti}\varepsilon_{t+s,j}|\ge \tau_n)\\
\le &n\max_{1\le i,j\le p}\max_{1\le s\le K}\left\{\operatorname{P}(|\varepsilon_{ti}|\ge \tau_n^{1/2})+\operatorname{P}(|\varepsilon_{t+s,j}|\ge \tau_n^{1/2})\right\}\\
=&O(p^{-1}+n^{-\epsilon/8}).
\end{align*}

Note that
\begin{align}
\left|\max _{1 \leq k \leq q} \nu_{k}^{2}-\max _{1 \leq k \leq q} \breve{\nu}_{k}^{2}\right|\leq 2 \max _{1 \leq k \leq q}| \breve{\nu}_{k}|\max _{1 \leq k \leq q}| \nu_{k}-\breve{\nu}_{k}|+\max _{1 \leq k \leq q}\left|\nu_{k}-\breve{\nu}_{k}\right|^{2}.
\end{align}
Therefore, to prove this theorem, it is enough to show
$$
\lim _{N \rightarrow \infty} \operatorname{P}\left(\max _{i \in D_{N}}\left|\breve\nu_{i}\right|>z\right)=1-\exp \left(-\frac{1}{\sqrt{\pi}} e^{-x / 2}\right)
$$
as $N \rightarrow \infty$. Then, by Bonferroni inequality,
$$
\sum_{t=1}^{2 k}(-1)^{t-1} \alpha_{t} \leq \textrm{P}\left(\max _{i \in D_{N}}\left|\breve{\nu}_{i}\right|>z\right) \leq \sum_{t=1}^{2 k+1}(-1)^{t-1} \alpha_{t}
$$
for any $k \geq 1$, where
$$
\alpha_{t}\doteq\sum^* \textrm{P}\left(\left|\breve{\nu}_{i_{1}}\right|>z, \cdots,\left|\breve{\nu}_{i_{t}}\right|>z\right)
$$
for $1 \leq t \leq N,$ and the sum runs over all $i_{1}<\cdots<i_{t}$ and $i_{1}, \cdots, i_{t} \in D_{N} .$
First, we will prove that
\begin{align}\label{at}
\lim _{N \rightarrow \infty} \alpha_{t}=\frac{1}{t !} \pi^{-t / 2} e^{-t y / 2}
\end{align}
for each $t \geq 1 .$
All the assumptions in Theorem 1.1 in {\color{black} \citet{z1987}} are satisfied. Thus, we have
\begin{align*}
 &\sum^* \operatorname{P}\left\{\left|Z_{i_{1}}\right|>z+\zeta_n(\log N)^{-1/2}, \cdots,\left|Z_{i_{t}}\right|>z+\zeta_n(\log N)^{-1/2}\right\}\\
 &-\left(\begin{array}{c} |D_N|\\ t\end{array}\right) c_1t^{5/2}\exp\left\{-\frac{n^{1/2}\zeta_n}{c_2t^{3}(\log N)^{1/2}}\right\}\\
\le &\sum^* \operatorname{P}\left\{\left|\breve{\nu}_{i_{1}}\right|>z, \cdots,\left|\breve{\nu}_{i_{t}}\right|>z\right\}\\
\le &\sum^* \operatorname{P}\left\{\left|Z_{i_{1}}\right|>z-\zeta_n(\log N)^{-1/2}, \cdots,\left|Z_{i_{t}}\right|>z-\zeta_n(\log N)^{-1/2}\right\}\\
&+\left(\begin{array}{c} |D_N|\\ t\end{array}\right)c_1t^{5/2}\exp\left\{-\frac{n^{1/2}\zeta_n}{c_2t^{3}(\log N)^{1/2}}\right\},
\end{align*}
where $(Z_{i_1},\cdots,Z_{i_t})^\top$ follows a multivariate normal distribution with mean zero and the same covariance matrix with $(\breve{\nu}_{i_1},\cdots,\breve{\nu}_{i_t})^\top$. By Lemma \ref{atd}, we have
\begin{align*}
\sum^* \operatorname{P}\left\{\left|Z_{i_{1}}\right|>z+\zeta_n(\log N)^{-1/2}, \cdots,\left|Z_{i_{t}}\right|>z+\zeta_n(\log N)^{-1/2}\right\}\to \frac{1}{t !} \pi^{-t / 2} e^{-t y / 2},\\
\sum^* \operatorname{P}\left\{\left|Z_{i_{1}}\right|>z-\zeta_n(\log N)^{-1/2}, \cdots,\left|Z_{i_{t}}\right|>z-\zeta_n(\log N)^{-1/2}\right\}\to \frac{1}{t !} \pi^{-t / 2} e^{-t y / 2},
\end{align*}
with $\zeta_n\to 0$ and $N\to \infty$.
Additionally,
\begin{align*}
&\left(\begin{array}{c} |D_N|\\ t\end{array}\right)c_1t^{5/2}\exp\left\{-\frac{n^{1/2}\zeta_n}{c_2t^{3}(\log N)^{1/2}}\right\}\\
\le &C \left(\begin{array}{c} N\\ t\end{array}\right)t^{5/2}\exp\left\{-\frac{n^{1/2}\zeta_n}{c_2t^{3}(\log N)^{1/2}}\right\}\to 0
\end{align*}
for $\zeta_n\to 0$ sufficiently slow. Thus, we have
\begin{align*}
\sum^* \operatorname{P}\left(\left|\breve{\nu}_{i_{1}}\right|>z, \cdots,\left|\breve{\nu}_{i_{t}}\right|>z\right)\to \frac{1}{t !} \pi^{-t / 2} e^{-t y / 2}.
\end{align*}
Let $N \rightarrow \infty$, we have
$$
\begin{aligned}
&\sum_{t=1}^{2 k}(-1)^{t-1} \frac{1}{t !}\left(\frac{1}{\sqrt{\pi}} e^{-x / 2}\right)^{t} \\
 \leq &\liminf _{N \rightarrow \infty} \operatorname{P}\left(\max _{i \in D_{N}}\left|\breve{\nu}_{i}\right|>z\right) \\
 \leq &\limsup _{N \rightarrow \infty} \operatorname{P}\left(\max _{i \in D_{N}}\left|\breve{\nu}_{i}\right|>z\right) \leq \sum_{t=1}^{2 k+1}(-1)^{t-1} \frac{1}{t !}\left(\frac{1}{\sqrt{\pi}} e^{-x / 2}\right)^{t}
\end{aligned}
$$
for each $k \geq 1 .$ By letting $k \rightarrow \infty$ and using the Taylor expansion of the function $1-e^{-x}$, we obtain the result. \hfill $\square$

\subsection{Proof of Theorem \ref{opr0}}
Define $(i_0,j_0,k_0)=\arg\max_{1\le i<j\le p, 1\le k\le K}|\rho_{ij}(k)|$. Let $\gamma_{il}=\mathbb{E}(\varepsilon_{it}^2\varepsilon_{i,t+l}^2)-\sigma_{i}^4$. By the condition of Theorem \ref{opr0}, the long-run variance $\gamma_i^L=\lim_{n\to\infty}\{\gamma_{i0}+2\sum_{l=1}^n (1-l/n)\gamma_{il}\}$ is bounded. Hence,
\begin{align*}
\mathbb{E}\left(\frac{1}{n}\sum_{t=1}^{n-k_0}\varepsilon_{i_0t}^2\right)=&\sigma_{i_0}^2,\\
\var\left(\frac{1}{n}\sum_{t=1}^{n-k_0}\varepsilon_{i_0t}^2\right)=&\frac{1}{n}\{\mathbb{E}(\varepsilon_{i_0t}^4)-\sigma_{i_0}^4\}+\frac{2}{n^2}\sum_{s<t}\{\mathbb{E}(\varepsilon_{i_0t}^2\varepsilon_{i_0s}^2)-\sigma_{i_0}^4\}\\
=&\frac{1}{n}\gamma_{i_0}+\frac{2}{n}\sum_{l=1}^{n-1}(1-l/n)\gamma_{i_0l}\to 0.
\end{align*}
Thus, we have $\frac{1}{n}\sum_{t=1}^{n-k_0}\varepsilon_{i_0t}^2\cp \sigma_{i_0}^2$. Similarly, we have
\begin{center}
$\frac{1}{n}\sum_{t=1}^{n-k_0}\varepsilon_{j_0t}^2\cp \sigma_{j_0}^2$ and $\frac{1}{n}\sum_{t=1}^{n-k_0}\varepsilon_{i_0t}\varepsilon_{j_0,t+k_0}\cp \sigma_{i_0j_0}(k_0)$.
\end{center}
Thus, $\hat{\rho}_{i_0j_0}(k_0)\cp \rho_{i_0j_0}(k_0)$. As $n,p\to \infty$, we have
\begin{align*}
&\operatorname{P}\left\{\max_{1\le k\le K}\max_{1\le i<j\le p}n\hat{\rho}^2_{ij}(k)-2\log(Kp^2)+\log\log(Kp^2)\ge q_{\alpha}\right\}\\
\ge &\operatorname{P}\left\{n\hat{\rho}^2_{i_0j_0}(k_0)-2\log(Kp^2)+\log\log(Kp^2)\ge q_{\alpha}\right\}\\
\to &\operatorname{P}\left\{n{\rho}^2_{i_0j_0}(k_0)-2\log(Kp^2)+\log\log(Kp^2)\ge q_{\alpha}\right\}=1
\end{align*}
by the condition ${\rho}_{i_0j_0}(k_0)\ge3\sqrt{\log p/n}>\sqrt{2\log (Kp^2)/n}$.\hfill $\square$

\subsection{Proof of Theorem \ref{opr}}
Define $\bm X=(\bmv_{\cdot 1}^\top,\cdots,\bmv_{\cdot p}^\top)^\top\in \mathbb{R}^{d}$, $d=np$ and $\bmv_{\cdot i}=(\varepsilon_{1i},\cdots,\varepsilon_{ni})^\top$.
Consider the Gaussian setting and a simple alternative set of parameters
$$
\mathcal{F}(\rho)\doteq\left\{{\bf \Xi}: {\bf \Xi}=\diag\{\underbrace{\I_n,\cdots,\I_n}_{k-1},\bms(\rho),\I_n,\cdots,\I_n\}, 1\le k\le p\right\},
$$
where
\begin{align*}
\bms(\rho)=\left(\begin{array}{llllll}
1 & \rho & 0 & \cdots &0& 0 \\
\rho & 1 & \rho & \cdots&0 & 0 \\
0 & \rho & 1 & \cdots&0 & 0 \\
\vdots & \vdots & \vdots & & \vdots& \vdots\\
0 &0 &0 & \cdots &1& \rho\\
0 &0 &0 & \cdots &\rho& 1
\end{array}\right).
\end{align*}

Let $\mu_{\rho}$ be the uniform measure on $\mathcal{F}(\rho)$ and $\rho=c_{0}(\log d / n)^{1 / 2}$ for some small enough constant $c_{0}<$
1. Let $\mathrm{pr}_{{\bf \Xi}}$ denote the probability measure of $N_{d}(\bm 0, {\bf \Xi})$ and $\mathrm{pr}_{\mu_{\rho}}=\int \mathrm{pr}_{{\bf \Xi}} \mathrm{d} \mu_{\rho}({\bf \Xi}).$ Let $\mathrm{pr}_{0}$ denote the
probability measure of $N_{d}\left(0, \I_{d}\right) .$ Note that, for any set $A,$ we have
$$
\sup _{{\bf \Xi} \in \mathcal{F}(\rho)} \operatorname{pr}_{{\bf \Xi}}\left(A^{C}\right) \geq \operatorname{pr}_{\mu_{\rho}}\left(A^{C}\right), ~~ 1=\operatorname{pr}_{\mu_{\rho}}\left(A^{C}\right)+\operatorname{pr}_{\mu_{\rho}}(A)
$$
and
$$
\operatorname{pr}_{\mu_{\rho}}(A) \leq \operatorname{pr}_{0}(A)+\left|\operatorname{pr}_{\mu_{\rho}}(A)-\operatorname{pr}_{0}(A)\right|.
$$
Letting $A =\left\{T_{\alpha}=1\right\},$ the above equations yield
\begin{align*}
\inf _{T_{\alpha} \in \mathcal{T}_{\alpha},{\bf \Xi} \in \mathcal{F}(\rho)} \sup _{\operatorname{pr}_{{\bf \Xi}}}\left(T_{\alpha}=0\right) \geq & 1-\alpha-\sup _{A: \mathrm{pr}_{0}(A) \leq \alpha}\left|\operatorname{pr}_{\mu_{\rho}}(A)-\operatorname{pr}_{0}(A)\right|\\
 \geq & 1-\alpha-\frac{1}{2}\left\|\mathrm{pr}_{\mu_{\rho}}-\mathrm{pr}_{0}\right\|_{T V},
\end{align*}
where $\|\cdot\|_{T V}$ denotes the total variation norm. Setting $L_{\mu_{\rho}}(y) =\operatorname{dpr}_{\mu_{\rho}}(y) / \mathrm{d} \mathrm{pr}_{0}(y),$ and by Jensen's inequality, we have
\begin{align*}
\left\|\mathrm{pr}_{\mu_{\rho}}-\mathrm{pr}_{0}\right\|_{T V}=& \int\left|L_{\mu_{\rho}}(y)-1\right| \mathrm{d} \mathrm{pr}_{0}(y)\\
=& \mathbb{E}_{\mathrm{pr}_{0}}\left|L_{\mu_{\rho}}(Y)-1\right| \leq\left[\mathbb{E}_{\mathrm{pr}_{0}}\left\{L_{\mu_{\rho}}^{2}(Y)\right\}-1\right]^{1 / 2}.
\end{align*}
Therefore, as long as $\mathbb{E}_{\mathrm{pr}_{0}}\left\{L_{\mu_{\rho}}^{2}(Y)\right\}=1+o(1),$ we have
$$
\inf _{T_{\alpha} \in \mathcal{T}_{\alpha}, {\bf \Xi} \in \mathcal{F}(\rho)} \sup _{\mathrm{pr}_{{\bf \Xi}}}\left(T_{\alpha}=0\right) \geq 1-\alpha-o(1)>0.
$$
We then prove that $\mathbb{E}_{\mathrm{pr}_{0}}\left\{L_{\mu_{\rho}}^{2}(Y)\right\}=1+o(1) .$ By construction, we have
$$
L_{\mu_{\rho}}=\frac{1}{p} \sum_{{\bf \Xi} \in \mathcal{F}(\rho)}\left[ \frac{1}{|{\bf \Xi}|^{1 / 2}} \exp \left\{-\frac{1}{2} Z_{i, \cdot}^{\mathrm{T}}\left({\bf \Omega}-\I_{d}\right) {\color{black}Z_{i,\cdot}}\right\}\right],
$$
where ${\bf \Omega} = {\bf \Xi}^{-1}$ and {\color{black} $\left\{Z_{i,\cdot}: 1 \leq i \leq n\right\}$ are independent and identically distributed as $\mathcal{N}_{d}\left(0, \I_{d}\right)$}. We have
\begin{align*}
&\mathbb{E}_{\mathrm{pr}_{0}}\left\{L_{\mu_{\rho}}^{2}(Y)\right\}\\
=&\frac{1}{p^{2}} \sum_{{\bf \Xi}_{1}, {\bf \Xi}_{2} \in \mathcal{F}(\rho)} \mathbb{E}\left[\frac{1}{\left|{\bf \Xi}_{1}\right|^{1 / 2}} \frac{1}{\left|{\bf \Xi}_{2}\right|^{1 / 2}} \exp \left\{-\frac{1}{2} Z_{i, \cdot}^{\mathrm{T}}\left({\bf \Omega}_{1}+{\bf \Omega}_{2}-2 I_{d}\right) Z_{i, \cdot}\right\}\right],
\end{align*}
where ${\bf \Omega}_{i}={\bf \Xi}_{i}^{-1}$ for $i=1,2$. We write
\begin{align*}
\mathbb{E}_{\mathrm{pr}_{0}}\left\{L_{\mu_{\rho}}^{2}(Y)\right\}=&\frac{p-1}{p}\mathbb{E}\left[\frac{1}{\left|{\bf \Xi}_{1}\right|^{1 / 2}} \frac{1}{\left|{\bf \Xi}_{2}\right|^{1 / 2}} \exp \left\{-\frac{1}{2} Z_{i, \cdot}^{\mathrm{T}}\left({\bf \Omega}_{1}+{\bf \Omega}_{2}-2 \I_{d}\right) Z_{i, \cdot}\right\}\right]\\
&+\frac{1}{p}\mathbb{E}\left[\frac{1}{\left|{\bf \Xi}\right|} \exp \left\{-\frac{1}{2} Z_{i, \cdot}^{\mathrm{T}}\left(2{\bf \Omega}-2 \I_{d}\right) Z_{i, \cdot}\right\}\right]\\
\doteq & E_1+E_2,
\end{align*}
where $E_{1}$ represents the set of $\left({\bf \Xi}_{1}, {\bf \Xi}_{2}\right)$ with ${\bf \Xi}_{1} \neq {\bf \Xi}_{2},$ and $E_{2}$ represents the set of $\left({\bf \Xi}_{1}, {\bf \Xi}_{2}\right)$ with ${\bf \Xi}_{1}={\bf \Xi}_2$. By standard argument in moment generating functions of the Gaussian quadratic form, we have
\begin{align*}
\mathbb{E}\left\{\exp\left(-\frac{1}{2}\W^\top \A\W\right)\right\}=&{\color{black}\left[\{1+\lambda_1(\A)\}\cdots \{1+\lambda_q(\A)\}\right]^{-1/2}}\\
=&\{\det(\I_q+\A)\}^{-1/2}
\end{align*}
if $\W\sim \mathcal{N}(\bm 0, \I_q)$ and $\A\in \mathbb{R}^{q\times q}$. Without loss of generality, define ${\bf \Xi}_1=\diag\{\bms(\rho),\I_n,\cdots,\I_n\}$ and ${\bf \Xi}_2=\diag\{\I_n,\bms(\rho),\I_n,\cdots,\I_n\}$. Thus, $|{\bf \Xi}_1|=|{\bf \Xi}    _2|=|\bms(\rho)|$. Additionally, define $\O_1=\diag\{\bms(\rho)^{-1},\I_n,\cdots,\I_n\}$ and $\O_2=\diag\{\I_n,\bms(\rho)^{-1},\I_n,\cdots,\I_n\}$.
\begin{align*}
&\mathbb{E}\left[\frac{1}{\left|{\bf \Xi}_{1}\right|^{1 / 2}} \frac{1}{\left|{\bf \Xi}_{2}\right|^{1 / 2}} \exp \left\{-\frac{1}{2} Z_{i, \cdot}^{\mathrm{T}}\left({\bf \Omega}_{1}+{\bf\Omega}_{2}-2 I_{d}\right) Z_{i, \cdot}\right\}\right]\\
=&\mathbb{E}\left(\frac{1}{\left|{\bf \Xi}_{1}\right|^{1 / 2}} \frac{1}{\left|{\bf \Xi}_{2}\right|^{1 / 2}} \exp \left[-\frac{1}{2} Z_{1, \cdot}^{\mathrm{T}}\left\{\bms(\rho)^{-1}-\I_n\right\} Z_{1, \cdot}-\frac{1}{2} Z_{2, \cdot}^{\mathrm{T}}\left\{\bms(\rho)^{-1}-\I_n\right\} Z_{2, \cdot}\right]\right)\\
=&\frac{1}{\left|\bms(\rho)\right|}|\bms(\rho)^{-1}-\I_n+\I_n|^{-1/2}|\bms(\rho)^{-1}-\I_n+\I_n|^{-1/2}=1.
\end{align*}
Thus,
\begin{align}\label{e1}
E_1=\frac{p-1}{p}=1+o(1)
\end{align} as $p\to\infty$. Similarly,
\begin{align*}
&\frac{1}{p}\mathbb{E}\left[\frac{1}{\left|{\bf \Xi}\right|} \exp \left\{-\frac{1}{2} Z_{i, \cdot}^{\mathrm{T}}\left(2{\bf \Omega}-2 I_{d}\right) Z_{i, \cdot}\right\}\right]\\
=&\frac{1}{p}\frac{1}{|\bms(\rho)|}\mathbb{E}\left[ \exp \left\{-\frac{1}{2} Z_{1, \cdot}^{\mathrm{T}}\left(2\bms(\rho)^{-1}-2 \I_{n}\right) Z_{1, \cdot}\right\}\right]\\
=&\frac{1}{p}\frac{1}{|\bms(\rho)|}|2\bms(\rho)^{-1}- \I_{n}|^{-1/2}.
\end{align*}
By (10.1) in {\color{black} \cite{d1956}}, we have $|\bms(\rho)|=(1-\rho^2)^{n-1}$.
Define
\begin{align*}
\A=\left(\begin{array}{llllll}
0 & 1 & 0 & \cdots &0& 0 \\
1 & 0 & 1 & \cdots&0 & 0 \\
0 & 1 & 0 & \cdots&0 & 0 \\
\vdots & \vdots & \vdots & & \vdots& \vdots\\
0 &0 &0 & \cdots &0& 1\\
0 &0 &0 & \cdots &1& 0
\end{array}\right).
\end{align*}
Then, $\bms(\rho)=\I_n+\rho\A$. By the Taylor expansion, we have $(\I_n+\rho\A)^{-1}=\sum_{k=0}^\infty (-\rho)^k\A^k$ and
\begin{align*}
2\bms(\rho)^{-1}- \I_{n}=&2(\I_n+\rho\A)^{-1}-\I_n =2(\I_n-\rho\A+\rho^2\A^2-\rho^3\A^3+\cdots)-\I_n\\
=&\I_n-2\rho\A+2\rho^2\A^2 \sum_{k=0}^\infty (-\rho)^k\A^k\\
=&\I_n-2\rho\A+2\rho^2\A^2(\I_n+\rho\A)^{-1}.
\end{align*}
Because $\rho^2\A^2(\I_n+\rho\A)^{-1}$ is positive definite,  we have
\begin{align*}
|2\bms(\rho)^{-1}- \I_{n}|\ge|\I_n-2\rho\A|=(1-4\rho^2)^{n-1}.
\end{align*}
Thus,
\begin{align}\label{e2}
E_2\le p^{-1}(1-\rho^2)^{-n+1}(1-4\rho^2)^{-(n-1)/2}=p^{-1}\exp(3c_0^2\log p)\{1+o(1)\}\to 0
\end{align}
if $\rho=c_0(\log p/n)^{1/2}$ and $c_0^2<1/3$. Combining (\ref{e1}) and (\ref{e2}), we have $\mathbb{E}_{\mathrm{pr}_{0}}\left\{L_{\mu_{\rho}}^{2}(Y)\right\}=1+o(1).$
Lastly, we can easily show that  for $\rho=c_0(\log p/n)^{1/2}$,
$$\left [F(\bmv): \cor_F\{(\bmv_{\cdot 1}^\top,\cdots,\bmv_{\cdot p}^\top)^\top\}\in\mathcal{F}(\rho), F(\bmv) \mbox{ is Gaussian}\right]\subset \left [F(\bmv): R\{F(\bmv)\}\in\mathcal{U}(c)\right],$$
where $\bmv=\{\bmv_1,\cdots,\bmv_n\}$ and $R\{F(\bmv)\}=\big\{\cor_F(\bmv_{t+1},\bmv_t),\cdots,\cor_F(\bmv_{t+K},\bmv_t)\big\}$. Thus,
$$
\inf _{T_{\alpha} \in \mathcal{T}_{\alpha}} \sup_{R\{F(\bmv)\}\in \mathcal{U}(c)} \operatorname{pr}_{{\bf \Xi}}\left(T_{\alpha}=0\right) \geq \inf _{T_{\alpha} \in \mathcal{T}_{\alpha}}  \sup _{{\bf \Xi} \in \mathcal{F}(\rho)}\operatorname{pr}_{{\bf \Xi}}\left(T_{\alpha}=0\right) \geq 1-\alpha-o(1)>0.
$$
This completes the proof. \hfill $\square$

\subsection{Proof of Theorem \ref{thsum}}

Firstly, we restate Lemma 2.1 in \citet{s2009} on the quadratic forms.
\begin{lemma}\label{zm}
Under Condition (C4), for any $m \times m$ symmetric matrix $A=\{a_{ij}\}_{1\leq i,j\leq m}$ and $B=\{b_{ij}\}_{1\leq i,j\leq m}$ of constants, we have
\begin{align*}
\mathbb{E}\{(\z_t^TA\z_t)^2\}=&\Delta\sum_{i=1}^m a_{ii}^2+2\tr(A^2)+\{\tr(A)\}^2,\\
\var(\z_t^TA\z_t)=&\Delta\sum_{i=1}^p a_{ii}^2+2\tr(A^2),\\
\mathbb{E}\{(\z_t^TA\z_t)(\z_t^TB\z_t)\}=&\Delta\sum_{i=1}^m
a_{ii}b_{ii}+2\tr(AB)+\tr(A)\tr(B),
\end{align*}
where $\Delta=\mathbb{E}(z_{it}^4)-3$.
\end{lemma}

Now, we are ready to present the proof of Theorem \ref{thsum}.

\proof Let
\begin{align*}
T_{\operatorname{SUM}}=\sum_{l=1}^K\frac{2}{n(n-1)}\underset{t<s}{\sum\sum}\bmv_t^\top \bmv_s\bmv_{t+l}^\top \bmv_{s+l}\doteq\sum_{l=1}^K T_{l}.
\end{align*}
We will show that for each $l\in \{1,\cdots,K\}$,
\begin{align}\label{clt}
\frac{T_l}{\sqrt{\frac{2}{n(n-1)}\tr^2(\bms^2)}}\cd \mathcal{N}(0,1).
\end{align}

Define $V_{nj}=n^{-1}(n-1)^{-1}\sum_{i=l+1}^{j-1} \bmv_{i-l}^\top\bmv_{j-l}\bmv_i^T \bmv_j$, $j\in \{l+2,\cdots,n\}$ and $W_{nk}=\sum_{i=l+2}^k V_{ni}$, $k\in \{l+2,\cdots,n\}$. Let $\mathcal{F}_{i}\doteq\sigma\{\bmv_1,\cdots,\bmv_i\}$ be the $\sigma$-field generated by $\{\bmv_j\}_{j\le i}$. It is easy to show that $\mathbb{E}(V_{ni}|\mathcal{F}_{i-1})=0$ and it follows that $\{W_{nk},\mathcal{F}_k: l+2\le k\le n\}$ is a zero mean martingale. Let $v_{ni}=\mathbb{E}(V_{ni}^2|\mathcal{F}_{i-1})$, $l+2\le i \le n$ and $V_n=\sum_{i=l+2}^n v_{ni}$. The central limit theorem \citep{hh1980} will hold if we can show
\begin{align}\label{clt1}
\frac{V_n}{\var(W_{nn})}\cp 1,
\end{align}
and for any $\epsilon>0$,
\begin{align}\label{clt2}
\sum_{i=l+2}^n n^{2}\tr^{-2}(\bms^2)
\mathbb{E}\left[V_{ni}^2\mathbb{I}\left\{|V_{ni}|>\epsilon\sqrt{n^{-2}\tr^2(\bms^2)}\right\}|\mathcal{F}_{i-1}\right]
\cp 0.
\end{align}
It can be shown that
\begin{align*}
&v_{ni}\\
=&\frac{1}{n^2(n-1)^2}\left\{\sum_{j=l+1}^{i-1}
(\bmv_{i-l}^\top\bmv_{j-l})^2\bmv_j^T \bms \bmv_j
+2\sum_{l+1\le j<k<i}\bmv_{i-l}^\top\bmv_{j-l}\bmv_{i-l}^\top\bmv_{k-l}\bmv_j^T
 \bms \bmv_k \right\}.
\end{align*}
Then,
\begin{align*}
&\frac{V_n}{\var(W_{nn})}\\
=&\frac{2}{n(n-1)\tr^2(\bms^2)} \Big\{\sum_{i=l+2}^n\sum_{j=l+1}^{i-1}(\bmv_{i-l}^\top\bmv_{j-l})^2\bmv_j^T \bms \bmv_j\\
&~~~~~~~~~~~~~~~~~~~~~~~~+2\sum_{i=l+2}^n\sum_{l+1\le j<k\le i}\bmv_{i-l}^\top\bmv_{j-l}\bmv_{i-l}^\top\bmv_{k-l}\bmv_j^T \bms \bmv_k \Big\}\\
\doteq & C_{n1}+C_{n2}.
\end{align*}
Simple algebras lead to
\begin{align*}
\mathbb{E}(C_{n1})&=\frac{(n-l)(n-l-1)}{n(n-1)},\\
\var(C_{n1})&=\frac{4}{n^2(n-1)^2\tr^4(\bms^2)}
\mathbb{E}\left[\sum_{i=l+2}^n\sum_{j=l+1}^{n-1} \left\{(\bmv_{i-l}^\top\bmv_{j-l})^4(\bmv_j^T
\bms \bmv_j)^2-\tr^4(\bms^2)\right\}\right].
\end{align*}
By Lemma \ref{zm}, we have $\mathbb{E}\left\{(\bmv_j^T
\bms \bmv_j)^2-\tr^2(\bms^2)\right\}=O\{\tr(\bms^4)\}$. Next, we will show that  $\mathbb{E}\left\{(\bmv_{i-l}^\top\bmv_{j-l})^4-\tr^2(\bms^2)\right\}=O\{\tr(\bms^4)\}$.
Define $\bms^{1/2}\bms\bms^{1/2}\doteq\{\omega_{kl}\}_{1\le k,l\le p}$.
\begin{align*}
&\mathbb{E}\{(\bmv_i^\top\bmv_s)^4\}=\mathbb{E}\left\{(\z_i^T\bms\z_s)^4\right\}=\mathbb{E}\left\{\left(\sum_{k,l=1}^m \sigma_{kl}z_{ik}z_{jl}\right)^4\right\}\\
=&\sum_{k,l=1}^m \sigma_{kl}^4\mathbb{E}(z_{ik}^4)\mathbb{E}(z_{jl}^4)+\sum_{k\not=l}^m \sum_{s\not=t}^m \sigma_{kl}^2\sigma_{st}^2\mathbb{E}(z_{ik}^2)\mathbb{E}(z_{is}^2)\mathbb{E}(z_{jl}^2)\mathbb{E}(z_{jt}^2)\\
&+2\sum_{k=1}^m\sum_{s\not=t}^m
\sigma_{ks}^2\sigma_{kt}^2\mathbb{E}(z_{ik}^4)\mathbb{E}(z_{js}^2z_{jt}^2)+\sum_{k\not=l}^m
\sum_{s\not=t}^m
\sigma_{kl}\sigma_{kt}\sigma_{st}\sigma_{sl}\mathbb{E}(z_{ik}^2)\mathbb{E}(z_{jl}^2)\mathbb{E}(z_{is}^2)\mathbb{E}(z_{jt}^2).
\end{align*}
Note that $\tr^2(\bms^2)=(\sum_{s,t}\sigma_{st}^2)^2=\sum_{k,l,s,t}\sigma_{st}^2\sigma_{kl}^2$ and
\begin{align*}
\sum_{k,l=1}^m \sigma_{kl}^4 &\le \left(\sum_{k,l}\sigma_{kl}^2\right)^2,\\
\sum_{k=1}^m\sum_{s\not=t}^m \sigma_{ks}^2\sigma_{kt}^2 &\le \left(\sum_{k,l}\sigma_{kl}^2\right)^2,\\
\sum_{k\not=l}^m \sum_{s\not=t}^m \sigma_{kl}^2\sigma_{st}^2 &\le \sum_{k,l,s,t}\sigma_{st}^2\sigma_{kl}^2,\\
 \sum_{k\not=l}^m \sum_{s\not=t}^m \sigma_{kl}\sigma_{kt}\sigma_{st}\sigma_{sl} &\le \sum_{k\not=l}\omega_{kl}^2 \le \sum_{k,l}\omega_{kl}^2
=\tr(\bms^4).
\end{align*}
Thus, we have
\begin{align}\label{v4}
\mathbb{E}\{(\bmv_i^\top\bmv_s)^4\}-\tr^2(\bms^2)=O\{\tr(\bms^4)\}.
\end{align}
Hence, $\var(C_{n1})\to 0$ due to $\tr(\bms^4)=o\{\tr^2(\bms^2)\}$. Then, $C_{n1}\cp 1$. Similarly,
$\mathbb{E}(C_{n2})=0$ and
\begin{align*}
\var(C_{n2})=O(n^{-2})\frac{\tr^2(\bms^4)}{\tr^4(\bms^2)}
\to 0,
\end{align*}
which implies $C_{n2} \cp 0$. Thus, (\ref{clt1}) holds.

It remains to show (\ref{clt2}). Since
\begin{align*}
\mathbb{E}\left[V_{ni}^2\mathbb{I}\left\{|V_{ni}|>\epsilon \sqrt{n^{-2}\tr^2(\bms^2)}\right\}|\mathcal{F}_{i-1}\right]\le \mathbb{E}(V_{ni}^4|\mathcal{F}_{i-1})/\{\epsilon^2n^{-2}\tr^2(\bms^2)\},
\end{align*}
we only need to show that
\begin{align*}
\sum_{i=l+2}^n\mathbb{E}(V_{ni}^4)=o\{n^{-4}\tr^4(\bms^2)\}.
\end{align*}
Note that
\begin{align*}
\sum_{i=l+2}^n\mathbb{E}(V_{ni}^4)=O(n^{-4})\sum_{i=l+2}^n
\mathbb{E}\left\{\left(\sum_{j=l+1}^{i-1}\bmv_{i-l}^\top\bmv_{j-l}
\bmv_i^T\bmv_j\right)^4\right\},
\end{align*}
which can be decomposed as $3Q+P$ with
\begin{align*}
Q&=O(n^{-8})\sum_{i=l+2}^n \sum_{s\not=t}^{i-1}\mathbb{E}\left(\bmv_{i-l}^T \bmv_{s-l}\bmv_{s-l}^T \bmv_{i-l}\bmv_{i-l}^T \bmv_{t-l}\bmv_{t-l}^T \bmv_{i-l}\bmv_i^T \bmv_s\bmv_s^T \bmv_i\bmv_i^T \bmv_t\bmv_t^T \bmv_i\right),\\
P&=O(n^{-8})\sum_{i=l+2}^n
\sum_{s=1}^{i-1}\mathbb{E}\left\{(\bmv_{i-l}^\top\bmv_{j-l})^4(\bmv_i^T \bmv_s)^4\right\}.
\end{align*}
Note that $Q={O}(n^{-4})\mathbb{E}^2\{(\bmv_i^T \bmv_i)^2\}=o\{n^{-4}\tr^4(\bms^2)\}$ by
Lemma \ref{zm}. By (\ref{v4}), we have $P=O\{n^{-4}\tr^2(\bms^4)\}=o\{n^{-4}\tr^4(\bms^2)\}$.
And then (\ref{clt2}) follows immediately. This completes the proof of
(\ref{clt}). Finally, after some simple algebras, we have $\mathbb{E}(T_lT_k)=0$ if $l\not=k$. Thus, we have
\begin{align}\label{clt4}
\frac{\sum_{l=1}^KT_l}{\sqrt{\frac{2K}{n(n-1)}\tr^2(\bms^2)}}\cd \mathcal{N}(0,1).
\end{align}
Here, we complete the proof.  \hfill $\square$

\subsection{Proof of Proposition \ref{prop1} }
\proof
Under $H_0$, due to {\color{black}Proposition A.2 in \citet{czz2010}} and the condition $\tr(\bms^4)=o(\tr^2(\bms^2))$, we have
\begin{align*}
  \mathbb{E}\{\widehat{\tr(\bms^2)}\}=& \tr(\bms^2),\\
  \var\{\widehat{\tr(\bms^2)}\}=&4 n^{-2} \operatorname{tr}^{2}\left(\bms^{2}\right)+8 n^{-1} \operatorname{tr}\left(\bms^{4}\right)+4 \Delta n^{-1} \operatorname{tr}\left({\color{black}\bms^{2} \circ \bms^{2}}\right)\\
  &+O\left\{n^{-3} \operatorname{tr}^{2}\left(\bms^{2}\right)+n^{-2} \operatorname{tr}\left(\bms^{4}\right)\right\}=o\{\tr^2(\bms^2)\},
\end{align*}
where ${\bf A}\circ {\bf B}=\{a_{ij}b_{ij}\}$ for two matrix ${\bf A}=\{a_{ij}\}$ and ${\bf B}=\{b_{ij}\}$ .
Hence, we complete the proof of this proposition. \hfill $\square$

\subsection{Proof of Theorem \ref{sh1}}
\proof
Recall that $\bmv_t=\A_0\z_t+\A_1\z_{t-1}$ and $K=1$.
Actually,
\begin{align*}
G_{1}=&\frac{1}{n(n-1)} \sum_{s\not=t}^{T}\left(\A_{0} \z_{s}+\A_{1} \z_{s-1}\right)^{\top}\left(\A_{0} \z_{t}+\A_{1} \z_{t-1}\right)\\
& ~~~~~~~~~~~~~~~~~\left(\A_{0} \z_{t-1}+\A_{1} \z_{t-2}\right)^{\top}\left(\A_{0} \z_{s-1}+\A_{1} \z_{s-2}\right)\\
= &G(I)+G(II)+G(III),
\end{align*}
where
\begin{align*}
&G(I) \\
\doteq&\frac{1}{n(n-1)} \sum_{s\not=t}^{T}\left(\z_{s}^{\top} \A_{0}^{\top} \A_{0} \z_{t} \z_{t-1}^{\top} \A_{0}^{\top} \A_{0} \z_{s-1}+\z_{s-1}^{\top} \A_{1}^{\top} \A_{1} \z_{t-1} \z_{t-2}^{\top} \A_{1}^{\top} \A_{1} \z_{s-2}\right.\\
&\left.+\z_{s}^{\top} \A_{0}^{\top} \A_{0} \z_{t} \z_{t-2}^{\top} \A_{1}^{\top} \A_{1} \z_{s-2}+\z_{s-1}^{\top} \A_{1}^{\top} \A_{1} \z_{t-1} \z_{t-1}^{\top} \A_{0}^{\top} \A_{0} \z_{s-1}\right)\\
&G(I I) \\
\doteq&\frac{1}{n(n-1)} \sum_{s\not=t}^{T}\left(\z_{s}^{\top} \A_{0}^{\top} \A_{1} \z_{t-1} \z_{t-1}^{\top} \A_{0}^{\top} \A_{0} \z_{s-1}+\z_{s-1}^{\top} \A_{1}^{\top} \A_{0} \z_{t} \z_{t-1}^{\top} \A_{0}^{\top} \A_{0} \z_{s-1}\right.\\
&+\z_{s-1}^{\top} \A_{1}^{\top} \A_{1} \z_{t-1} \z_{t-1}^{\top} \A_{0}^{\top} \A_{1} \z_{s-2}+\z_{s-1}^{\top} \A_{1}^{\top} \A_{1} \z_{t-1} \z_{t-2}^{\top} \A_{1}^{\top} \A_{0} \z_{s-1} \\
&+\z_{s}^{\top} \A_{0}^{\top} \A_{0} \z_{t} \z_{t-2}^{\top} \A_{1}^{\top} \A_{0} \z_{s-1}+\z_{s}^{\top} \A_{0}^{\top} \A_{0} \z_{t} \z_{t-1}^{\top} \A_{0}^{\top} \A_{1} \z_{s-2} \\
&\left.+\z_{s}^{\top} \A_{0}^{\top} \A_{1} \z_{t-1} \z_{t-2}^{\top} \A_{1}^{\top} \A_{1} \z_{s-2}+\z_{s-1}^{\top} \A_{1}^{\top} \A_{0} \z_{t} \z_{t-2}^{\top} \A_{1}^{\top} \A_{1} \z_{s-2}\right)\\
&G(I I I) \\
\doteq&\frac{1}{n(n-1)} \sum_{s\not=t}^{T}\left(\z_{s}^{\top} \A_{0}^{\top} \A_{1} \z_{t-1} \z_{t-1}^{\top} \A_{0}^{\top} \A_{1} \z_{s-2}+\z_{s-1}^{\top} \A_{1}^{\top} \A_{0} \z_{t} \z_{t-2}^{\top} \A_{1}^{\top} \A_{0} \z_{s-1}\right.\\
&\left.+\z_{s}^{\top} \A_{0}^{\top} \A_{1} \z_{t-1} \z_{t-2}^{\top} \A_{1}^{\top} \A_{0} \z_{s-1}+\z_{s-1}^{\top} \A_{1}^{\top} \A_{0} \z_{t} \z_{t-1}^{\top} \A_{0}^{\top} \A_{1} \z_{s-2}\right).
\end{align*}
After some tedious algebra, we have
\begin{align*}
\mathbb{E}\{G(I)\}=\tr(\tilde{\bms}_0\tilde{\bms}_1), ~\mathbb{E}\{G(II)\}=0,~\mathbb{E}\{G(III)\}=\frac{2}{T}\tr^2(\tilde{\bms}_{01})
\end{align*}
and
\begin{align*}
\var\{G(I)\}=&\frac{2}{T^2}\tr^2(\tilde\bms_0^2+\tilde\bms_1^2)+\frac{6}{T^2}\tr^2(\tilde\bms_0\tilde\bms_1)\\
&+\frac{4}{T}\left[2 \operatorname{tr}\left(\widetilde{\bms}_{0} \tilde{\bms}_{1}\right)^{2}+\left(\nu_{4}-3\right) \operatorname{tr}\left\{D^{2}\left(\tilde{\bms}_{0} \tilde{\bms}_{1}\right)\right\}\right]+r_n,\\
\var\{G(II)\}=&\frac{8}{T^2}\tr(\tilde\bms_{01}\tilde{\bms}_{01}^\top)\tr(\tilde\bms_0^2+\tilde\bms_1^2)+\frac{16}{T^2}\tr(\tilde\bms_{01}\tilde\bms_1)\tr(\tilde\bms_{01}\tilde\bms_0)\\
&+\frac{16}{T^2}\tr(\tilde\bms_0+\tilde{\bms}_1)\left\{\tr(\tilde\bms_{01}^\top\tilde\bms_{01}\tilde\bms_0)+\tr(\tilde\bms_{01}\tilde\bms_{01}^\top\tilde\bms_1)\right\}\\
&+\frac{16}{T^2}\tr(\tilde\bms_{01} )\left\{\operatorname{tr}\left(\tilde{\bms}_{0}^{2} \tilde{\bms}_{01}^{\top}\right)+\operatorname{tr}\left(\tilde{\bms}_{1}^{2} \tilde{\bms}_{01}\right)+2 \operatorname{tr}\left(\tilde{\bms}_{1} \tilde{\bms}_{01} \tilde{\bms}_{0}\right)\right\}\\
&+\frac{4}{T} \operatorname{tr}\left(\tilde{\bms}_{01}^{\top} \tilde{\bms}_{01} \tilde{\bms}_{0}^{2}+\tilde{\bms}_{01} \tilde{\bms}_{01}^{\top} \tilde{\bms}_{1}^{2}+2 \tilde{\bms}_{01}^{\top} \tilde{\bms}_{1} \tilde{\bms}_{01} \tilde{\bms}_{0}\right)+r_n,\\
\var\{G(I I I)\}=&\frac{4}{T} \operatorname{tr}\left(\widetilde{\bms}_{01} \tilde{\bms}_{01}^{\top} \tilde{\bms}_{01}^{\top} \tilde{\bms}_{01}\right)+\frac{12}{T^{2}} \operatorname{tr}^{2}\left(\widetilde{\bms}_{01} \tilde{\bms}_{01}^{\top}\right)\\
&+\frac{16}{T^{2}} \operatorname{tr}\left(\tilde{\bms}_{01}\right) \operatorname{tr}\left(\tilde{\bms}_{01} \tilde{\bms}_{01}^{\top} \tilde{\bms}_{01}^{\top}\right)+r_n,\\
\cov\{G(I),G(III)\}=&\frac{4}{T^{2}} \operatorname{tr}^{2}\left(\tilde{\bms}_{0} \tilde{\bms}_{01}\right)+\frac{4}{T^{2}} \operatorname{tr}^{2}\left(\tilde{\bms}_{1} \tilde{\bms}_{01}\right)+r_n,\\
\cov\{G(I), G(I I)\}=&r_n, ~~\cov\{G(I I), G(I I I)\}=r_n.
\end{align*}
Similar to the proof of (\ref{clt}),  we have
\begin{align*}
\frac{T_{\operatorname{SUM}}-\mu_{1}}{\sigma_{S1}}\cd \mathcal{N}(0,1).
\end{align*} \hfill $\square$

\subsection{Proof of Theorem \ref{ms}}

First, we present some technical results for the proof of Theorem \ref{ms}.

The following is a well-known formula for conditional distributions of multivariate normal distributions; see, for example, p.12 from {\color{black}\citet{m1982}}.
\begin{lemma}\lbl{sunny_coffee}
Let $\bd{X} \sim \mathcal{N}(\bm{\mu}, \bd{\Sigma})$ with $\bd{\Sigma}$ being invertible. Partition $\bd{X}, \bm{\mu}$ and $\bd{\Sigma}$ as
\begin{align*}\lbl{similar_beer}
\bd{X}=
\begin{pmatrix}
\bd{X}_1\\
\bd{X}_2
\end{pmatrix}
,\ \ \ \
\bd{\bmu}=
\begin{pmatrix}
\bd{\bmu}_1\\
\bd{\bmu}_2
\end{pmatrix}
,\ \ \ \
\bd{\Sigma}=
\begin{pmatrix}
\bd{\Sigma}_{11} & \bd{\Sigma}_{12}\\
\bd{\Sigma}_{21} & \bd{\Sigma}_{22}
\end{pmatrix},
\end{align*}
where $\bd{X}_2 \sim \mathcal{N}(\bd{\bmu}_2, \bd{\Sigma}_{22})$. Set $\bd{\Sigma}_{22\cdot 1}=\bd{\Sigma}_{22}-\bd{\Sigma}_{21}\bd{\Sigma}_{11}^{-1}\bd{\Sigma}_{12}$. Then
$\bd{X}_2-\bd{\Sigma}_{21}\bd{\Sigma}_{11}^{-1}\bd{X}_1 \sim \mathcal{N}_k(\bd{\bmu}_2-\bd{\Sigma}_{21}\bd{\Sigma}_{11}^{-1}\bd{\bmu}_1, \bd{\Sigma}_{22\cdot 1})$ and is independent of $\bd{X}_1$.
%
\end{lemma}


\begin{lemma}\lbl{akxbcvgf} For any $x\in \mathbb{R}$ and $y\in \mathbb{R}$, define $A_p=\{\frac{T_{\operatorname{SUM}}}{\sigma_S}\leq x\}$ and  $l_p= (2\log N -\log\log N+y)^{1/2}$  and  $B_{i}=\{|Z_i|>l_p\}.$ Then, for each $1\leq d\leq p$,
\beaa
\sum_{1\leq i_1<  \cdots < i_{d}\leq p}\big| \textrm{P}(A_pB_{i_1}\cdots B_{i_{d}}) - \textrm{P}(A_p)\cdot \textrm{P}(B_{i_1}\cdots B_{i_{d}}) \big|\to 0
\eeaa
as $p\to\infty$.
\end{lemma}

\proof The argument is divided into two steps.

{\it Step 1: appealing independence from normal distributions}.\\
 Note that $(\varepsilon_{t1}, \cdots, \varepsilon_{tp})^\top\sim \mathcal{N}(\bd{0}, \bd{\Sigma})$. Take $\bd{X}_{t1}=(\varepsilon_{t1}, \cdots, \varepsilon_{td})^\top$ and $\bd{X}_{t2}=(\varepsilon_{t,d+1}, \cdots, \varepsilon_{tp})^\top$. Recall the notation in Lemma \ref{sunny_coffee}.
Write $\bd{X}_{t2}=\bd{U}_{t} + \bd{V}_t$,
where $\bd{U}_t =\bd{X}_{t2}-\bd{\Sigma}_{21}\bd{\Sigma}_{11}^{-1}\bd{X}_{t1}\sim \mathcal{N}(\bd{0}, \bd{\Sigma}_{22\cdot 1})$ and $\bd{V}_t =\bd{\Sigma}_{21}\bd{\Sigma}_{11}^{-1}\bd{X}_{t1}\sim \mathcal{N}(\bd{0}, \bd{\Sigma}_{21}\bd{\Sigma}_{11}^{-1}\bd{\Sigma}_{12})$. By Lemma \ref{sunny_coffee},
\begin{equation}\label{old_man}
\bd{U}_t\ \mbox{and}\ \{\varepsilon_{t1}, \cdots, \varepsilon_{td}\}\ \mbox{are independent}.
\end{equation}
Write
\begin{align*}
&T_{\operatorname{SUM}}\\
=&\frac{1}{n(n-1)}\sum_{l=1}^K\underset{t\not=s}{\sum\sum}\bmv_t^\top \bmv_s\bmv_{t+l}^\top \bmv_{s+l}\\
=&\frac{1}{n(n-1)}\sum_{l=1}^K\underset{t\not=s}{\sum\sum}(\X_{t1}^\top\X_{s1}+\X_{t2}^\top\X_{s2})(\X_{t+l,1}^\top\X_{s+l,1}+\X_{t+l,2}^\top\X_{s+l,2})\\
=&\frac{1}{n(n-1)}\sum_{l=1}^K\underset{t\not=s}{\sum\sum}(\U_{t}^\top\U_{s}+\U_t^\top\V_s+\U_s\V_t+\V_t\V_s+\X_{t1}^\top\X_{s1})\\
&\times(\U_{t+l}^\top\U_{s+l}+\U_{t+l}^\top\V_{s+l}+\U_{s+l}\V_{t+l}+\V_{t+l}\V_{s+l}+\X_{t+l,1}^\top\X_{s+l,1})\\
=&\frac{1}{n(n-1)}\sum_{l=1}^K\underset{t\not=s}{\sum\sum}\U_{t}^\top\U_{s}\U_{t+l}^\top\U_{s+l}+\frac{2}{n(n-1)}\sum_{l=1}^K\underset{t\not=s}{\sum\sum}\U_{t}^\top\U_{s}\U_{t+l}^\top\V_{s+l}\\
&+\frac{2}{n(n-1)}\sum_{l=1}^K\underset{t\not=s}{\sum\sum}\U_{t}^\top\U_{s}\V_{t+l}^\top\V_{s+l}+\frac{2}{n(n-1)}\sum_{l=1}^K\underset{t\not=s}{\sum\sum}\U_{t}^\top\U_{s}\X_{t+l,1}^\top\X_{s+l,1}\\
&+\frac{2}{n(n-1)}\sum_{l=1}^K\underset{t\not=s}{\sum\sum}\U_{t}^\top\V_{s}\U_{t+l}^\top\U_{s+l}+\frac{4}{n(n-1)}\sum_{l=1}^K\underset{t\not=s}{\sum\sum}\U_{t}^\top\V_{s}\U_{t+l}^\top\V_{s+l}\\
&+\frac{2}{n(n-1)}\sum_{l=1}^K\underset{t\not=s}{\sum\sum}\U_{t}^\top\V_{s}\V_{t+l}^\top\V_{s+l}+\frac{2}{n(n-1)}\sum_{l=1}^K\underset{t\not=s}{\sum\sum}\U_{t}^\top\V_{s}\X_{t+l,1}^\top\X_{s+l,1}\\
&+\frac{2}{n(n-1)}\sum_{l=1}^K\underset{t\not=s}{\sum\sum}\X_{t1}^\top\X_{s1}\U_{t+l}^\top\U_{s+l}+\frac{4}{n(n-1)}\sum_{l=1}^K\underset{t\not=s}{\sum\sum}\X_{t1}^\top\X_{s1}\U_{t+l}^\top\V_{s+l}\\
&+\frac{2}{n(n-1)}\sum_{l=1}^K\underset{t\not=s}{\sum\sum}\X_{t1}^\top\X_{s1}\V_{t+l}^\top\V_{s+l}+\frac{2}{n(n-1)}\sum_{l=1}^K\underset{t\not=s}{\sum\sum}\X_{t1}^\top\X_{s1}\X_{t+l,1}^\top\X_{s+l,1}\\
\doteq& \frac{1}{n(n-1)}\sum_{l=1}^K\underset{t\not=s}{\sum\sum}\U_{t}^\top\U_{s}\U_{t+l}^\top\U_{s+l}+\sum_{q=1}^{11}\Theta_q\\
\doteq& S_p+R_p.
\end{align*}
Next, we will show that, for any $d\geq 1$ and $\iota>0$, there exists $t=t_p>0$ with  $\lim_{N\to \infty}t_p=\infty$ and integer $p_0\geq 1$ such that
\begin{align}\lbl{when_what}
\textrm{P}(|\Theta_q|\geq \iota\sigma_S)\leq \frac{1}{p^{t}}
\end{align}
as $p\geq p_0$.  Here we only consider $\Theta_1$. The proof of the other parts are similar to $\Theta_1$.

By the decomposition, we know that $\{\V_t\}_{t=1}^n$ is independent of $\{\U_t\}_{t=1}^n$. Thus, conditional on $\{\U_t\}_{t=1}^n$, $\Theta_1$ has the normal distribution. Hence, we have
\begin{align*}
&\operatorname{P}(|\Theta_1|\ge  \iota\sigma_S)\le \sum_{l=1}^K\operatorname{P}\left(\left|\frac{2}{n(n-1)}\underset{t\not=s}{\sum\sum}\U_{t}^\top\U_{s}\U_{t+l}^\top\V_{s+l}\right|\ge \iota\sigma_S/K\right)\\
&=\sum_{l=1}^K\mathbb{E}\left(\mathbb{E}\left[\mathbb{I}\left\{\left|\frac{2}{n(n-1)}\underset{t\not=s}{\sum\sum}\U_{t}^\top\U_{s}\U_{t+l}^\top\V_{s+l}\right|\ge \iota\sigma_S/K\right\}|\{\U_t\}_{t=1}^n\right]\right)\\
&=2\sum_{l=1}^K\mathbb{E}\left\{1-\Phi\left(\frac{\iota \sigma_S}{K\tilde\sigma_l}\right)\right\}\simeq \sum_{l=1}^K\mathbb{E}\left\{\frac{2}{\sqrt{2\pi}K^{-1}\iota\tilde\sigma_{l}^{-1}\sigma_S}e^{-(K^{-1}\iota\tilde\sigma_{l}^{-1}\sigma_S)^2/2}\right\},
\end{align*}
where
\begin{align*}
\tilde{\sigma}^2_l=\frac{4}{n^2(n-1)^2}\sum_{t,m\not=s}\U_t^\top\U_s\U_{t+l}^\top\bms_{21}\bms_{11}^{-1}\bms_{12}\U_{m+l}\U_m^\top\U_s.
\end{align*}
Here, $a_n\simeq b_n$ denotes that $a_n/b_n\to 1$.
Similar to the proof of Proposition \ref{prop1}, we have
\begin{align*}
\frac{\tilde{\sigma}^2_l}{\frac{4}{n(n-1)}\tr(\bms_{22\cdot1}^2)\tr(\bms_{22\cdot1}\cdot \bms_{21}\bms_{11}^{-1}\bms_{12})}\cp 1.
\end{align*}
Thus,
\begin{align*}
K^{-1}\iota\tilde\sigma_{l}^{-1}\sigma_S\cp \frac{\iota}{K}\frac{\tr(\bms^2)}{\tr^{1/2}(\bms_{22\cdot1}^2)\tr^{1/2}(\bms_{22\cdot1}\cdot \bms_{21}\bms_{11}^{-1}\bms_{12})}
\end{align*}
and
{\color{black}
\begin{align*}
&\operatorname{tr}\left\{\boldsymbol{\Sigma}_{22 \cdot 1} \cdot\left(\boldsymbol{\Sigma}_{21} \boldsymbol{\Sigma}_{11}^{-1} \boldsymbol{\Sigma}_{12}\right)\right\}
\leq  \lambda_{\max }\left(\boldsymbol{\Sigma}_{22 \cdot 1}\right) \cdot \operatorname{tr}\left(\boldsymbol{\Sigma}_{21} \boldsymbol{\Sigma}_{11}^{-1} \boldsymbol{\Sigma}_{12}\right) \\
\leq & \lambda_{\max }(\boldsymbol{\Sigma}) \cdot \operatorname{tr}\left(\boldsymbol{\Sigma}_{21} \boldsymbol{\Sigma}_{11}^{-1} \boldsymbol{\Sigma}_{12}\right)
\leq  \lambda_{\max }(\boldsymbol{\Sigma})\lambda_{\max }\left(\bms_{11}^{-1}\right) \cdot \operatorname{tr}\left(\bms_{12} \bms_{21}\right) \\
=& \lambda_{\max }(\boldsymbol{\Sigma})\frac{1}{\lambda_{\min }\left(\bms_{11}\right)} \cdot \operatorname{tr}\left(\bms_{12} \bms_{21}\right)
\leq  {d M_{p}\lambda_{\max }(\boldsymbol{\Sigma})}.
\end{align*}
In fact, in the above, we use the assertion $\lambda_{\min }\left(\bms_{11}\right)$ is bounded by Condition (C2) and the fact that
$$
\operatorname{tr}\left(\boldsymbol{\Sigma}_{12} \boldsymbol{\Sigma}_{21}\right)=\sum_{i=1}^{d} \sum_{j=d+1}^{p} \sigma_{i j}^{2} \leq d M_{p}.
$$
Additionally, $\boldsymbol{\Sigma}_{22}=\boldsymbol{\Sigma}_{22 \cdot 1}+\boldsymbol{\Sigma}_{21} \boldsymbol{\Sigma}_{11}^{-1} \boldsymbol{\Sigma}_{12}$ and all three matrices are non-negative definite.  Hence, we have $\tr(\bms^2)\ge \tr(\bms^2_{22\cdot1})$. Thus,
\begin{align*}
\frac{\iota}{K}\frac{\tr(\bms^2)}{\tr^{1/2}(\bms_{22\cdot1}^2)\tr^{1/2}(\bms_{22\cdot1}\cdot \bms_{21}\bms_{11}^{-1}\bms_{12})}\ge \frac{\iota}{K} \frac{\tr^{1/2}(\bms^2)}{\sqrt{dM_{p}\lambda_{\max}(\bms)}}.
\end{align*}
Then,
\begin{align*}
&\sum_{l=1}^K\mathbb{E}\left\{\frac{2}{\sqrt{2\pi}K^{-1}\iota\tilde\sigma_{l}^{-1}\sigma_S}e^{-(K^{-1}\iota\tilde\sigma_{l}^{-1}\sigma_S)^2/2}\right\}\\
\le& K\frac{1}{\sqrt{2\pi}}\frac{2}{\frac{\iota}{K} \frac{\tr^{1/2}(\bms^2)}{\sqrt{dM_p\lambda_{\max}(\bms)}}}\exp\left\{-\frac{\iota^2}{K^2} \frac{\tr(\bms^2)}{dM_p\lambda_{\max}(\bms)}\right\}\le p^{-t_p}
\end{align*}
by Condition (C6) with $t_p=\frac{\iota^2}{2K^2}(\log p)^{\gamma-1}$. Note that $\tr(\bms^2_{11})\le C_d$ is bounded.

Similarly, it can be proved that for the other parts $\Theta_2,\cdots,\Theta_{11}$,
\begin{align*}
&\mathrm{P}(|\Theta_2|\ge  \iota\sigma_S)\\
\le & K\frac{1}{\sqrt{2\pi}}\frac{2}{\frac{\iota}{K}\frac{\tr(\bms^2)}{\tr^{1/2}(\bms^2_{22\cdot 1})\tr^{1/2}\left\{(\bms_{21}\bms_{11}^{-1}\bms_{12})^2\right\}} }\exp\left[-\frac{\iota^2}{K^2} \frac{\tr^2(\bms^2)}{\tr(\bms^2_{22\cdot 1})\tr\left\{(\bms_{21}\bms_{11}^{-1}\bms_{12})^2\right\}} \right]\\
\le & K\frac{1}{\sqrt{2\pi}}\frac{2}{\frac{\iota}{K} \frac{\tr^{1/2}(\bms^2)}{dM_p}}\exp\left\{-\frac{\iota^2}{K^2} \frac{\tr(\bms^2)}{d^2M^2_p}\right\}\le p^{-t_p},
\end{align*}
\begin{align*}
&\mathrm{P}(|\Theta_3|\ge  \iota\sigma_S)\\
\le & K\frac{1}{\sqrt{2\pi}}\frac{2}{\frac{\iota}{K}\frac{\tr(\bms^2)}{\tr^{1/2}(\bms^2_{22\cdot 1})\tr^{1/2}(\bms_{11}^2)} }\exp\left\{-\frac{\iota^2}{K^2} \frac{\tr^2(\bms^2)}{\tr(\bms^2_{22\cdot 1})\tr(\bms_{11}^2)} \right\}\\
\le & K\frac{1}{\sqrt{2\pi}}\frac{2}{\frac{\iota}{K} \frac{\tr^{1/2}(\bms^2)}{C_d^{1/2}}}\exp\left\{-\frac{\iota^2}{K^2} \frac{\tr(\bms^2)}{C_d}\right\}\le p^{-t_p},
\end{align*}
\begin{align*}
\mathrm{P}(|\Theta_4|\ge  \iota\sigma_S)& \simeq \mathrm{P}(|\Theta_1|\ge  \iota\sigma_S), ~~\mathrm{P}(|\Theta_5|\ge  \iota\sigma_S)\simeq \mathrm{P}(|\Theta_2|\ge  \iota\sigma_S),
\end{align*}
\begin{align*}
&\mathrm{P}(|\Theta_6|\ge  \iota\sigma_S)\\
\le & K\frac{1}{\sqrt{2\pi}}\frac{2}{\frac{\iota}{K}\frac{\tr(\bms^2)}{\tr^{1/2}(\bms_{22\cdot 1}\bms_{21}\bms_{11}^{-1}\bms_{12})\tr^{1/2}\left\{(\bms_{21}\bms_{11}^{-1}\bms_{12})^2\right\}} }\\
 & \times \exp\left[-\frac{\iota^2}{K^2} \frac{\tr^2(\bms^2)}{\tr(\bms_{22\cdot 1}\bms_{21}\bms_{11}^{-1}\bms_{12})\tr\left\{(\bms_{21}\bms_{11}^{-1}\bms_{12})^2\right\}} \right]\\
\le & K\frac{1}{\sqrt{2\pi}}\frac{2}{\frac{\iota}{K} \frac{\tr(\bms^2)}{\left\{d^{3/2}K^{3/2}_p\lambda_{\max}(\bms)\right\}^{1/2}}}\exp\left\{-\frac{\iota^2}{K^2} \frac{\tr^2(\bms^2)}{d^3M^3_p\lambda_{\max}(\bms)}\right\}\le p^{-t_p},
\end{align*}
\begin{align*}
&\mathrm{P}(|\Theta_7|\ge  \iota\sigma_S)\\
\le & K\frac{1}{\sqrt{2\pi}}\frac{2}{\frac{\iota}{K}\frac{\tr(\bms^2)}{\tr^{1/2}(\bms_{22\cdot 1}\bms_{21}\bms_{11}^{-1}\bms_{12})\tr^{1/2}(\bms_{11}^2)} }\exp\left\{-\frac{\iota^2}{K^2} \frac{\tr^2(\bms^2)}{\tr(\bms_{22\cdot 1}\bms_{21}\bms_{11}^{-1}\bms_{12})\tr(\bms_{11}^2)} \right\}\\
\le & K\frac{1}{\sqrt{2\pi}}\frac{2}{\frac{\iota}{K} \frac{\tr(\bms^2)}{\left\{C_ddM_p\lambda_{\max}(\bms)\right\}^{1/2}}}\exp\left\{-\frac{\iota^2}{K^2} \frac{\tr^2(\bms^2) }{C_ddM_p\lambda_{\max}(\bms)}\right\}\le p^{-t_p},
\end{align*}
\begin{align*}
\mathrm{P}(|\Theta_8|\ge  \iota\sigma_S)& \simeq \mathrm{P}(|\Theta_3|\ge  \iota\sigma_S), ~~\mathrm{P}(|\Theta_9|\ge  \iota\sigma_S)\simeq \mathrm{P}(|\Theta_7|\ge  \iota\sigma_S),
\end{align*}
\begin{align*}
&\mathrm{P}(|\Theta_{10}|\ge  \iota\sigma_S)\\
\le & K\frac{1}{\sqrt{2\pi}}\frac{2}{\frac{\iota}{K}\frac{\tr(\bms^2)}{\tr^{1/2}\left\{(\bms_{21}\bms_{11}^{-1}\bms_{12})^2\right\}\tr^{1/2}(\bms_{11}^2)} }\exp\left[-\frac{\iota^2}{K^2} \frac{\tr^2(\bms^2)}{\tr\left\{(\bms_{21}\bms_{11}^{-1}\bms_{12})^2\right\}\tr(\bms_{11}^2)} \right]\\
\le & K\frac{1}{\sqrt{2\pi}}\frac{2}{\frac{\iota}{K} \frac{\tr(\bms^2)}{C_d^{1/2}dM_p}}\exp\left\{-\frac{\iota^2}{K^2} \frac{\tr^2(\bms^2)}{C_dd^2M^2_p}\right\}\le p^{-t_p},
\end{align*}
\begin{align*}
\mathrm{P}(|\Theta_{11}|\ge  \iota\sigma_S)\le & K\frac{1}{\sqrt{2\pi}}\frac{2}{\frac{\iota}{K}\frac{\tr(\bms^2)}{\tr(\bms_{11}^2)} }\exp\left\{-\frac{\iota}{K} \frac{\tr^2(\bms^2)}{\tr^2(\bms_{11}^2)} \right\}\\
\le & K\frac{1}{\sqrt{2\pi}}\frac{2}{\frac{\iota}{K} \frac{\tr(\bms^2)}{C_d^{1/2}}}\exp\left\{-\frac{\iota^2}{K^2} \frac{\tr^2(\bms^2)}{C_d}\right\}\le p^{-t_p}.
\end{align*}
}

Thus, we have
\begin{align}\lbl{when_what2}
\textrm{P}(|R_p|\geq \iota\sigma_S)\leq \frac{1}{p^{t_p}}.
\end{align}

Now, for clarity, we will revise the definition of $A_p$  as follows
\begin{align*}
A_p(x)=\Big\{\frac{T_{\operatorname{SUM}}}{\sigma_S}\leq x\Big\},\ \ x \in \mathbb{R},
\end{align*}
for $p\geq 1$. Due to the fact that $T_{\operatorname{SUM}}=S_p+R_p$, we see that
\beaa
 \operatorname{P}\{A_p(x)B_1\cdots B_d\}
&\leq & \mathrm{P}\Big\{A_p(x)B_1\cdots B_d,\ \frac{|R_p|}{\sigma_S}<  \iota\Big\} + \frac{1}{p^t}\\
& \leq & \mathrm{P}\Big(\frac{S_p}{\sigma_S}\leq x+\iota
,\ B_1\cdots B_d\Big) +\frac{1}{p^{t}}\\
& = & \mathrm{P}\Big(\frac{S_p}{\sigma_S}\leq x+\iota\Big)\cdot P\big(
B_1\cdots B_d\big) +\frac{1}{p^{t}}
\eeaa
by the independence appeared in \eqref{old_man}. Hence,
\beaa
 \mathrm{P}\Big(\frac{S_p}{\sigma_S}\leq x+\iota\Big)
& \leq & \mathrm{P}\Big(\frac{S_p}{\sigma_S}\leq x+\iota,\ \frac{|R_p|}{\sigma_S}<  \iota\Big) + \frac{1}{p^{t}} \\
& \leq & \mathrm{P}\Big\{\frac{1}{\sigma_S}(S_p+R_p)\leq  x+2\iota\Big\} + \frac{1}{p^{t}}
 \leq  \textrm{P}\big\{A_p(x+2\iota)\big\} + \frac{1}{p^{t}}.
\eeaa
Combine the two inequalities to get
\begin{align}\lbl{shakings}
 \operatorname{P}\{A_p(x)B_1\cdots B_d\}
\leq \operatorname{P}\big\{A_p(x+2\iota)\big\}\cdot \operatorname{P}\big(
B_1\cdots B_d\big)  + \frac{2}{p^{t}}.
\end{align}
Similarly,
\beaa
&&\mathrm{P}\Big(\frac{S_p}{\sigma_S}\leq x-\iota,\
B_1\cdots B_d\Big)\\
&\leq & \mathrm{P}\Big(\frac{S_p}{\sigma_S}\leq x-\iota,
B_1\cdots B_d, \frac{|R_p|}{\sigma_S}<  \iota\Big)  +\frac{1}{p^{t}} \leq  \mathrm{P}\Big(\frac{S_p}{\sigma_S}\leq x,\ B_1\cdots B_d\Big) +\frac{1}{p^{t}}.
\eeaa
By the independence from \eqref{old_man},
\beaa
 \mathrm{P}\{A_p(x)B_1\cdots B_d\} \geq
\mathrm{P}\Big(\frac{S_p}{\sigma_S}\leq x-\iota\Big)\cdot \mathrm{P}(
B_1\cdots B_d)-\frac{1}{p^{t}}.
\eeaa
Furthermore,
$$
\mathrm{P}\Big(\frac{T_{\operatorname{SUM}}}{\sigma_S}\leq x-2\iota\Big)
 \leq  \mathrm{P}\Big(\frac{T_{\operatorname{SUM}}}{\sigma_S}\leq x-2\iota,\ \frac{|R_p|}{\sigma_S}<  \iota\Big) + \frac{1}{p^{t}}
 \leq  \mathrm{P}\Big(\frac{S_p}{\sigma_S}\leq x-\iota\Big) +\frac{1}{p^{t}},
$$
where the fact $T_{\operatorname{SUM}}=S_p+R_p$ is used again.
Combining the above two inequalities, we get
\beaa
\mathrm{P}\{A_p(x)B_1\cdots B_d\}
\geq  \mathrm{P}\{A_p(x-2\iota)\}\cdot \mathrm{P}(B_1\cdots B_d)-\frac{2}{p^{t}}.
\eeaa
This together with \eqref{shakings} concludes
\begin{align}\lbl{hei_po}
\left|\mathrm{P}\{A_p(x)B_1\cdots B_d\}-\mathrm{P}\{A_p(x)\}\cdot \mathrm{P}(B_1\cdots B_d)\right|
\leq \Delta_{p, \iota}\cdot  \mathrm{P}(B_1\cdots B_d)+\frac{2}{p^{t}}
\end{align}
as $p\geq p_0$, where
\beaa
\Delta_{p, \iota}&\doteq &|\mathrm{P}\{A_p(x)\}-\mathrm{P}\{A_p(x+2\iota)\}| + |\mathrm{P}\{A_p(x)\}-\mathrm{P}\{A_p(x-2\iota)\}|\\
&=& \mathrm{P}\{A_p(x+2\iota)\}-\mathrm{P}\{A_p(x-2\iota)\}
\eeaa
since $\mathrm{P}\{A_p(x)\}$ is increasing in $x\in \mathbb{R}.$
An important observation is that the derivation of \eqref{when_what} is based on three key facts:  inequality \eqref{when_what}, the identity  $T_{\operatorname{SUM}}=S_p+R_p$ and the fact $\bd{U}_t$ and $ \{\varepsilon_{t1}, \cdots, \varepsilon_{td}\}$ are independent from \eqref{old_man}.

Thus, the three corresponding key facts aforementioned also hold  for the quantities related to $\Lambda=\{i_1, \cdots, i_d\}$. Therefore, similar to the derivation of \eqref{hei_po},
we have
\beaa
&&\big|\mathrm{P}\{A_p(x)B_{i_1}\cdots B_{i_d}\}-\mathrm{P}\{A_p(x)\}\cdot \mathrm{P}(B_{i_1}\cdots B_{i_d})\big|\\
& \leq & \Delta_{p, \iota}\cdot  \mathrm{P}(B_{i_1}\cdots B_{i_d})+\frac{2}{p^{t}}
\eeaa
as $p\geq p_0$.   As a result,
\begin{align}\lbl{last_orange}
\zeta(p,d) \doteq& \sum_{1\leq i_1< \cdots < i_{d}\leq p}\big|\mathrm{P}\{A_p(x) B_{i_1}\cdots B_{i_{d}}\} - \mathrm{P}\{A_p(x)\}\cdot \mathrm{P}(B_{i_1}\cdots B_{i_{d}})\big| \nonumber\\
 \leq & \sum_{1\leq i_1< \cdots < i_{d}\leq N}\Big\{\Delta_{p, \iota}\cdot  \mathrm{P}(B_{i_1}\cdots B_{i_{d}})+\frac{2}{p^{t}}\Big\} \nonumber\\
 \leq & \Delta_{p, \iota}\cdot H(d, p)+ \binom{p}{d}\cdot \frac{2}{p^{t}},
\end{align}
where
\beaa
H(d, N)\doteq \sum_{1\leq i_1< \cdots < i_{d}\leq p}\mathrm{P}(B_{i_1}\cdots B_{i_{d}}).
\eeaa
In the following, we will show $\lim_{\iota\downarrow 0}\limsup_{p\to\infty}\Delta_{p, \iota}=0$ and $\limsup_{p\to\infty}H(d, p)<\infty$ for each $d\geq 1$. Assuming these are true, by using  $\binom{p}{d} \leq p^d$ and \eqref{last_orange}, for fixed $d\geq 1$, sending $p\to\infty$ first, then sending $\iota\downarrow 0$, we get $\lim_{p\to \infty}\zeta(p,d)= 0$ for each $d\geq 1$. The proof is then completed.

{\it Step 2: the proofs of ``\,$\lim_{\iota\downarrow 0}\limsup_{p\to\infty}\Delta_{p, \iota}=0$" and ``\,$\limsup_{p\to\infty}H(d, p)<\infty$ for each $d\geq 1$"}.\\
Under Condition (C5), Theorem \ref{thsum} holds  and we have
\begin{align}\lbl{hunshui_bainian}
\frac{T_{\operatorname{SUM}}}{\sigma_S}\to \mathcal{N}(0, 1)\ \mbox{weakly}
\end{align}
as $p\to\infty$ and hence
\begin{align}\lbl{864123}
\Delta_{p, \iota} \to  \Phi(x+2\iota)-\Phi(x-2\iota)
\end{align}
as $p\to\infty$, where $\Phi(x)=\frac{1}{\sqrt{2\pi}}\int_{-\infty}^xe^{-t^2/2}\,dt$.  This implies that  $$\lim_{\iota\downarrow 0}\limsup_{p\to\infty}\Delta_{p, \iota}=0.$$

Second, under the normality assumption, by Conditions (C2) and (C3), Theorem \ref{maxnull} holds. Hence, by identifying ``$H(t, p)$" here as ``$\alpha_t$" in (\ref{at}) for each $t\geq 1$, we obtain
\begin{align}\lbl{Baoxian}
\lim_{p\to\infty}H(d, p)=\frac{1}{d!}\pi^{-d/2}e^{-nx/2}
\end{align}
for each $d\geq 1.$ The proof is finished.  \hfill $\square$

Now, we are ready to present the proof of Theorem \ref{ms}.

\proof By Conditions (C2), (C3) and (C5), Theorems \ref{maxnull} and \ref{thsum}  hold.
By Theorem \ref{thsum},
\begin{align}
\mathrm{P}\Big(\frac{T_{\operatorname{SUM}}}{\sigma_S}\leq x\Big)=\Phi(x) \lbl{gan_doufu}
\end{align}
as $p\to \infty$ for any $x\in \mathbb{R}$, where $\sigma_S=\{2Kn^{-2}\mbox{tr}^2(\bd{\Sigma}^2)\}^{1/2}$ and $\Phi(x)=\frac{1}{\sqrt{2\pi}}\int_{-\infty}^xe^{-t^2/2}\,dt$.
Define $N=Kp^2$ and $Z_{i+(j-1)p+(k-1)p^2}=n^{1/2}\tilde{\rho}_{ij}(k)$, $i,j=1,\cdots,p,~k=1,\cdots,K$.
From Theorem \ref{maxnull}, we have
\begin{align}
\operatorname{P}\big(\max_{1\leq i \leq N}Z_i^2-2\log N +\log\log N \leq y\big) \to
G(y)=\exp\Big(-\frac{1}{\sqrt{\pi}}e^{-y/2}\Big)\  \lbl{jian_jiao}
\end{align}
as $N\to \infty$ for any $y \in \mathbb{R}.$  To show asymptotic independence,  it is enough to prove
$$
\lim_{N\to \infty}\mathrm{P}\Big(\frac{T_{\operatorname{SUM}}}{\sigma_S}\leq x,\ \max_{1\leq i \leq N}Z_i^2-2\log N +\log\log N\leq y\Big)= \Phi(x)\cdot G(y)
$$
for any $x\in \mathbb{R}$ and $y \in \mathbb{R}$. Set
\begin{align}\lbl{whale}
L_N=\max_{1\leq i \leq N}|Z_i|\ \ \mbox{and}\ \ l_N= (2\log N -\log\log N+y)^{1/2},
\end{align}
where the latter one  makes sense for large $N$. Because of  \eqref{gan_doufu}, the above is equivalent to that
\begin{align}\lbl{wealth_no}
\lim_{N\to \infty}\mathrm{P}\Big(\frac{T_{\operatorname{SUM}}}{\sigma_S}\leq x,\ L_N>l_N\Big)= \Phi(x)\cdot \{1-G(y)\}
\end{align}
for any $x\in \mathbb{R}$ and $y \in \mathbb{R}$.  Recalling the notation in Lemma \ref{akxbcvgf}, we have
\begin{align}
A_p=\Big\{\frac{T_{\operatorname{SUM}}}{\sigma_S}\leq x\Big\}\ \ \ \mbox{and}\ \ \ B_{i}=\big\{|Z_i|>l_N\big\} \lbl{tea_red}
\end{align}
for $1\leq i\leq N$. Therefore,
\begin{align}\lbl{abci}
\mathrm{P}\Big(\frac{T_{\operatorname{SUM}}}{\sigma_S}\leq x,\ L_N>l_N\Big)=\mathrm{P}\Big(\bigcup_{i=1}^NA_pB_{i}\Big).
\end{align}
From the inclusion-exclusion principle,
\begin{align}
\mathrm{P}\Big(\bigcup_{i=1}^NA_pB_{i}\Big)  \leq & \sum_{1\leq i_1 \leq N}\mathrm{P}(A_pB_{i_1})-\sum_{1\leq i_1< i_2\leq N}\mathrm{P}(A_pB_{i_1}B_{i_2})+\cdots+\nonumber\\
 & \sum_{1\leq i_1<  \cdots < i_{2k+1}\leq N}\mathrm{P}(A_pB_{i_1}\cdots B_{i_{2k+1}}) \lbl{Upper_bound}
\end{align}
and
\begin{align}
\mathrm{P}\Big(\bigcup_{i=1}^NA_pB_{i}\Big)  \geq & \sum_{1\leq i_1 \leq N}\mathrm{P}(A_pB_{i_1})-\sum_{1\leq i_1< i_2\leq N}\mathrm{P}(A_pB_{i_1}B_{i_2})+\cdots- \nonumber\\
 & \sum_{1\leq i_1<  \cdots < i_{2k}\leq  N}\mathrm{P}(A_pB_{i_1}\cdots B_{i_{2k}})   \lbl{Lower_bound}
\end{align}
for any integer $k\geq 1$. Define
$$
H(N, d)=\sum_{1\leq i_1<  \cdots < i_{d}\leq N}\mathrm{P}(B_{i_1}\cdots B_{i_{d}})
$$
for $d\geq 1$. From \eqref{Baoxian} we know
%
%
\begin{align}\lbl{Maya1}
\lim_{d\to\infty}\limsup_{p\to\infty}H(N, d)=0.
\end{align}
Set
$$
\zeta(N,d)=\sum_{1\leq i_1<  \cdots < i_d\leq N}\Big\{\mathrm{P}(A_pB_{i_1}\cdots B_{i_d}) - \mathrm{P}(A_p)\cdot \mathrm{P}(B_{i_1}\cdots B_{i_d})\Big\}
$$
for $d\geq 1.$ By Lemma \ref{akxbcvgf},
\begin{align}\lbl{back_campus}
\lim_{p\to\infty}\zeta(N,d)=0
\end{align}
for each $d\geq 1$. The assertion \eqref{Upper_bound} implies that
\begin{align}\lbl{639475}
\operatorname{P}\Big(\bigcup_{i=1}^NA_pB_{i}\Big)
 \leq & \mathrm{P}(A_p)\Big\{\sum_{1\leq i_1 \leq N}\mathrm{P}(B_{i_1})-\sum_{1\leq i_1< i_2\leq N}\mathrm{P}(B_{i_1}B_{i_2})+\cdots-  \nonumber\\
& \sum_{1\leq i_1<  \cdots < i_{2k} \leq N}\mathrm{P}(B_{i_1}\cdots B_{i_{2k}})\Big\}+ \sum_{d=1}^{2k}\zeta(N,d) + H(N, 2k+1)  \nonumber\\
\leq & \mathrm{P}(A_p)\cdot P\Big(\bigcup_{i=1}^NB_{i}\Big)+ \sum_{d=1}^{2k}\zeta(N,d) + H(N, 2k+1),
\end{align}
where the inclusion-exclusion formula is used again in the last inequality, that is,
\beaa
\textrm{P}\Big(\bigcup_{i=1}^NB_{i}\Big) &\geq & \sum_{1\leq i_1 \leq N}\mathrm{P}(B_{i_1})-\sum_{1\leq i_1< i_2\leq N}\mathrm{P}(B_{i_1}B_{i_2})\\
& &+\cdots - \sum_{1\leq i_1<  \cdots < i_{2k}\leq N}\mathrm{P}(B_{i_1}\cdots B_{i_{2k}})
\eeaa
for all $k\geq 1$.
 By the definition of $l_N$ and \eqref{jian_jiao},
\beaa
 \mathrm{P}\Big(\bigcup_{i=1}^NB_{i}\Big)=\mathrm{P}\big(L_N>l_N\big)=\mathrm{P}\big(L_N^2-2\log N +\log\log N> y\big) \to 1-G(y)
\eeaa
as $p\to\infty$. By \eqref{gan_doufu}, $\mathrm{P}(A_p)\to \Phi(x)$ as $p\to\infty.$ From \eqref{abci}, \eqref{back_campus} and \eqref{639475}, by fixing $k$ first and sending $p\to \infty$, we obtain that
\beaa
\limsup_{p\to\infty}\mathrm{P}\Big(\frac{T_{\operatorname{SUM}}}{\sigma_S}\leq x,\ L_N>l_N\Big)\leq \Phi(x) \{1-G(y)\} +\lim_{p\to\infty}H(N, 2k+1).
\eeaa
Now, by letting $k\to \infty$ and using \eqref{Maya1} we have
\begin{align}\lbl{vskdnti}
\limsup_{p\to\infty}\mathrm{P}\Big(\frac{T_{\operatorname{SUM}}}{\sigma_S}\leq x,\ L_N>l_N\Big)\leq \Phi(x)  \{1-G(y)\}.
\end{align}
By applying the same argument to \eqref{Lower_bound}, we see that the counterpart of \eqref{639475} becomes
\beaa
\mathrm{P}\Big(\bigcup_{i=1}^NA_pB_{i}\Big)
& \geq & \mathrm{P}(A_p)\Big\{\sum_{1\leq i_1 \leq N}\mathrm{P}(B_{i_1})-\sum_{1\leq i_1< i_2\leq N}\mathrm{P}(B_{i_1}B_{i_2})+\cdots + \nonumber\\
&& \sum_{1\leq i_1<  \cdots < i_{2k-1}\leq N}\mathrm{P}(B_{i_1}\cdots B_{i_{2k-1}})\Big\} + \sum_{d=1}^{2k-1}\zeta(N,d) - H(N, 2k)  \nonumber\\
&\geq & \mathrm{P}(A_p)\cdot \mathrm{P}\Big(\bigcup_{i=1}^NB_{i}\Big) + \sum_{d=1}^{2k-1}\zeta(N,d) - H(N, 2k),
\eeaa
where in the last step we use the inclusion-exclusion principle such that
\beaa
\mathrm{P}\Big(\bigcup_{i=1}^NB_{i}\Big) &\leq & \sum_{1\leq i_1 \leq N}\mathrm{P}(B_{i_1})-\sum_{1\leq i_1< i_2\leq N}\mathrm{P}(B_{i_1}B_{i_2})\\
& & +\cdots +  \sum_{1\leq i_1<  \cdots < i_{2k-1}\leq N}\mathrm{P}(B_{i_1}\cdots B_{i_{2k-1}})
\eeaa
for all $k\geq 1$. Review \eqref{abci} and repeat the earlier procedure to see
\begin{align}\label{d40}
\liminf_{p\to\infty}\mathrm{P}\Big(\frac{T_{\operatorname{SUM}}}{\sigma_S}\leq x,\ L_N>l_N\Big)\geq \Phi(x) \{1-G(y)\}
\end{align}
by sending $p\to \infty$ and then sending $k\to\infty.$
Here \eqref{d40} and \eqref{vskdnti} yield \eqref{wealth_no}. The proof is then completed. \hfill$\Box$

{\color{black}
\subsection{Proof of Theorem \ref{ms2}}

\proof By the Slutsky's Theorem, we only need to show that
\begin{align}\label{ai1}
\operatorname{P}\left\{\max_{1\le k\le K}\max_{1\le i,j\le p}n \tilde{\rho}^2_{ij}(k)-2\log(Kp^2)+\log \log
  (Kp^2) \leq x,T_{\operatorname{SUM}} / {\sigma}_{S} \leq
  y\right\} \rightarrow G(x)\cdot \Phi(y),
\end{align}
Define
\begin{align*}
W(\x_1,\cdots,\x_n)=\frac{1}{n(n-1)\sigma_S}\sum_{l=1}^K\underset{t\not=s}{\sum\sum} \x_t^\top \x_s \x_{t+l}^\top \x_{s+l}.
\end{align*}
Hence, $T_{\operatorname{SUM}}/\sigma_S=W(\bmv_1,\cdots,\bmv_n)$.

For $z=\left(z_{1}, \ldots, z_{p}\right)^{\prime} \in \mathbb{R}^{p}$, consider the function
$$
F_{\beta}(z)\doteq \beta^{-1} \log \left(\sum_{j=1}^{p} \exp \left(\beta z_{j}\right)\right),
$$
where $\beta>0$ is the smoothing parameter that controls the level of approximation. An elementary calculation shows that for all $z \in \mathbb{R}^{p}$,
$$
0 \leq F_{\beta}(z)-\max _{1 \leq j \leq p} z_{j} \leq \beta^{-1} \log p.
$$
In the following, we define $\beta=n^{1/12}\log p$.
Define
\begin{align*}
V(\x_1,\cdots,\x_n)=\beta^{-1}\log \left(\sum_{k=1}^K\sum_{1\le i,j\le p} \exp\left(\beta n^{1/2}\sigma_i^{-1}\sigma_j^{-1}\left(n^{-1}\sum_{t=1}^{n-k}x_{t+k,i}x_{tj}\right)\right)\right).
\end{align*}
Then, $V(\bmv_1,\cdots,\bmv_n)=\beta^{-1} \log \left(\sum_{k=1}^K\sum_{1\le i,j\le p} \exp \left(\beta n^{1/2} \tilde{\rho}_{ij}(k)\right)\right)$.
 Because $\beta^{-1} \log p=n^{-1/12}\to 0$, we only need to show that
\begin{align}\label{ai12}
\operatorname{P}\left\{V^2(\bmv_1,\cdots,\bmv_n)-2\log(Kp^2)+\log \log
  (Kp^2) \leq x,W(\bmv_1,\cdots,\bmv_n) \leq
  y\right\} \rightarrow G(x)\cdot \Phi(y).
\end{align}
Suppose $\bxi_1,\cdots,\bxi_n $ are independent and identical distributed as $\mathcal{N}(\bm 0,\bms)$ and independent of $(\bmv_1,\cdots,\bmv_n)$. Next, we will show that $(W(\bmv_1,\cdots,\bmv_n),V(\bmv_1,\cdots,\bmv_n))$ has the same limited distribution as $(W(\bxi_1,\cdots,\bxi_n),V(\bxi_1,\cdots,\bxi_n))$.
Then, according to Theorem 6, we will obtain the result.

It is known that a sequence of random variables $\left\{\xi_{n}\right\}_{n=1}^{\infty}$ converges weakly to a random variable $\xi$ if and only if for every $f \in \mathscr{C}_{b}^{3}(\mathbb{R}^2)$, $\mathbb{E} f\left(\xi_{n}\right) \rightarrow \mathrm{E} f(\xi)$; see, e.g., \citet{Pollard}, Chapter III, Theorem 12. We use this property to give a metrization of the weak convergence in $\mathbb{R}^2$.


Thus, we only need to show that
\begin{align*}
\mathbb{E}[f(W(\bmv_1,\cdots,\bmv_n),V(\bmv_1,\cdots,\bmv_n))]-\mathbb{E}[f(W(\bxi_1,\cdots,\bxi_n),V(\bxi_1,\cdots,\bxi_n))]\to 0
\end{align*}
for every $f\in \mathscr{C}_{b}^{3}(\mathbb{R}^2)$ as $n,p\to\infty$. Define
\begin{align*}
W_d=W(\bmv_1,\cdots,\bmv_{d-1},\bxi_d,\cdots,\bxi_n),~V_d=V(\bmv_1,\cdots,\bmv_{d-1},\bxi_d,\cdots,\bxi_n).
\end{align*}
We have
\begin{align*}
&\left|\mathbb{E}[f(W(\bmv_1,\cdots,\bmv_n),V(\bmv_1,\cdots,\bmv_n))]-\mathbb{E}[f(W(\bxi_1,\cdots,\bxi_n),V(\bxi_1,\cdots,\bxi_n))]\right|\\
&\le \sum_{d=1}^n\left|\mathbb{E}(f(W_d,V_d))-\mathbb{E}(f(W_{d+1},V_{d+1}))\right|.
\end{align*}
In the following, we only proof the result with $K=1$. For the other fixed integer $K$, the proof are very similar.

Define
\begin{align*}
W_{d,0}=&\frac{1}{n(n-1)\sigma_S}\underset{1\le t\not=s\le d-2}{\sum\sum} \x_t^\top \x_s \x_{t+1}^\top \x_{s+1}+\frac{1}{n(n-1)\sigma_S}\underset{d+1\le t\not=s\le n}{\sum\sum} \x_t^\top \x_s \x_{t+1}^\top \x_{s+1}\\
&+\frac{2}{n(n-1)\sigma_S}\sum_{t=1}^{d-2}\sum_{s=d+1}^n \x_t^\top \x_s \x_{t+1}^\top \x_{s+1},
\end{align*}
which only relies on $\mathcal{F}_d=\sigma\{\bmv_1,\cdots,\bmv_{d-2},\bxi_{d+1},\cdot,\bxi_n\}$.
Hence,
\begin{align*}
W_{d}-W_{d,0}=&\frac{1}{n(n-1)\sigma_S}\sum_{t=1}^{d-2}\bmv_{d-1}^\top \bmv_{t}\bmv_{d}^\top \bmv_{t+1}+\frac{1}{n(n-1)\sigma_S}\sum_{t=d}^{n}\bmv_{d-1}^\top \bxi_{t}\bmv_{d}^\top \bxi_{t+1}\\
&+\frac{1}{n(n-1)\sigma_S}\sum_{t=1}^{d-2}\bxi_{d}^\top \bmv_{t}\bxi_{d+1}^\top \bmv_{t+1}\\
&+\frac{1}{n(n-1)\sigma_S}\sum_{t=d}^{n}\bxi_{d}^\top \bxi_{t}\bxi_{d+1}^\top \bxi_{t+1}+\frac{1}{n(n-1)\sigma_S}\bmv_{d-1}^\top \bxi_d\bmv_{d}^\top \bxi_{d+1},\\
W_{d+1}-W_{d,0}=&\frac{1}{n(n-1)\sigma_S}\sum_{t=1}^{d-2}\bmv_{d-1}^\top \bmv_{t}\bmv_{d}^\top \bmv_{t+1}+\frac{1}{n(n-1)\sigma_S}\sum_{t=d}^{n}\bmv_{d-1}^\top \bxi_{t}\bmv_{d}^\top \bxi_{t+1}\\
&+\frac{1}{n(n-1)\sigma_S}\sum_{t=1}^{d-2}\bmv_{d}^\top \bmv_{t}\bmv_{d+1}^\top \bmv_{t+1}+\frac{1}{n(n-1)\sigma_S}\sum_{t=d}^{n}\bmv_{d}^\top \bxi_{t}\bmv_{d+1}^\top \bxi_{t+1}\\
&+\frac{1}{n(n-1)\sigma_S}\bmv_{d-1}^\top \bmv_d\bmv_{d}^\top \bmv_{d+1}.
\end{align*}
Without loss of generality, we assume that $\sigma_i=1$, $i=1,\cdots,p$.
Define
\begin{align*}
V_{d,0}=\beta^{-1}\log \left(\sum_{1\le i,j\le p} \exp\left(\beta \left(n^{-1/2}\sum_{t=1}^{d-2}\varepsilon_{t+1,i}\varepsilon_{tj}+n^{-1/2}\sum_{t=d+1}^{n-1}\xi_{t+1,i}\xi_{tj}\right)\right)\right),
\end{align*}
which also only relies on $\mathcal{F}_d$. For simplicity, we define $l=i+(j-1)p$ and $\breve \rho^{(d,0)}_l=n^{-1}\sum_{t=1}^{d-2}\varepsilon_{t+1,i}\varepsilon_{tj}+n^{-1}\sum_{t=d+1}^{n-1}\xi_{t+1,i}\xi_{tj}$ for all pairs $(i,j)$. Then,
\begin{align*}
V_{d,0}=\beta^{-1}\log \left(\sum_{l=1}^{p^2} \exp\left(\beta n^{1/2}\breve \rho^{(d,0)}_l\right)\right).
\end{align*}
Similarly, we define define $l=i+(j-1)p$ and
$$
\breve \rho^{(d)}_l=\breve \rho^{(d,0)}_l+n^{-1}\varepsilon_{d-1,i}\varepsilon_{d-2,j}+n^{-1}\varepsilon_{d}\xi_{d-1}+n^{-1}\xi_{d+1}\xi_{d}
$$ for all pairs $(i,j)$. Then,
\begin{align*}
V_{d}=\beta^{-1}\log \left(\sum_{l=1}^{p^2} \exp\left(\beta n^{1/2}\breve \rho^{(d)}_l\right)\right).
\end{align*}

Define $f=f(x,y)$ and $\frac{\partial f}{\partial x}=f_1(x,y)$, $\frac{\partial f}{\partial y}=f_2(x,y)$, $\frac{\partial f^2}{\partial^2 x}=f_{11}(x,y)$, $\frac{\partial f^2}{\partial^2 y}=f_{22}(x,y)$,  $\frac{\partial f^2}{\partial x \partial y}=f_{12}(x,y)$. By Taylor's expansion, we have
\begin{align*}
f(W_d,V_d)-f(W_{d,0},V_{d,0})=&f_1(W_{d,0},V_{d,0})(W_d-W_{d,0})+f_2(W_{d,0},V_{d,0})(V_d-V_{d,0})\\
&+\frac{1}{2}f_{11}(W_{d,0},V_{d,0})(W_d-W_{d,0})^2+\frac{1}{2}f_{22}(W_{d,0},V_{d,0})(V_d-V_{d,0})^2\\
&+\frac{1}{2}f_{12}(W_{d,0},V_{d,0})(W_d-W_{d,0})(V_d-V_{d,0})\\
&+O(|(V_d-V_{d,0})|^3)+O(|(W_d-W_{d,0})|^3)
\end{align*}
and
\begin{align*}
&f(W_{d+1},V_{d+1})-f(W_{d,0},V_{d,0})\\
=&f_1(W_{d,0},V_{d,0})(W_{d+1}-W_{d,0})+f_2(W_{d,0},V_{d,0})(V_{d+1}-V_{d,0})\\
&+\frac{1}{2}f_{11}(W_{d,0},V_{d,0})(W_{d+1}-W_{d,0})^2+\frac{1}{2}f_{22}(W_{d,0},V_{d,0})(V_{d+1}-V_{d,0})^2\\
&+\frac{1}{2}f_{12}(W_{d,0},V_{d,0})(W_{d+1}-W_{d,0})(V_{d+1}-V_{d,0})\\
&+O(|(V_{d+1}-V_{d,0})|^3)+O(|(W_{d+1}-W_{d,0})|^3)
\end{align*}
Because $\mathbb{E}(\bmv_t)=\mathbb{E}(\bxi_t)=\bm 0$ and $\mathbb{E}(\bmv_t\bmv_t^\top)=\mathbb{E}(\bxi_t\bxi_t^\top)$, we can verify that
\begin{align*}
\mathbb{E}(W_{d}-W_{d,0}|\mathcal{F}_d)=\mathbb{E}(W_{d+1}-W_{d,0}|\mathcal{F}_d),~\mathbb{E}((W_{d}-W_{d,0})^2|\mathcal{F}_d)=\mathbb{E}((W_{d+1}-W_{d,0})^2|\mathcal{F}_d).
\end{align*}
Thus,
\begin{align*}
&\mathbb{E}(f_1(W_{d,0},V_{d,0})(W_{d}-W_{d,0}))=\mathbb{E}(f_1(W_{d,0},V_{d,0})(W_{d+1}-W_{d,0})), \\ &\mathbb{E}(f_{11}(W_{d,0},V_{d,0})(W_{d}-W_{d,0})^2)=\mathbb{E}(f_{11}(W_{d,0},V_{d,0})(W_{d+1}-W_{d,0})^2).
\end{align*}

Next, we consider $V_{d}-V_{d,0}$. Define $\bm z_{d,0}=(n^{1/2}\breve \rho_1^{(d,0)},\cdots,n^{1/2}\breve \rho_{p^2}^{(d,0)})^\top$\\ and $\bm z_{d}=(n^{1/2}\breve \rho_1^{(d)},\cdots,n^{1/2}\breve \rho_{p^2}^{(d)})^\top$.   By Taylor's expansion, we have
\begin{align}\label{vd}
V_d-V_{d,0}=&n^{1/2}\sum_{l=1}^{p^2}\partial_l F_\beta(\bm z_{d,0})(\breve \rho_l^{d}-\breve\rho_l^{d,0})+\frac{n}{2}\sum_{l=1}^{p^2}\sum_{k=1}^{p^2}\partial_k\partial_l F_\beta(\bm z_{d,0})(\breve \rho_l^{d}-\breve\rho_l^{d,0})(\breve \rho_k^{d}-\breve\rho_k^{d,0})\nonumber\\
&+\frac{1}{6}n^{3/2}\sum_{l=1}^{p^2}\sum_{k=1}^{p^2}\sum_{q=1}^{p^2}\partial_q\partial_k\partial_l F_\beta(\bm z_{d,0}+\delta(\bm z_d-\bm z_{d,0}))(\breve \rho_l^{d}-\breve\rho_l^{d,0})(\breve \rho_k^{d}-\breve\rho_k^{d,0})(\breve \rho_q^{d}-\breve\rho_q^{d,0}).
\end{align}
By $\mathbb{E}(\bmv_t)=\mathbb{E}(\bxi_t)=\bm 0$ and $\mathbb{E}(\bmv_t\bmv_t^\top)=\mathbb{E}(\bxi_t\bxi_t^\top)$, we can also verify that
\begin{align*}
&\mathbb{E}((\breve \rho_l^{d}-\breve\rho_l^{d,0})|\mathcal{F}_d)=\mathbb{E}((\breve \rho_l^{d+1}-\breve\rho_l^{d,0})|\mathcal{F}_d),\\
&\mathbb{E}((\breve \rho_l^{d}-\breve\rho_l^{d,0})^2|\mathcal{F}_d)=\mathbb{E}((\breve \rho_l^{d+1}-\breve\rho_l^{d,0})^2|\mathcal{F}_d),
\end{align*}
By Lemma A.2 in \citet{2012Gaussian}, we have
\begin{align*}
\left|\sum_{l=1}^{p^2}\sum_{k=1}^{p^2}\sum_{q=1}^{p^2}\partial_q\partial_k\partial_l F_\beta(\bm z_{d,0}+\delta(\bm z_d-\bm z_{d,0}))\right| \le C\beta^2
\end{align*}
for some positive constant $C$. By Condition (C1$'$), if $\bmv_t$ has polynomial-type tails, we have
\begin{align*}
\textrm{P}(\max_{1\le t \le n, 1\le i\le p}|\varepsilon_{it}|>C n^{\frac{1}{6}-\delta})\le np (C n^{\frac{1}{6}-\delta})^{6\gamma_0+6+\epsilon}\mathbb{E}(|\varepsilon_{it}/\sigma_i|^{6\gamma_0+6+\epsilon})\to 0,
\end{align*}
where $0<\delta<\frac{\epsilon}{6(6\gamma_0+6+\epsilon)}$. And for random variables $\xi_{it}\sim \mathcal{N}(0,1)$, we also have $$
\textrm{P}\left(\max_{1\le t\le n,1\le i\le p}|\xi_{it}|>C \log (np)\right)\to 0.$$
Thus, we have
\begin{align*}
&\left|\frac{1}{6}n^{3/2}\sum_{l=1}^{p^2}\sum_{k=1}^{p^2}\sum_{q=1}^{p^2}\partial_q\partial_k\partial_l F_\beta(\bm z_{d,0}+\delta(\bm z_d-\bm z_{d,0}))(\breve \rho_l^{d}-\breve\rho_l^{d,0})(\breve \rho_k^{d}-\breve\rho_k^{d,0})(\breve \rho_q^{d}-\breve\rho_q^{d,0})\right|\\
&\le C\beta^2 n^{-7/6-2\delta}
\end{align*}
as probability tending to one.
Hence, we have
\begin{align*}
\left|\mathbb{E}(f_2(W_{d,0},V_{d,0})(V_d-V_{d,0}))-\mathbb{E}(f_2(W_{d,0},V_{d,0})(V_{d+1}-V_{d,0}))\right|\le  \beta^2 n^{-7/6-2\delta}.
\end{align*}
Similarly, we can show that
\begin{align*}
&\left|\mathbb{E}(f_{22}(W_{d,0},V_{d,0})(V_d-V_{d,0})^2)-\mathbb{E}(f_{22}(W_{d,0},V_{d,0})(V_{d+1}-V_{d,0})^2)\right|\le  \beta^2 n^{-7/6-2\delta},\\
&\left|\mathbb{E}(f_{12}(W_{d,0},V_{d,0})(W_{d}-W_{d,0})(V_{d}-V_{d,0}))
-\mathbb{E}(f_{12}(W_{d,0},V_{d,0})(W_{d+1}-W_{d,0})(V_{d+1}-V_{d,0}))\right|\\&\le  \beta^2 n^{-7/6-2\delta}.
\end{align*}
Hence, we can have
\begin{align*}
&\sum_{d=1}^n\left|\mathbb{E}(f(W_d,V_d))-\mathbb{E}(f(W_{d+1},V_{d+1}))\right|\\
\le &C\beta^2 n^{-1/6-2\delta}+2\sum_{d=1}^n[\mathbb{E}(|(V_{d}-V_{d,0})|^3)+\mathbb{E}(|(W_{d}-W_{d,0})|^3)].
\end{align*}
By (\ref{vd}), we have $\mathbb{E}(|(V_{d}-V_{d,0})|^3)=O(n^{-1-3\delta})$ and
\begin{align*}
\sum_{d=1}^n\mathbb{E}|(W_{d}-W_{d,0})|^3\le \sum_{d=1}^n \{\mathbb{E}((W_{d}-W_{d,0})^4)\}^{3/4}.
\end{align*}
Similar to the proof of Theorem 4, we have $\mathbb{E}((W_{d}-W_{d,0})^4)=O(n^{-2})$. So we have
\begin{align*}
\sum_{d=1}^n\left|\mathbb{E}(f(W_d,V_d))-\mathbb{E}(f(W_{d+1},V_{d+1}))\right|\le C\beta^2 n^{-1/6-2\delta}+Cn^{-3\delta}+Cn^{-1/2}\to 0
\end{align*}
as $n\to \infty$. Then, we obtain the result. If $\bmv_t$ has sub-gaussian-type tails, we can also prove the result by the similar arguments.

}

\subsection{Proof of Theorem 8}
Similar to the proof of Theorem 7, we only need to prove the result under the normality assumption. Without loss of generality, under the assumption of Theorem 8, we assume that
$$
\A_0=\left(
\begin{array}{cc}
  {\bf A}_{011} & {\bf 0} \\
  {\bf 0} & {\A_{022}}
\end{array}
\right),~~
\A_1=\left(
\begin{array}{cc}
  {\A}_{111} & {\bf 0} \\
  {\bf 0} & {\bf 0}
\end{array}
\right).
$$
 Under the alternative hypothesis, we have $\bd{X}_{t1}=\A_{011}\bm y_t+\A_{111}\bm y_{t-1}$, where $\bm y_t=(z_{t1},\cdots,z_{td})^\top$ is independent of $\X_{t2}=\A_{022}(z_{td+1},\cdots,z_{tp})^\top$. As $\bmv_t=(\X_{t1}^\top,\X_{t2}^\top)^\top$,
we can decompose $T_{{\rm SUM}}$  as follows
\begin{align*}
&T_{\operatorname{SUM}}\\
=&\frac{1}{n(n-1)} \underset{t\not=s}{\sum\sum}\bmv_t^\top \bmv_s\bmv_{t+1}^\top \bmv_{s+1}\\
=&\frac{1}{n(n-1)} \underset{t\not=s}{\sum\sum}(\X_{t1}^\top\X_{s1}+\X_{t2}^\top\X_{s2})(\X_{t+1,1}^\top\X_{s+1,1}+\X_{t+1,2}^\top\X_{s+1,2})\\
=&\frac{1}{n(n-1)} \underset{t\not=s}{\sum\sum}\X_{t1}^\top\X_{s1}\X_{t+1,1}^\top\X_{s+1,1}+\frac{1}{n(n-1)} \underset{t\not=s}{\sum\sum}\X_{t1}^\top\X_{s1}\X_{t+1,2}^\top\X_{s+1,2}\\
&+\frac{1}{n(n-1)} \underset{t\not=s}{\sum\sum}\X_{t2}^\top\X_{s2}\X_{t+1,1}^\top\X_{s+1,1}+\frac{1}{n(n-1)} \underset{t\not=s}{\sum\sum}\X_{t2}^\top\X_{s2}\X_{t+1,2}^\top\X_{s+1,2}.
\end{align*}
Similar to the proof of Theorem 5, we have
\begin{align*}
&\sigma_{S1}^{-2}\var\left(\frac{1}{n(n-1)}\underset{t\not=s}{\sum\sum}\X_{t1}^\top\X_{s1}\X_{t+1,1}^\top\X_{s+1,1}\right)=O\left(\frac{d^2}{p^2}\right),\\
&\sigma_{S1}^{-2}\var\left(\frac{1}{n(n-1)} \underset{t\not=s}{\sum\sum}\X_{t1}^\top\X_{s1}\X_{t+1,2}^\top\X_{s+1,2}\right)=O\left(\frac{d}{p}\right),\\
&\sigma_{S1}^{-2}\var\left(\frac{1}{n(n-1)} \underset{t\not=s}{\sum\sum}\X_{t2}^\top\X_{s2}\X_{t+1,1}^\top\X_{s+1,1}\right)=O\left(\frac{d}{p}\right),
\end{align*}
by the condition that the eigenvalues of $\bms$ are all bounded. Thus, we have
\begin{align*}
T_{{\rm SUM}}=&\frac{1}{n(n-1)} \underset{t\not=s}{\sum\sum}\X_{t2}^\top\X_{s2}\X_{t+1,2}^\top\X_{s+1,2}+E(T_{{\rm SUM}})+o_p(\sigma_{S1})\\
\doteq&T_{{\rm SUM}}^{(2)}+E(T_{{\rm SUM}})+o_p(\sigma_{S1}).
\end{align*}
Furthermore, taking the same procedure as the proof of Theorem 1, we have
\begin{align*}
&T_{{\rm MAX}}\\
=&\max_{1\le i,j\le p}|n^{1/2}\tilde{\rho}_{ij}(1)|+o_p(1)\\
=&\max\{\max_{1\le i,j\le d}|n^{1/2}\tilde{\rho}_{ij}(1)|,\max_{1\le i\le d,d+1\le j\le p}|n^{1/2}\tilde{\rho}_{ij}(1)|,\max_{d+1\le i,j\le p}|n^{1/2}\tilde{\rho}_{ij}(1)|\}+o_p(1).
\end{align*}
By the independence between $\X_{t1}$ and $\X_{t2}$, we know that $\max_{1\le i,j\le d}|n^{1/2}\tilde{\rho}_{ij}(1)|$ is independent of $T_{{\rm SUM}}^{(2)}$. Due to Theorem 6, we have $\max_{d+1\le i,j\le p}|n^{1/2}\tilde{\rho}_{ij}(1)|$ is asymptotically independent of $T_{{\rm SUM}}^{(2)}$.
Because $\X_{t1}$ is independent of $\X_{t2}$, we also can prove that $\max_{1\le i\le d,d+1\le j\le p}|n^{1/2}\tilde{\rho}_{ij}(1)|$ is asymptotically independent of $T_{{\rm SUM}}^{(2)}$ by taking the same procedure as Theorem 6.  Thus, we can prove that $T_{\rm MAX}$ is asymptotically independent of $T_{{\rm SUM}}^{(2)}$. By Lemma 7.10 in \cite{FJLLX2021}, we can complete the proof.


\end{document}